\documentclass[fleqn,usenatbib]{mnras}

\usepackage[T1]{fontenc}
\usepackage{ae,aecompl}

\usepackage{dutchcal}

\usepackage{graphicx}	%
\usepackage{amsmath}	%
\usepackage{amssymb}	%
\usepackage{ulem}

\newcommand{\code}[1]{\texttt{#1}}

\newcommand{\mesa}{\code{MESA}}

\newcommand{\ppmstar}{\code{PPMstar}}

\newcommand{\sprn}{\mbox{$s$ process}}

\newcommand{\dhydro}{\ensuremath{D_\mathrm{hydro}}}
\newcommand{\digwhydro}{\ensuremath{D_\mathrm{IGW-hydro}}}
\newcommand{\dconvhydro}{\ensuremath{D_\mathrm{conv-hydro}}}
\newcommand{\digwtherm}{\ensuremath{D_\mathrm{IGW-thermal}}}
\newcommand{\digw}{\ensuremath{D_\mathrm{IGW}}}

\newcommand{\khydro}{\ensuremath{K_\mathrm{hydro}}}
\newcommand{\komega}{$l-\nu$ diagram}

\newcommand{\brunt}{Brunt-V\"ais\"al\"a}

\newcommand{\numberspace}{\ensuremath{\;}}

\newcommand{\unitstyle}[1]{\ensuremath{\mathrm{#1}}}
\newcommand{\hour}{\unitstyle{h}}

\newcommand{\power}[2]{\ensuremath{{#1}^{#2}}}
\newcommand{\natlog}[2]{\ensuremath{#1\times 10^{#2}}} 

\newcommand{\centi}{\unitstyle{c}}
\newcommand{\kilo}{\unitstyle{k}}
\newcommand{\Mega}{\unitstyle{M}}

\newcommand{\meter}{\unitstyle{m}}

\newcommand{\second}{\unitstyle{s}}
\newcommand{\Kelvin}{\unitstyle{K}}
\newcommand{\K}{\Kelvin}  

\newcommand{\cm}{\centi\meter}
\newcommand{\gram}{\unitstyle{g}}

\newcommand{\Msun}{\ensuremath{\unitstyle{M}_\odot}}

\newcommand{\Msunyr}{\Msun\,\power{\yr}{-1}}

\newcommand{\minute}{\unitstyle{min}} 
\newcommand{\yr}{\unitstyle{yr}}        
\newcommand{\km}{\kilo\meter}   
\newcommand{\Mm}{\Mega\meter}   
\newcommand{\Hz}{\unitstyle{Hz}}        

\newcommand{\Hp}{\ensuremath{\unitstyle{H_P}}}
\newcommand{\Hpzero}{\ensuremath{\unitstyle{H_{P0}}}}

\newcommand{\unit}[2]{\ensuremath{#1\numberspace\mathrm{#2}}}

\input{vectors}
\newcommand{\lSect}[1]{{\label{sec:#1}}}
\newcommand{\lFig}[1]{{\label{fig:#1}}}
\newcommand{\lEq}[1]{{\label{eq:#1}}}
\newcommand{\lTab}[1]{{\label{tab:#1}}}

\newcommand{\Tabff}[1]{{\ref{tab:#1}}}
\newcommand{\Tab}[1]{{Table~\Tabff{#1}}}

\newcommand{\pan}[1]{{\textit{#1}}}

\newcommand{\FIGFF}[2]{{\ref{fig:#2}\pan{#1}}}

\newcommand{\FIG}[2]{{Fig.~\FIGFF{#1}{#2}}}
\newcommand{\Fig}[1]{{\FIG{}{#1}}}
\newcommand{\FigTwo}[2]{{\FIGS{}{#1} and \FIGFF{}{#2}}}
\newcommand{\FIGS}[2]{{Figs.~\FIGFF{#1}{#2}}}

\newcommand{\Sectff}[1]{{\ref{sec:#1}}}
\newcommand{\Sect}[1]{{\S\Sectff{#1}}}

\newcommand{\Eqref}[1]{{\ref{eq:#1}}}
\newcommand{\Eqff}[1]{{(\Eqref{#1})}}

\newcommand{\Eq}[1]{{Eq.~\Eqff{#1}}}

\newcommand{\isofont}[1]{{\mathrm{#1}}}
\newcommand{\isomass}[1]{{\ensuremath{\isofont{^{#1}}}}}
\newcommand{\isocharge}[1]{{\ensuremath{\isofont{_{#1}}}}}
\newcommand{\isotope}[3]{{\ensuremath{\isocharge{#1}\isomass{#2}\isofont{#3}}}}

\newcommand{\papertwo}{Paper~II} 
\newcommand{\paperthree}{Paper~III} 
\newcommand{\paperfour}{Paper~IV}

\newcommand{\Mach}{\ensuremath{\mathcal{M}}} 

\usepackage[usenames,dvipsnames]{color}
\usepackage[]{color}
\usepackage{afterpage}

\newcommand{\mindudr}[0]{$\min d|U_\mathrm{t}|/dr$}
\newcommand{\maxdfvdr}[0]{$\max d FV/dr$}
\newcommand{\npeak}[0]{$N^2$-peak}

\title[Hydrodynamic simulations of main-sequence stars I]{3D hydrodynamic simulations of massive main-sequence stars. I. Dynamics and mixing of convection and internal gravity waves}

\author[F. Herwig et al.]{Falk Herwig$^{1,\dagger}$\thanks{E-mail: fherwig@uvic.ca},
 Paul R. Woodward$^{2,\dagger}$,
 Huaqing Mao$^{2,\dagger}$,
 William R. Thompson$^{1}$
\newauthor
 Pavel Denissenkov,$^{1,\dagger}$
 Josh Lau$^{1,\dagger}$,
 Simon Blouin$^{1,\dagger}$,
 Robert Andrassy$^{3,1}$,
\newauthor
 Adam Paul$^{1,\dagger}$\\
$^{1}$Department of Physics \& Astronomy, University of Victoria, Victoria, B.C., V8W 2Y2, Canada\\
$^{2}$LCSE and Department of Physics and Astronomy, University of Minnesota, Minneapolis, MN 55455, USA\\
$^{3}$Heidelberger Institut für Theoretische Studien, 69118 Heidelberg, Germany\\
$^{\dagger}$Joint Institute for Nuclear Astrophysics - Center for the Evolution of the Elements (JINA-CEE)
}

\date{Submitted: July 30, 2022; accepted: July 12, 2023}

\pubyear{2023}

\begin{document}
\label{firstpage}
\pagerange{\pageref{firstpage}--\pageref{lastpage}}
\maketitle
\normalem
\begin{abstract}
We performed 3D hydrodynamic simulations of the inner $\approx 50\%$
radial extent of a \unit{25}{\Msun} star in the early phase of the
main sequence and investigate core convection and internal gravity
waves in the core-envelope boundary region. Simulations for different
grid resolutions and driving luminosities establish scaling relations
to constrain models of mixing for 1D applications. As in previous
works, the turbulent mass entrainment rate extrapolated to nominal
heating is unrealistically high (\unit{\natlog{1.58}{-4}}{\Msunyr}),
which is discussed in terms of the non-equilibrium response of the
simulations to the initial stratification. We measure quantitatively
the effect of mixing due to internal gravity waves excited by core
convection interacting with the boundary in our simulations. The wave
power spectral density as a function of frequency and wavelength
agrees well with the \code{GYRE} eigenmode predictions based on the 1D
spherically averaged radial profile. A diffusion coefficient profile
that reproduces the spherically averaged abundance distribution
evolution is determined for each simulation. Through a combination of
eigenmode analysis and scaling relations it is shown that in the
\npeak\ region, mixing is due to internal gravity waves and follows
the scaling relation $\digwhydro \propto L^{4/3}$ over a $\goa
\unit{2}{\mathrm{dex}}$ range of heating factors. Different
extrapolations of the mixing efficiency down to nominal heating are
discussed. If internal gravity wave mixing is due to
thermally-enhanced shear mixing, an upper limit is $\digw \loa 2$ to
$\unit{\natlog{3}{4}}{cm^2/s}$ at nominal heating in the
\npeak\ region above the convective core.
\end{abstract}


\section{Introduction}
\lSect{intro} The properties of core convection determine the
observational properties of main-sequence stars individually and as a
population and set the stage for all subsequent evolutionary phases of
intermediate mass and massive stars. Convective boundary mixing
(CBM)\footnote{Following the reasoning of \citet{Denissenkov:2012cu}
the broad term \emph{convective boundary mixing} is meant to include a
wide range of mixing processes at a deep-interior convective boundary
irrespective of physical origin, such as overshooting, penetration, or
entrainment in rapidly evolving convection zones.} infuences the
main-sequence lifetime and the internal stratification for later
evolutionary phases. It has been calibrated in 1D stellar evolution
models by comparing model predictions with the observed width of the
main sequence either from photometry or spectroscopy
\citep[e.g.][]{Schaller:1992vq,KozhurinaPlatais:1997if} or from
eclipsing binaries
\citep[e.g.][]{Stancliffe:2015fq,Claret:2019ez,Tkachenko2020a}. It is
now also possible to constrain CBM through asteroseismology
observations \citep[e.g.][]{Moravveji:2015kr,noll:21}, and even more
detailed model properties such as the temperature gradient or mixing
in the stable layer may be constrained in the future
\citep{Pedersen:2018ew,Michielsen:2019ht,michielsen:21,Bowman2021c}. For
massive stars, the observed width of the main sequence appears to
require more efficient mixing beyond the convective core compared to
the range of values typically calibrated with the methods mentioned
above \citep{Castro:2014fm,Schneider:2018cq}.

Observations of massive main-sequence stars also show clear
observational evidence of mixing in the stable layers all the way to
the surface. \cite{Venn:2002ih} reported depletion of B and
simultaneous enrichment of N in B-type stars that generally matched
the predictions of rotating stars. However, more recent work has
revealed a picture that appears to be more complicated, such as larger
N enhancement and B depletion than predicted by rotating models
\citep{Mendel:2006iw,Martins:2015fd} and slowly rotating N-enriched
stars \citep{Morel:2008gj,Hunter:2008fb,Dufton:2018kz} in which
rotation-induced mixing predictions seem to fall short and additional
physics processes must be at work. In a careful statistical analysis,
\citet{Aerts:2014je} found that observed stellar rotation rates have
no predictive power regarding the observed N enhancement \citep[see
  also][]{Markova:2018fi}. This adds to the motivation to investigate
and possibly identify and quantify mixing processes that are unrelated
to rotation in the stable layers of massive stars.

One such possible transport mechanism is internal gravity waves (IGW)
\citep{press:81,Talon:2005iu} that would transport species
\citep{GarciaLopez:1991jq,Denissenkov:2003gx} and angular momentum
\citep{Kumar:1999fe}. Analytical approaches such as those mentioned
rely on a number of assumptions, such as the wave-generating mechanism
and power spectrum as well as the fundamental transport physics of
IGWs \citep{Lecoanet:2013hc}. Ultimately, realistic representations of
these complicated fluid-dynamics properties can be revealed by
multi-dimensional hydrodynamic simulations. IGWs have indeed been
observed and analyzed in numerous simulations, for example of
solar-type stars \citep{Dintrans:2005bq,Rogers:2006ks,Alvan:2015gs},
of He-shell flash convection \citep{Herwig:2006gk}, and of O-shell and
core convection
\citep{Meakin:2007dj,Gilet:2013bj,Browning:2004bx}. Based on 2D
simulations, \citet{Rogers:2013fl} investigated the role of IGWs in
transporting angular momentum in massive stars. Predicting
\citep{Aerts:2015jv} and indeed observing
\citep{Bowman:2019ka,Bowman:2019ib,Bowman2020b} oscillations due to
stochastically excited IGWs in massive stars has triggered renewed
efforts to determine the quantitative properties of IGW spectra from
2D \citep{Horst:2020ds} and 3D \citep{Edelmann:2019jh} simulations.

Despite great progress in multi-dimensional convection simulations in
general and of core convection in massive stars specifically, the
computational cost of these simulations is still placing severe
limitations on obtaining quantitative and even qualitative
results. Attempts to determine quantitative mixing efficiencies of
IGWs from simulations are still in their infancy. As far as we are
aware, only \citet{Rogers:2017bd} derived a diffusion coefficient
profile for IGW mixing for the radiative envelope from anelastic 2D
simulations of a $3\Msun$ star based on a tracer-particle
post-processing approach \citep[similar to the approach
  by][]{Freytag:vw,Herwig:2006gk}. However, the actual magnitude could
not be reliably determined. When applied in 1D stellar evolution
calculations, the diffusion coefficients from the 2D simulations had
to be reduced by approximately four orders of magnitude in order to
match asteroseismic observations
\citep{Pedersen:2018ew,Pedersen2021a}.

The computational challenge is indeed substantial and
multi-faceted. In order to explore both convective boundary and IGW
mixing quantitatively, simulations must represent the core and a good
portion of the radiative envelope with sufficiently fine grid
resolution to resolve both convective and wave fluid motions. The
simulations should include the global morphology of the largest
core-convection modes, which requires $4\pi$ 3D domains of the
complete sphere. As the large-scale convective motions approach the
convective boundary, the spatial resolution should be sufficient to
capture the relative shift of spectral power to smaller scales and the
interaction of these convective boundary motions with the radiative
layer above, within a narrow interfacial region. Simulations need to
have sufficient resolution in the stable layer to capture the dominant
wavelength of IGWs \citep{Gilet:2013bj}. Even if simulations are
targeting only the dynamic response to a given thermal state, for
example the radial structure from a stellar evolution model, they need
to cover enough star time, so that any inevitable initial simulation
transients can be excluded from the analysis of a sufficiently long
subsequent quasi-steady state. Another challenge is the likely
discrepancy between a stellar evolution structure and the
thermal-dynamic equilibrium at which the 3D hydrodynamic simulation
would ultimately arrive. The computational demands for sufficient
spatial resolution are very significant. For example, even one 3D
simulation of adiabatic interior convection with ten convective
turn-over times with a heating boost factor ($1000\times$) resulting
in ten times higher convective velocities can take several tens of
millions of core hours \citep{Horst:2020ds}. Even with the most
efficient codes and on the largest available supercomputers, it is
therefore impractical to \emph{just run} such 3D simulations for years
of star times.

The aim of this work is to report the results of our initial set of 3D
hydrodynamic simulations of a $25\Msun$ main-sequence star with the
\code{PPMstar} code. We characterize the flow morphology of core
convection and boundary layers, the mixing processes in the
core-envelope interfacial region, and the excitation and mixing of
IGWs in the stable layers immediately adjacent to the convective core,
based on high-resolution simulations. We establish the behavior of our
simulation results under grid refinement and as a function of heating
factor. In order to establish a baseline for future work, we adopt
idealized input physics by assuming an ideal gas equation of
state. Additional physics ingredients, such as radiation pressure,
radiative diffusion, and rotation will be deferred to a later time at
which we plan to document the differential effect of adding those
physics processes one at a time. In this way we hope ultimately to get
a clearer understanding of the impact of each individual physics
aspect and their mutual interaction. This paper has a companion paper
\cite[][\papertwo]{Thompson:2023a} that focuses on the asteroseismic
properties and predictions of our simulations.

In \Sect{methods}, we present the simulation method and simulation
setup and assumptions, as well as our wave and mixing analysis
techniques. \Sect{general-flow} describes the general flow morphology
and the entrainment process. The convective boundary section
\Sect{conv-bound} includes a discussion on the location of the
boundary and the simultaneous presence of convective and wave motions
in the boundary region. In \Sect{mixing-1D}, we estimate the mixing
efficiency of IGWs and briefly demonstrate possible implications for
physical mixing process of IGWs. The paper closes with conclusions in
\Sect{conclusions}.

\section{Methods and assumptions}
\lSect{methods}
\subsection{Base state for 3D simulations from stellar evolution}
\lSect{stellar_evolution}
\subsubsection{Properties of the 1D model}
\lSect{s.prop_1D_model} One goal of this work is to establish how
convection and wave mixing based on a particular base state
stratification behave in three dimensions. In 3D simulations, the
complete fluid-dynamics equations are solved as opposed to the
one-dimensional picture that we obtain from stellar evolution, in
which convection is approximated by the mixing-length theory (MLT) and
supplemented by a convective boundary mixing model. Obviously, the
representation of the important boundary layer is qualitatively and
conceptually different in the two cases. The results of the 3D
simulations are the dynamic response to the given base state, and in
as much as the 1D base state is not realistic, the 3D dynamic response
will not be either. In fact, since the dynamics of the 3D simulation
are obviously going to be different compared to the picture of the
dynamics of the 1D model according to the MLT, we must expect the 3D
simulation to evolve on a local secular time scale toward a different
dynamic-thermal equilibrium. The diffusive time scale to equilibrate,
say one pressure scale height above the convective core, is of the
order of thousands of convective turn-over times, whereas in these
simulations we cover less than a hundred convective turn-overs (at
$1000\times$ heating factor). Thus we cannot expect to reach a new
thermal equilibrium, which justifies ignoring radiation diffusion
altogether in this initial investigation. By investigating the 3D
dynamic response, we aim to reveal the fundamental dynamic processes
of the configuration taken from the 1D model, which will hopefully
lead in turn to a better understanding of the complex dynamic
interactions and physical processes at the convective boundary, and
ultimately to more realistic CBM models for 1D stellar evolution.

As outlined in the introduction, in many 3D simulations enhanced
heating rates are assumed to accommodate various computational
limitations. As we will show in this paper, in simulations with larger
heating factors, for example $1000$, mass entrainment rates are large
enough and simulations can be followed long enough so that the
initially assumed boundary stratification will be completely
rearranged after an initial phase of a few hundred hours
(cf.\ \Sect{s.bnd_sph_ave}). The detail of the initial stratification
in the boundary region is then obviously no longer important. At lower
heating rates, for example a boost factor of $100$, the original
boundary interface will not be changed very much over months of
simulated star time (cf.\ \Sect{s.bnd_sph_ave}).
\begin{figure}
  \includegraphics[width=\columnwidth]{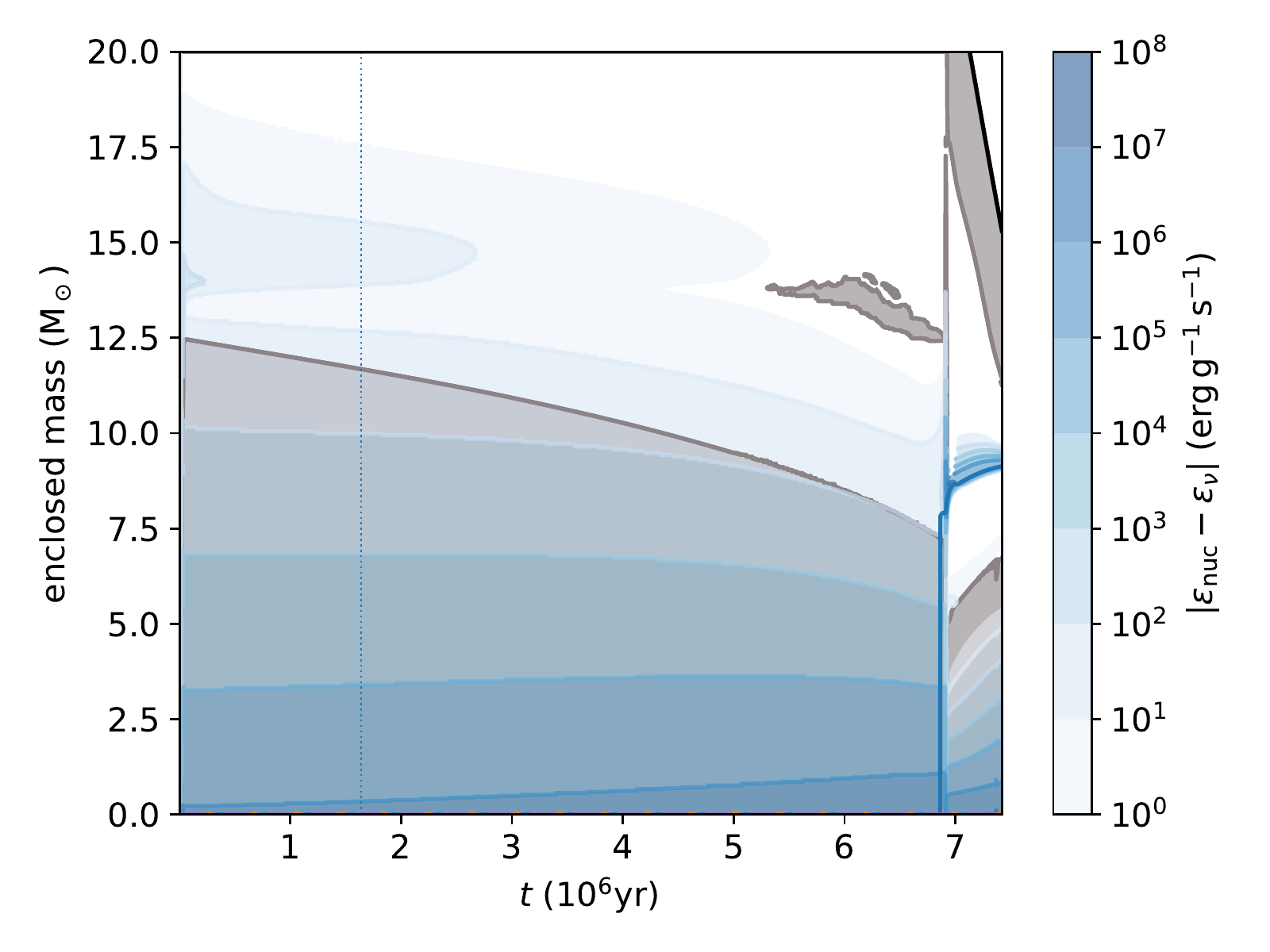}
  \caption{Kippenhahn diagram of the $25\Msun$ \code{MESA} stellar
    evolution model \emph{template} from \citet{Davis:2018jz} during
    H-burning core-convection phase. The vertical dotted line shows
    the time of model 4000 that has been used for the initial
    stratification of the 3D hydro simulations shown in this paper.}
  \lFig{fig:kippenhahn}
\end{figure}         
\begin{figure}
  \includegraphics[width=\columnwidth]{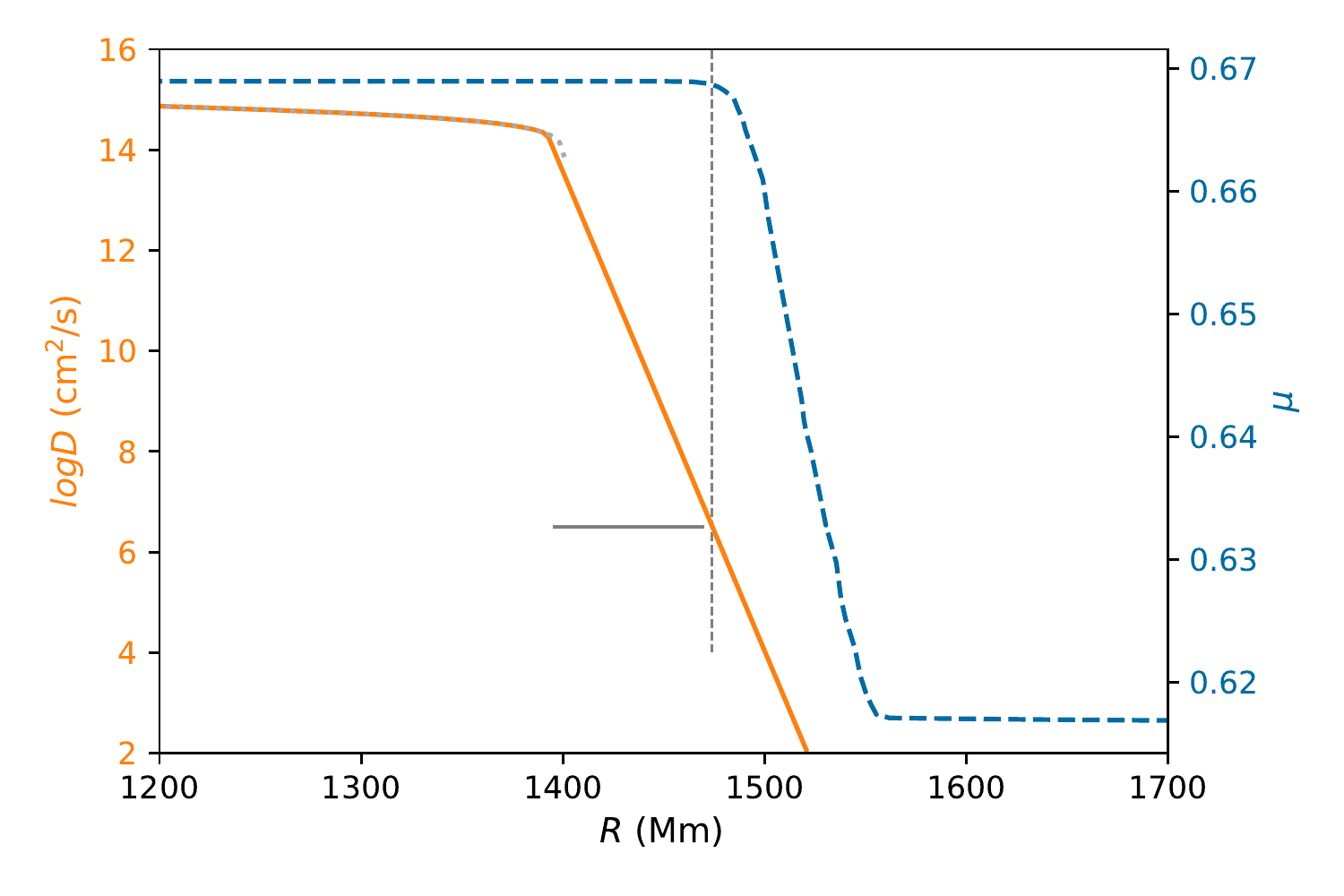}
  \caption{Profile of the diffusion coefficient and mean molecular
    weight $\mu$ of the \code{MESA} model from which the base state
    for the 3D hydrodynamic simulations is derived. Shown is the
    convective boundary region. The solid, horizontal black line
    indicates the range from the convective boundary according to the
    Schwarzschild condition ($r_\mathrm{SB} = 1395\Mm$) to
    $r_\mathrm{SB} + 0.18\Hpzero$ where $\Hpzero=417.8\Mm$ is the
    pressure scale height at $r_\mathrm{SB}$. The vertical dashed line
    marks the approximate radius to which the exponential CBM has
    mixed species essentially completely. The time of the model is
    indicated in \Fig{fig:kippenhahn} with a vertical dotted line.}
  \lFig{fig:1D-R-mu-D-MESA}
\end{figure}         

The base state is constructed from the $25\Msun$ \code{MESA} stellar
evolution model \citep[time step 4000 of the \emph{template} run
  from][]{Davis:2018jz} $\natlog{1.64}{6}\yr$ after the start of H
burning on the zero-age main sequence (\Fig{fig:kippenhahn}). The
central H mass fraction has decreased to $X(\mathrm{H})_\mathrm{c} =
0.606$ from initially $0.706$. In the \code{MESA} simulation H-core
burning ends after $\natlog{6.91}{6}\yr$.

The fact that the mass of the convective core of 1D main-sequence
models is decreasing throughout the H-core burning phase
(\Fig{fig:kippenhahn}) determines how the convective boundary mixing
model shapes the mean molecular weight $\mu$ profile at the boundary
(\Fig{fig:1D-R-mu-D-MESA}). The diffusion coefficient profile reflects
efficient convective species mixing inside the Schwarzschild boundary
and the decrease of the mixing coefficient according to the
exponential boundary mixing model outside the Schwarzschild
boundary. For the H-core burning phase the \emph{template} model
\citet{Davis:2018jz} adopted $f_\mathrm{ov} = 0.022$.  The
effectiveness of convective boundary mixing depends on how fast the
boundary is changing its location. For a given Lagrangian boundary
velocity a certain part of the exponential CBM region is essentially
instantaneously mixed. For a faster moving boundary this layer is
smaller. The relationship between the progression of the convective
boundary and the mixing properties of exponential convective boundary
mixing has already been described in detail by \citet{Herwig:2000ua}
in the context of the formation of the \isotope{}{13}{C} pocket for
the \sprn\ at the bottom of the convective envelope in AGB stars and
the modeling of the third dredge-up phenomenon. In
\Fig{fig:1D-R-mu-D-MESA} the instantaneously mixed layer outside the
formally convective core is indicated by the horizontal solid line and
this region has an extent of $\approx 0.18\Hpzero$. The boundary of
the instantaneously mixed layer is indicated by the vertical dashed
line. The $\mu$ gradient above the dashed line is due to two
processes. In the immediate vicinity of the dashed line the profile is
the result of the exponential mixing, but the bulk of this profile is
due to the receding core convection. It reflects the history of the
core shrinking, which in turn is impacted by the assumed value for
$f_\mathrm{ov}$ in the stellar evolution model. In this work we show
that the region of the $\mu$ gradient hosts a particular set of
internal gravity waves that is associated with mixing. The $\mu$
gradient is a dominant contribution to the $N^2$ profile, where $N$ is
the \brunt\ frequency. It is useful to keep in mind what the origin of
this $\mu$ profile and thus the \npeak\ profile in the underlying 1D
model is.

Another point deserves explanation. According to the exponential
convective boundary mixing model, the region outside the Schwarzschild
boundary obeys the radiative temperature gradient. This assumption
stems from the original work by \citet{Freytag:vw} who based their
analysis on the shallow surface convection of white dwarfs and A-type
stars. Whether or not this assumption is appropriate in the deep
interior is uncertain. Recent idealized simulations \citep{anders:21}
in plane-parallel geometry that ignore the $\mu$ gradient suggest that
if applied to a $25\Msun$ stellar model, a very large penetration zone
forms over timescales corresponding to $60,000$ convective turn-over
times, much longer than the thermal time scale of the envelope of this
$25\Msun$ stellar model ($\approx 5000$ convective turn-over
times). The question of the temperature gradient in the convective
boundary layer may be constrained in the future by asteroseismology
\citep[e.g.\ ][]{Michielsen:2019ht,michielsen:21}. But in any case it
requires including radiation diffusion. We plan to report on such
simulations in the future \cite[][\paperthree]{Mao:2023a} and just
note that the results reported here for the dynamic response to the
adopted \mesa\ base state do not change qualitatively when radiation
pressure and diffusion are added. In our 3D simulations we
\emph{assume} that the entire instantaneously mixed core up to the
location where $\nabla\mu \neq 0$ (indicated by the dashed vertical
line in \Fig{fig:1D-R-mu-D-MESA}) is adiabatically stratified. In
other words, \emph{initially} and contrary to the 1D model the entropy
and $\mu$ gradient are \emph{assumed} to be the same across the
convective boundary except where they have to diverge where the $\mu$
gradient becomes zero outside the core, yet the entropy gradient
remains positive, see \Fig{fig:1Dstratification} just above $\approx
\unit{1550}{\Mm}$. As the simulations progress the $\mu$ and entropy
gradients can decouple depending on the relative strength and physical
processes of species and heat mixing or diffusion, the latter due to
radiation (not included here) or numerical effects.

\subsubsection{Constructing the 3D base state}
\lSect{s.construction_base_state} The radiation pressure fraction is
$\approx 20\%$ throughout the stellar model in the core and the
envelope with the exception of the outermost $\approx 300\Mm$, where
it amounts to $\approx 50\%$. In the 3D simulations presented in this
work, we adopt the ideal gas equation of state and ignore the
radiation pressure. With this assumption we plan to establish a
baseline of idealized simulation results in the same spirit as
\citet{Jones:2017kc}, from which we can establish the impact of adding
additional physics, such as radiation pressure, in the future.
\begin{figure}
  \includegraphics[width=\columnwidth]{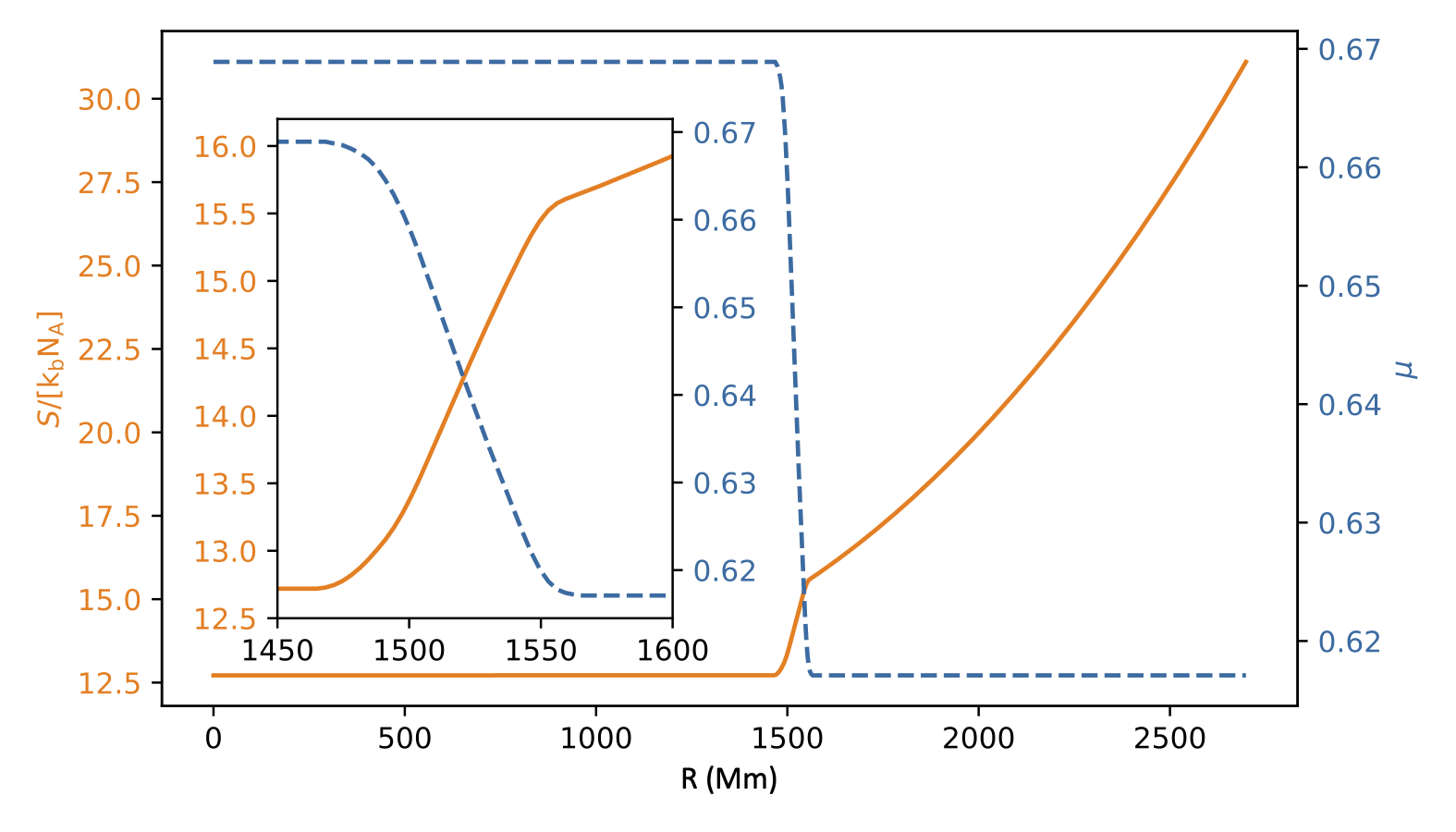}
  \caption{Entropy $S$ and mean molecular weight $\mu$ 1D
    stratifications (cf.\ \Fig{fig:1D-R-mu-D-MESA}) adopted as the
    base state for the 3D hydrodynamic simulations. Entropy is given
    in terms of code units. The \code{PPMstar} code units are length
    $L=10^8\cm$, mass $M=10^{27}\gram$, time $T=\second$, and
    temperature $\Theta=\K$.}  \lFig{fig:1Dstratification}
\end{figure}         

We construct the base state or initial stratification of our
simulations as follows. We use the \mesa\ density, temperature, and
mean molecular weight $\mu$ profile and calculate the entropy profile
according to the assumed ideal gas EOS. We then smooth the S and $\mu$
profiles (see below) and enforce a zero gradient for entropy in the
core. This determines the central pressure. The density and pressure
profiles of the base state then follow from requiring hydrostatic
equilibrium and mass conservation together with the equation of state.

Quantities that represent derivatives, such as the Brunt-V\"ais\"al\"a
frequency $N$, are usually quite noisy when using a default
\code{MESA} inlist file, unless special care is taken to optimize for
smooth stratification profiles \citep[e.g.][]{michielsen:21}. However,
such smoothing measures may reflect stellar physics assumptions that
are uncertain and that the 3D hydrodynamic simulation is supposed to
reveal. Any given 1D stellar evolution profile carries assumptions
about turbulent and IGW mixing, CBM mixing, and how the turbulence and
wave motions are interacting with radiative diffusion. Certain
properties of the 3D simulation may depend sensitively on the
assumptions made to construct the 1D base state from a stellar
evolution model, while others may be more robust. Ultimately an
iterative process involving parameter studies with different physics
assumptions and careful analysis is required to disentangle these
complex interrelations, which are beyond the scope of this paper. In
view of this complication to constructing a base state for the 3D
simulations, we adopt an approach in which we take the \code{MESA}
profile as it is calculated with standard assumptions and then apply a
smoothing procedure in a post-processing step (see
\Sect{s.append-constructbasestate} for details). The resulting base
state definition consists of the central pressure and the entropy $S$
and mean molecular weight $\mu$ profiles (\Fig{fig:1Dstratification}).

Because the radiation pressure contribution is ignored, not all state
variables can match the \code{MESA} profile. This is a common problem
when mapping between 1D and 3D simulations with different equation of
state assumptions. One can only select two variables to match the
other state. Our 3D base state matches the entropy and the $\mu$
profiles of the 1D \mesa\ profiles in the transition from the top of
the convection zone and throughout the envelope because these two
quantities determine the stability of the stratification. There is
still some freedom in selecting the central conditions. In our base
state, the central pressure and density are $20\%$ and $15\%$ smaller
while the temperature is $17\%$ larger than in the \code{MESA}
model. Additional technical details and a comparison with the
\code{MESA} profiles are given in appendix
\Sect{s.append-constructbasestate}.

\subsection{Stellar hydrodynamics simulations}
\lSect{hydro_sims} We use the \ppmstar\ gas dynamics code
\citep{Woodward:2013uf,Jones:2017kc,woodward:19,Andrassy:2020,andrassy:22},
with several important updates. This version solves the conservation
laws in terms of perturbations with respect to a base state. As a
result, the computation can be carried out in 32-bit precision and at
high accuracy. The other update relates to an improvement of how
accurately mixing at the convective boundary is treated. In the past,
our simulations had often focused on the ingestion of small amounts of
material from the stable layer into the convective layer, and great
care had been taken to advect correctly such small amounts of
entrained material. In the main-sequence simulations it is however
equally important to understand how much convective core material is
mixed outward into the stable layer. Therefore, envelope fluid
concentrations close to one are now treated equally accurately as
those close to zero.

The nuclear energy input from H burning that drives the convection is
represented by a constant volume heating with a Gaussian profile in
the radial direction that matches the heating profile in the
\code{MESA} model.

\begin{figure*}
  \includegraphics[width=0.495\textwidth]{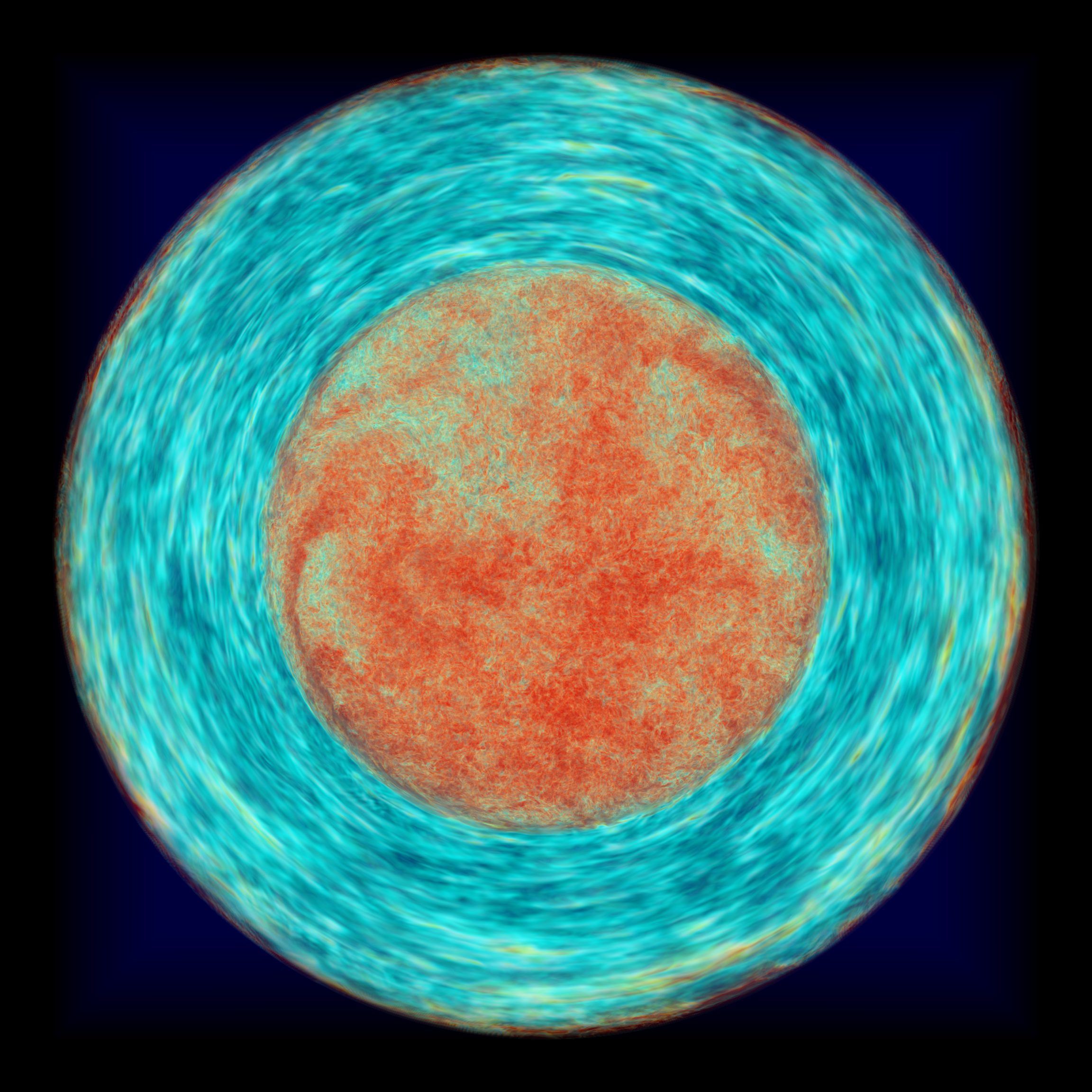}
  \includegraphics[width=0.495\textwidth]{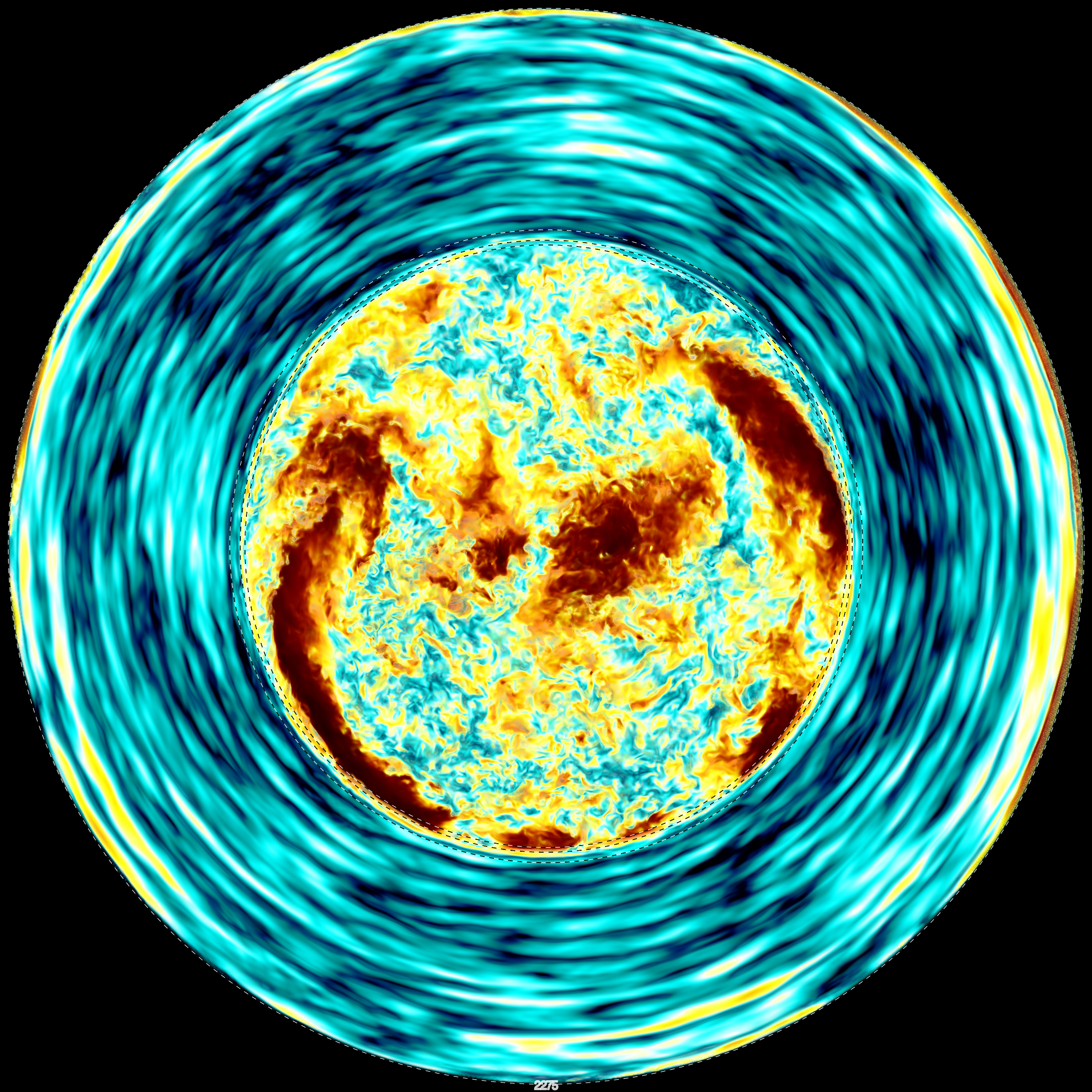}  
  \includegraphics[width=0.495\textwidth]{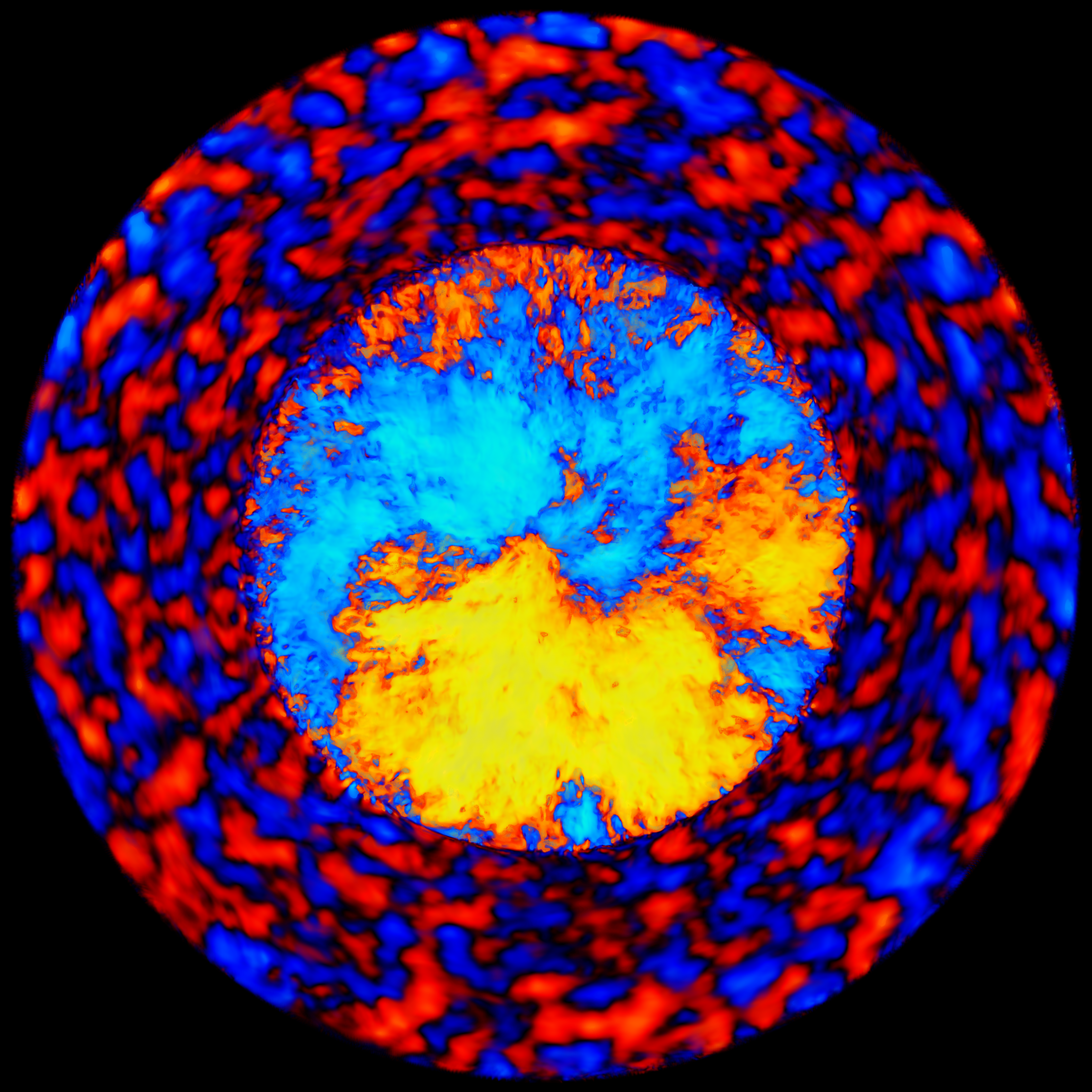}
  \includegraphics[width=0.495\textwidth]{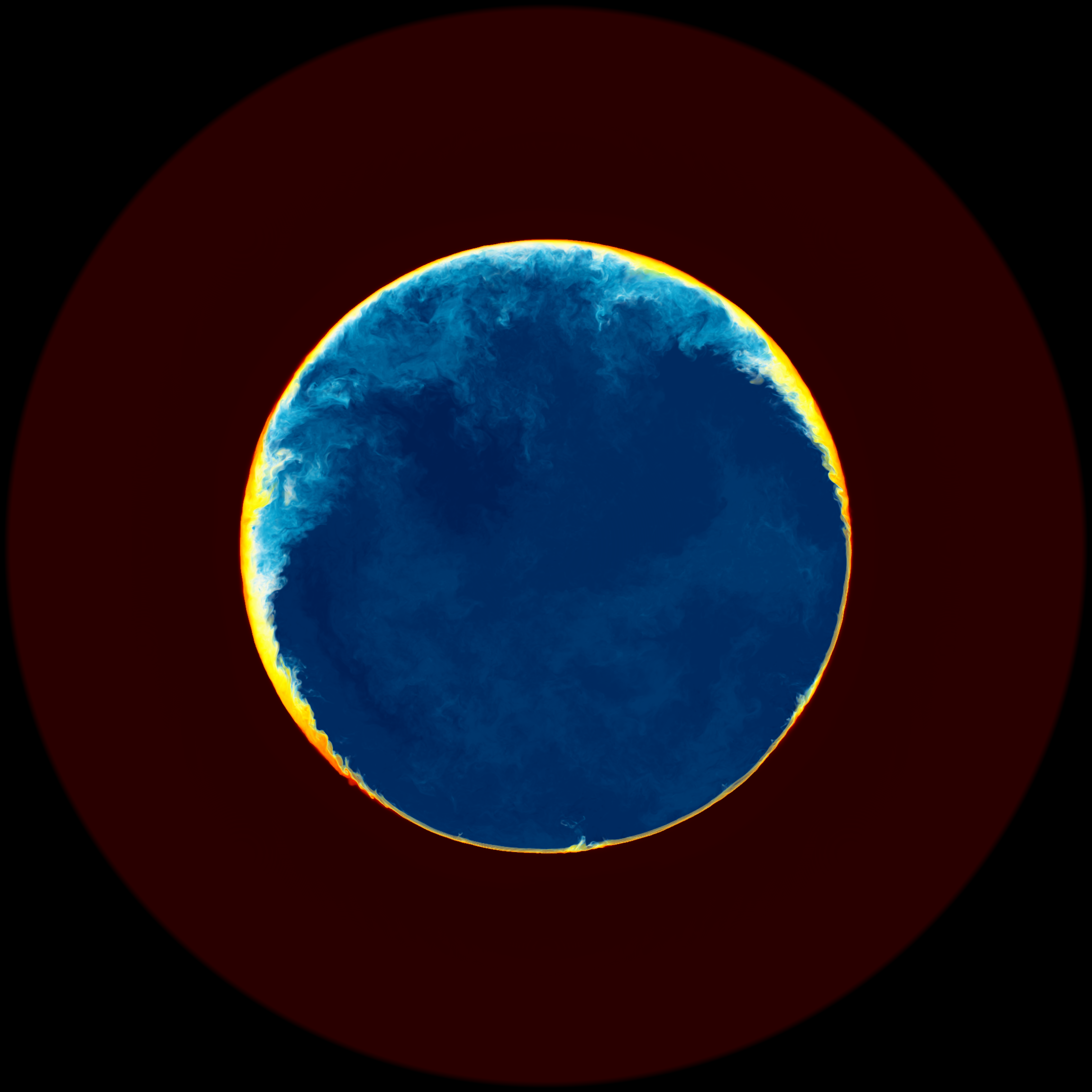}
  \caption{Renderings of vorticity (left top, from red to yellow,
    white, light and darker blue represents decreasing magnitudes of
    vorticity), horizontal velocity magnitude (right top, the same
    color sequence as for vorticity represents decreasing magnitudes
    of the horizontal velocity component), radial velocity magnitude
    (bottom left, blueish colors are inward motions, red-orange colors
    are outward directed flows), and fractional volume of the material
    initially only in the radiative zone of a central slice at
    $t=1615.25\hour$ (dump 2275) from simulation M115 ($1728^3$
    grid). These images have non-linear color and transparency maps
    intended to bring out important flow patterns as clearly as
    possible. The images are not intended to provide a quantitative
    scale, which can instead be derived from line plots such as
    \Fig{fig:M114-velocities_U} and
    \FigTwo{fig:M114-velr-evol}{fig:M114-velt-evol}. Movies and
    full-resolution images of these quantities are available at
    \url{https://www.ppmstar.org}.}
  \lFig{fig:M115-images-HcoreM025Z0}
\end{figure*}
\code{PPMstar} performs its computations in Cartesian coordinates
using a uniform 3-D grid of cubical grid cells.  This structure of the
computation optimizes numerical accuracy for a general fluid flow
problem.  It also gives rise to a simple and highly effective design
in which the computation proceeds in symmetrized sequences of 1-D
passes in the 3 coordinate directions, a procedure called directional
operator splitting. One consequence of our coordinate choice is that
the application of boundary conditions becomes more difficult.  We
generally place the grid boundaries at radii that are well removed
from the action that is under study.  Boundary conditions are
implemented at specific radii, an inner radius (or optionally no inner
radius) and an outer radius.  Because we place these bounding spheres
well away from the region of study, we handle them in ways that make
their implementation easy.  It is important to realize that we do not
attempt to apply our boundary conditions on truly spherical surfaces.
Instead, we approximate the sphere by the nearest set of cubical grid
cell faces.  This means that the bounding sphere is ragged at the
scale of the grid.  Since the grid is made fine enough to faithfully
compute the fluid flow, this raggedness of the bounding spheres is
usually not a concern, especially since they are located well away
from the convection zone. When studying core convection, we have no
inner bounding sphere, and the outer one, given the Cartesian grid, is
better resolved than any spherical surface inside it. At the bounding
sphere we impose a reflecting boundary condition.  This is imposed
using ghost cells that mirror the cells across the bounding surfaces.
This is done in each 1-D pass, and in each such pass the bounding
surface is perpendicular to the direction of the pass, but it is not
perpendicular to the gravitational acceleration vector.  For our
convenience, we therefore smoothly turn off gravity beginning a few
grid cell widths in radius before the bounding sphere is reached.
This allows us to implement a trivially simple boundary condition in
each 1-D pass.  The cost of this approach is that we introduce a very
thin layer in which the gravitational acceleration smoothly drops to
zero right next to the boundary.  In the simulations we have performed
with this code to date, this has caused no noticeable problems.  If
one is only interested in the convection flow and the behavior near
the convective boundary or boundaries, this approach is easily
defended.  If one is also interested in studying the internal gravity
waves that are excited by the convection and that propagate in the
stably stratified regions outside it, the reflection properties of the
gravity waves at the boundaries, if those boundaries are reached by
the waves, could matter. Any impact this approximation may have on our
results would be revealedin the resolution study we typically do on
any problem we work on.

We perform simulations for a range of heating rates and grid
resolutions. Simulations with different grid resolutions allow some
estimate about the numerical convergence properties. In most cases it
is sufficient to determine if simulations are \emph{approaching}
convergence under grid refinement, i.e.\ does the ratio of quantities
of interest become smaller for equal ratios of grid refinement. If
this is the case, a simulation series with different grid sizes gives
an indication of the accuracy of quantitative results.

All of our simulations have a larger driving luminosity compared to
the nominal energy generation rate of the $25\Msun$ stellar
model. This is necessary because at nominal heating the Mach number
\Mach\ of the convective flow is very small. According to the MLT
velocities from the MESA \emph{template} model the average is $\Mach
\approx \natlog{5}{-4}$. Prohibitively small computational grid cells
would be required for accurate simulations. Recall that the
\ppmstar\ code is an explicit gas dynamics code, and although it is
optimized to efficiently perform low-\Mach\ number stellar convection
simulations, there is a natural limit for what can be expected of any
such numerical approach. We vary the heating factor from $10^{1.5}$ to
$10^4$. Such a heating series allows us to extrapolate relevant
quantities to the nominal heating of the simulated star (\Tab{sims}).
\begin{table}
        \centering
        \caption{Summary of simulations used in this paper. Given are
          the run ID, number of grid zones in each dimension, the
          length of the simulated time in hours, the heating factor
          compared to the nominal luminosity of the 1D stellar model
          and the number of convective turnovers $N_\mathrm{conv}$
          computed. For the latter the length of the run from column
          three is divided by the convective turnover time for
          $1000\times$ heating runs of \unit{128}{\hour}
          (cf.\ \Sect{dipole}) scaled according to $U_\mathrm{conv}
          \propto L^{1/3}$ (cf.\ \Sect{digw-dynamic}).}  \lTab{sims}
        \begin{tabular}{lrrrr}  
        \hline
        ID   & grid  & $t_\mathrm{end}/\hour$ & $\log L/L_\mathrm{\star}$ & $N_\mathrm{conv}$ \\  
        \hline
        M109 &   768 &    885  &  4.0 &  14.9 \\
        M118 &  1152 &    905  &  4.0 &  15.2 \\
        M108 &   768 &   1414  &  3.5 &  16.2 \\
        M119 &  1152 &   1414  &  3.5 &  16.2 \\
        M107 &   768 &   7049  &  3.0 &  55.1 \\
        M114 &  1152 &   4189  &  3.0 &  32.7 \\
        M115 &  1728 &   2473  &  3.0 &  19.3 \\
        M111 &  2688 &   1355  &  3.0 &  10.6 \\
        M106 &   768 &   3472  &  2.5 &  18.5 \\
        M100 &  1152 &   1531  &  2.5 &   8.1 \\
        M105 &   768 &   2847  &  2.0 &  10.3 \\
        M116 &  1152 &   2885  &  2.0 &  10.5 \\
        M110 &   768 &   2155  &  1.5 &   5.3 \\
        M117 &  1152 &   3370  &  1.5 &   8.3 \\
	\hline
        \end{tabular}
        \lTab{tab:runs-summary}
\end{table}

As in previous work \citep{Andrassy:2020,stephens:21}, the analysis is
based on three different types of output from the \ppmstar\ code. In
these main-sequence simulations, detailed outputs that we call
\emph{dumps} are written to disk every $42.6\minute$ of star time (in
most of the simulations discussed here) which corresponds to $\approx
400$ time steps on a $768^3$ grid and correspondingly more on the
larger grids. For each dump, radial profiles of spherically averaged
quantities are written out as well as \emph{briquette} data that
contains relevant derived quantities, such as the vorticity,
calculated from the full-resolution grid and then averaged to a 3D
grid that is four times smaller in each Eulerian grid dimension. This
filtered data is of high quality and can usually be analyzed
conveniently in a post-processing step. The third output type are the
3D full resolution byte-sized data cubes used to generate images. A
number of default images are also written out for convenience during
the simulation at each dump.

\subsection{Wave analysis}
\lSect{s.wave_analysis} A key analysis of our simulations is to
determine the oscillation properties of our 3D
simulations. \papertwo\ is dedicated to a comprehensive wave analysis
of these simulations, including predictions of asteroseismic
observations. Here we use the wave analysis to identify fluid motions
due to IGWs in the layers immediately above the convection zone, as
these may be relevant for the convective boundary mixing as discussed
in \Sect{conv_bound_mix}.

In brief, our wave analysis consists of two parts. We determine the
vibrational modes present in the simulations to generate a
frequency-wavenumber diagram by post-processing the 3D briquette (cf.
\Sect{hydro_sims}) velocity data. The \code{GYRE} code
\citep{Townsend:2013ez,Townsend:2018} searches for eigenfrequencies of
standing wave modes for a specified value of $l$ and a range of
$n$. We calculate eigenfrequencies of IGWs for spherical harmonic
degrees $1\leq l \leq 50$ and radial orders $n \in [-1 \dots -20]$
according to the 1D spherically averaged stratification of the 3D
simulation using \code{GYRE}. By comparing the \code{GYRE} predictions
with the spherical harmonics decomposition of the 3D simulation, we
can identify the dominant presence of certain IGWs in the 3D
simulation and ascertain the wave nature of the fluid motions.

First, we decompose each of 2000 dumps into spherical harmonics using
the \code{SHTools} library. We then take the discrete Fourier
transform of each spherical harmonic coefficient after applying a
Hanning window to control spectral leakage. Finally, for each
frequency bin, we compute the spherical harmonic power spectrum
normalized by degree $l$. This results in a grid of power spectral
density as a function of temporal frequency and spherical harmonic
degree $l$, the \komega. It can then be compared to the theoretical
dispersion relations and calculations from \code{GYRE}.

To calculate properties of IGW modes from the radial profiles of the
spherically averaged stratification for a given dump of a 3D
simulation with the stellar oscillation code \code{GYRE}, the radial
profile data from the 3D simulation is transformed to the \mesa\ input
format readable by \code{GYRE} as explained in \papertwo. By tuning
the control parameters of the \code{GYRE} code, we have been able to
determine the spherical harmonic degree $l$ from 1 to 50, finding
g-mode oscillations of the orders $n$ from -1 to -20, f modes, and a
few low-radial order p modes. The results of this analysis are
described in \Sect{conv_bound_mix}. Throughout this paper, $N$ is the
angular \brunt\ frequency.

\subsection{Mixing analysis}
\lSect{mixing-analysis} We determine radial diffusion coefficient
profiles by feeding appropriately averaged radial abundance profiles
into the inversion of the diffusion equation and determine the profile
$D(r)$ that would have been needed in 1D to generate the observed
evolution of abundance profiles from spherically and time-averaged 3D
data as first introduced by \citet{Jones:2017kc}. The method is based
on comparing angle-averaged radial profiles of composition at two
points in time with some time averaging applied around both endpoints
to suppress statistical noise. The transformation from the first
profile to the second is then assumed to be due to a diffusive
process, and the diffusion coefficient is derived by inverting an
appropriate discrete diffusion equation. The most important updates as
compared to \citet{Jones:2017kc} are: (1) we formulate the diffusion
equation using the mass coordinate as an independent variable and (2)
we take the star's spherical geometry into account. The mapping from
Eulerian to mass coordinates becomes important where mixing is slow
and differences between the two composition profiles become dominated
by 1D compression and expansion of the stratification.

We will critically interpret the IGW mixing results obtained in this
way in terms of mixing due to IGWs inducing shear mixing
\citep[e.g.][and references therein]{Denissenkov:2003gx} in the
general framework of shear-induced mixing by small-scale turbulence
\citep[][]{GarciaLopez:1991jq,Zahn:92,Prat:2016coa}. In this picture,
IGWs are generating shear motions reflected in magnitude by the
vorticity acting against the stabilizing effect of positive
\brunt\ frequency. For $\omega \ll N$, the IGW fluid motion is nearly
horizontal with a vertical velocity shear
$$\frac{du_\mathrm{h}}{dr}\approx u_\mathrm{h}k_\mathrm{v}\approx
u_\mathrm{h}k_\mathrm{h}\left(\frac{N}{\omega}\right) \mathrm{\, ,}
$$ where $u_\mathrm{h(v)}$ is the horizontal (vertical) velocity
component, $k_\mathrm{h(v)}$ the horizontal (vertical) wave number,
$\omega$ the wave frequency, and $N$ the buoyancy frequency
\citep{GarciaLopez:1991jq}.

The diffusion coefficient for shear-induced
mixing by small-scale turbulence has been estimated by
\citet{Zahn:1992} as
\begin{equation}
	D_\mathrm{shear} \approx
        \eta \frac{\left(\frac{du_\mathrm{h}}{dr}\right)^2}{N^2}K
        \lEq{eq:D-Ri}
\end{equation} 
 where 
\begin{equation}
   K = \frac{4acT^3}{3\kappa \rho^2 C_\mathrm{P}}    \lEq{eq:K_therm}
\end{equation}
is the thermal diffusivity and $\eta \approx 0.1$. Using the
definition of the Richardson number $Ri$ \citep[Eqn.\,
  8.13,][\Sect{Richardson}]{shu:92}
\begin{equation}
  Ri = \frac{N^2}{(d u_\mathrm{h}/dr)^2}
  \lEq{eq:Ri-def}
\end{equation}
the diffusion coefficient is
\begin{equation}
	D_\mathrm{shear} \approx \eta \frac{K}{Ri} \lEq{eq:D-Ri-shear}
\end{equation} 
This estimate of shear-induced mixing is supported by 3D hydrodynamic
simulations by \citet{Prat:2016coa} for horizontal velocity shear
artificially set up in a box. If the Richardson number in
\Eq{eq:D-Ri-shear} exceeds its critical value $1/4$, the vertical
variation of the horizontal velocity of IGW oscillations is stable for
adiabatic fluid motion. Through radiative heat losses, perturbed fluid
elements can lose some of their entropy memory and shear instability
can develop even when $Ri > 1/4$ \citep{Townsend:1958}, provided that
the viscosity is too small to stabilize it.  \cite{zahn:74} proposed a
new instability criterion that takes into account a finite viscosity,
according to which shear mixing may occur when $Ri\,Pr \la 10^{-3}$
\citep{garaud:21}.  For extremely low Prandtl numbers $Pr = \nu /K \ll
1$ in stellar interiors, e.g. for $Pr\sim 10^{-6}$ in the radiative
envelope of our $25\Msun$ model, such instability and shear-induced
mixing may develop at relatively large values of $Ri\la 10^3$.

The horizontal components of the vorticity of the IGW fluid motion can
be estimated as
\begin{equation} (\nabla\times\mathbf{u})_y = \frac{\partial u_x}{\partial z} -
\frac{\partial u_z}{\partial x} \approx u_\mathrm{h}k_\mathrm{v} -
u_\mathrm{v}k_\mathrm{h} \approx
u_\mathrm{h}k_\mathrm{h}\frac{N}{\omega} \ ,
\lEq{eq:igwvort}
\end{equation}
or $(\nabla\times\mathbf{u})^2_y\approx
(u_\mathrm{h}k_\mathrm{h})^2\left(\frac{N}{\omega}\right)^2\approx
(u_\mathrm{h}k_\mathrm{v})^2$.  A similar estimate
can be obtained for
the x component of the vorticity. If the total vorticity magnitude
$|\nabla\times\mathbf{u}|$ is dominated by the horizontal vorticity
magnitude $|\nabla\times\mathbf{u}|_\mathrm{h}$ (as is the case in our
simulations, \Sect{vorticity}), then $
(u_\mathrm{h}k_\mathrm{v})^2\approx (\nabla\times\mathbf{u})^2$ and
\begin{equation}
D_\mathrm{IGW}\approx
\eta \frac{(\nabla\times\mathbf{u})^2}{N^2}K 
\lEq{eq:digwtherm}
\end{equation}
which implies in this case $ Ri \approx \frac{N^2}{\vortsq}$.  The
factor $\eta$ of order unity reflects the specific type of shear
motion. It has been determined for specific flow morphologies from
hydrodynamic simulations \citep[][]{Prat:2016coa,garaud:17}. The exact
morphology of IGW-induced instabilities is not yet clear, and it
therefore remains to be shown if these calibrations can be applied
directly in this case.  We apply these concepts to interpret IGW
mixing efficiencies measured from our 3D simulations in
\Sect{digw-thermal}.

\section{General flow morphology} 
\lSect{general-flow}
\subsection{The velocity field}
An initial impression of the general flow morphology is provided by
central plane slices of velocity components, vorticity, and
concentration of the fluid in the stable layer shown in
\Fig{fig:M115-images-HcoreM025Z0}. The highly turbulent convective
core with finely granulated three-dimensional distribution of
vorticity is clearly distinguished from the orderly and layered motion
patterns seen in the stable envelope. These regular fluid motions that
are well distinguished in both velocity components and in the
vorticity are dominantly IGWs, as we demonstrate in
detail in \papertwo.

\subsubsection{The dipole dominating the convective core}
\lSect{dipole} The largest scale motion in the convection zone is the
single dipole mode that is best seen from the radial velocity
image. For the dump shown in \Fig{fig:M115-images-HcoreM025Z0}, the
dipole is almost exactly aligned along the north-south direction. It
is well known that the largest scale mode of a convectively stratified
layer fills the largest vertical size of the convection zone. For a
fully convective non-rotating sphere, this mode is the dipole
\citep{Jacobs:1998,Porter:2000b,Kuhlen:2003}. Our core convection
simulations are no different in that regard. Only non-spherical macro
physics processes such as rotation may break up this symmetry
\cite[][\paperfour]{Woodward:2021a}.

\begin{figure}
  \includegraphics[width=\columnwidth]{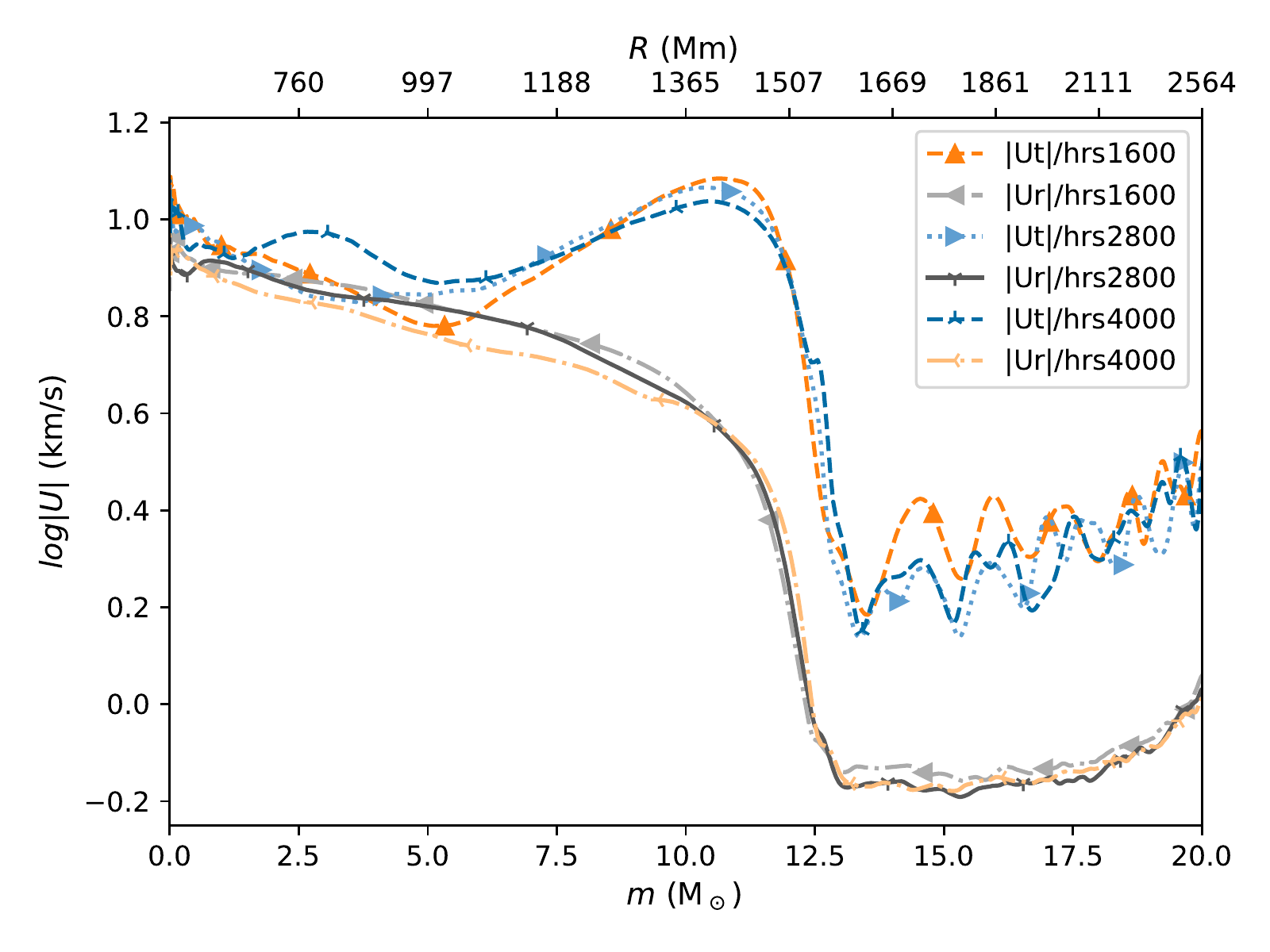}
  \caption{Radial and tangential velocity component of run M114
    ($1000\times$ heating, $1152^3$ grid). For each velocity component
    three different times are shown to demonstrate the typical
    amplitude of fluctuations.} \lFig{fig:M114-velocities_U}
\end{figure}         
The large-scale dipole flow passes right through the centre and
diverges when reaching the boundary in this case near the south
pole. The convective fluids return to the downflow origin at the
antipode located near the north pole in a sweeping tangential flow
along the convective boundary along both the east and west
meridians. The visualization of the tangential velocity magnitude
(\Fig{fig:M115-images-HcoreM025Z0}) resembles the shape of a horseshoe
(dark red indicating the largest tangential velocity magnitudes) that
is aligned with the convective boundary and open to the north. At
about the location of the equator for the flow along the western
meridian and about $30\deg$ further north along the eastern meridian,
the tangential boundary layer flow starts to separate from the
boundary and begins to develop an inward-directed velocity
component. The flow forms a characteristic wedge in these locations
that is also seen in the vorticity image.

\begin{figure}
  \includegraphics[width=\columnwidth]{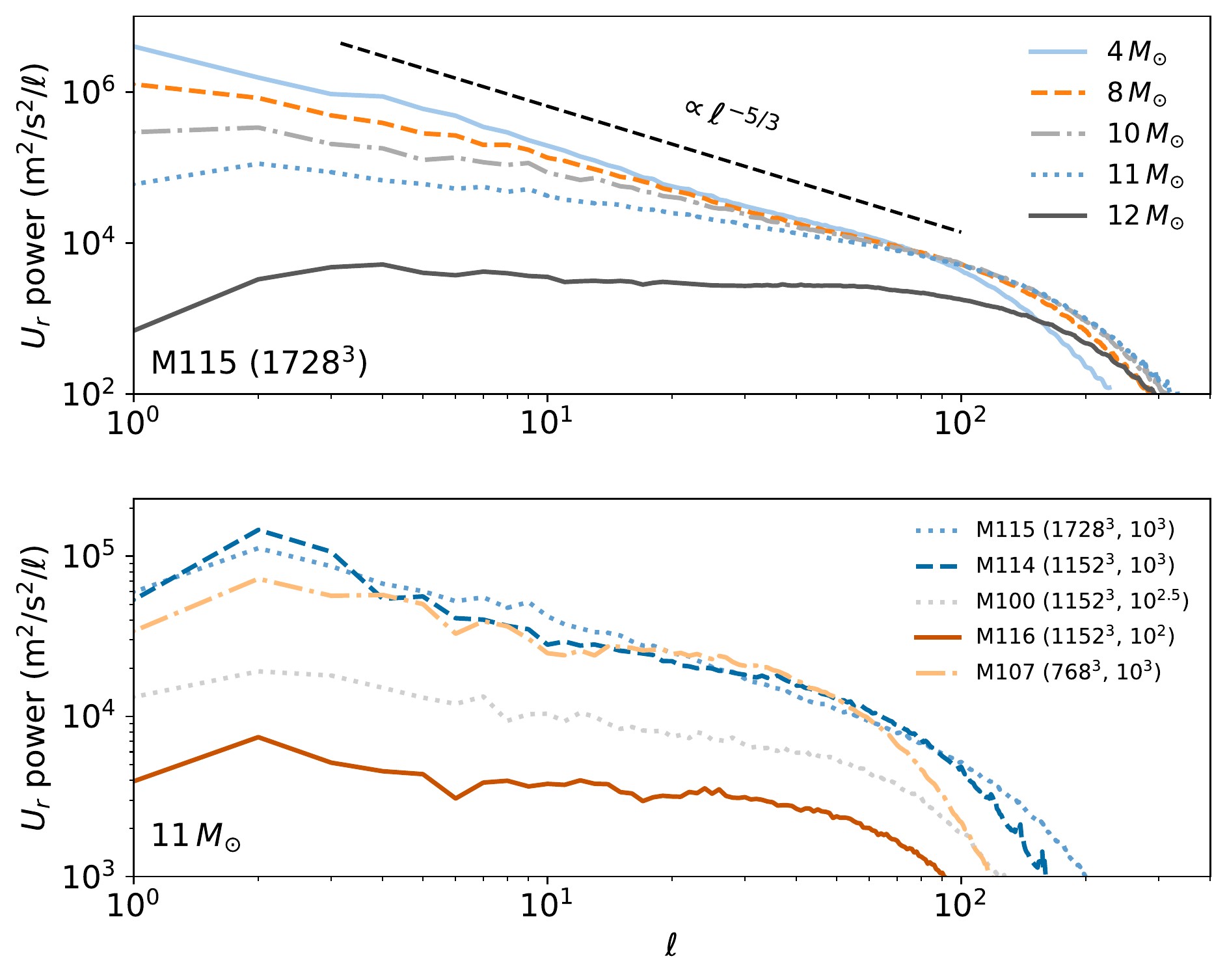}
  \caption{Top panel: Spectra of deep core and core close to boundary
    show how the turbulent spectrum changes from Kolmogorov to flat
    (the legend gives the mass coordinates where each spectrum was
    calculated). The power in the radial velocity component is binned
    as a function of the spherical harmonics angular degree $\ell$,
    and the spectra are averaged over dumps 1900 to 2300.  Bottom
    panel: Spectra at a fixed mass coordinate but with different grid
    resolutions (see legend) as well as for lower heating rates. The
    same dumps as in the top panel are used.}
  \lFig{fig:convective-spectrum}
\end{figure}
In \citet{Woodward:2013uf}, we have identified this boundary layer
separation region as the key feature that facilitates the entrainment
of fluid from the stable layer into the convection zone for the case
of the upper boundary of He-shell convection in a low-mass
star. Boundary-layer separation is a basic phenomenon in fluid
dynamics and described in introductory textbooks
\citep[e.g.][]{Kundu}.  Flow separating from a boundary experiences
additional nonlinear instabilities that resist analytical
descriptions. However, the reason for the separation of the flow from
the boundary is straightforward. The \Mach\ number of the simulated
convection, for example in simulation M114 ($1000\times$ heating
factor, $1152^3$ grid) is $\Mach \approx 0.015$ with maximum values
reaching $0.06$ in some locations near the convective boundary. At
such low \Mach\ numbers, the flow is nearly incompressible. Momentum
conservation dictates that a flow against an opposing pressure
gradient has to develop a perpendicular velocity component. In this
case, the opposing pressure gradient originates from the opposite
tangential boundary-layer flow heading toward the origin of the
downdraft. The outward-directed perpendicular flow direction is
prohibited by the stiff convective boundary and therefore only the
inward-directed perpendicular flow is possible. Because of the low
\Mach\ number, this boundary separation starts already at a large
distance away from the antipode near the north pole, where the centre
of the downdraft is located. As we discuss below, the boundary-layer
separation wedges are where the entrainment of material from the
stable layer into the convective core takes place. The wedges are also
a key mechanism in exciting IGWs (\papertwo).

Finally, for later use it is useful to specify a convective
timescale. If we adopt for the $1000\times$ runs an average convective
velocity of $U_\mathrm{conv} \approx 6.5\km/\second$
(\Fig{fig:M114-velocities_U}) and adopt as convective crossing
distance the diameter of the convective dipole to be $2R_\mathrm{conv}
\approx \unit{3000}{\Mm}$ then the convective time scale is $128\hour$
(or $\approx 180$ dumps).
\begin{figure}
  \includegraphics[width=\columnwidth]{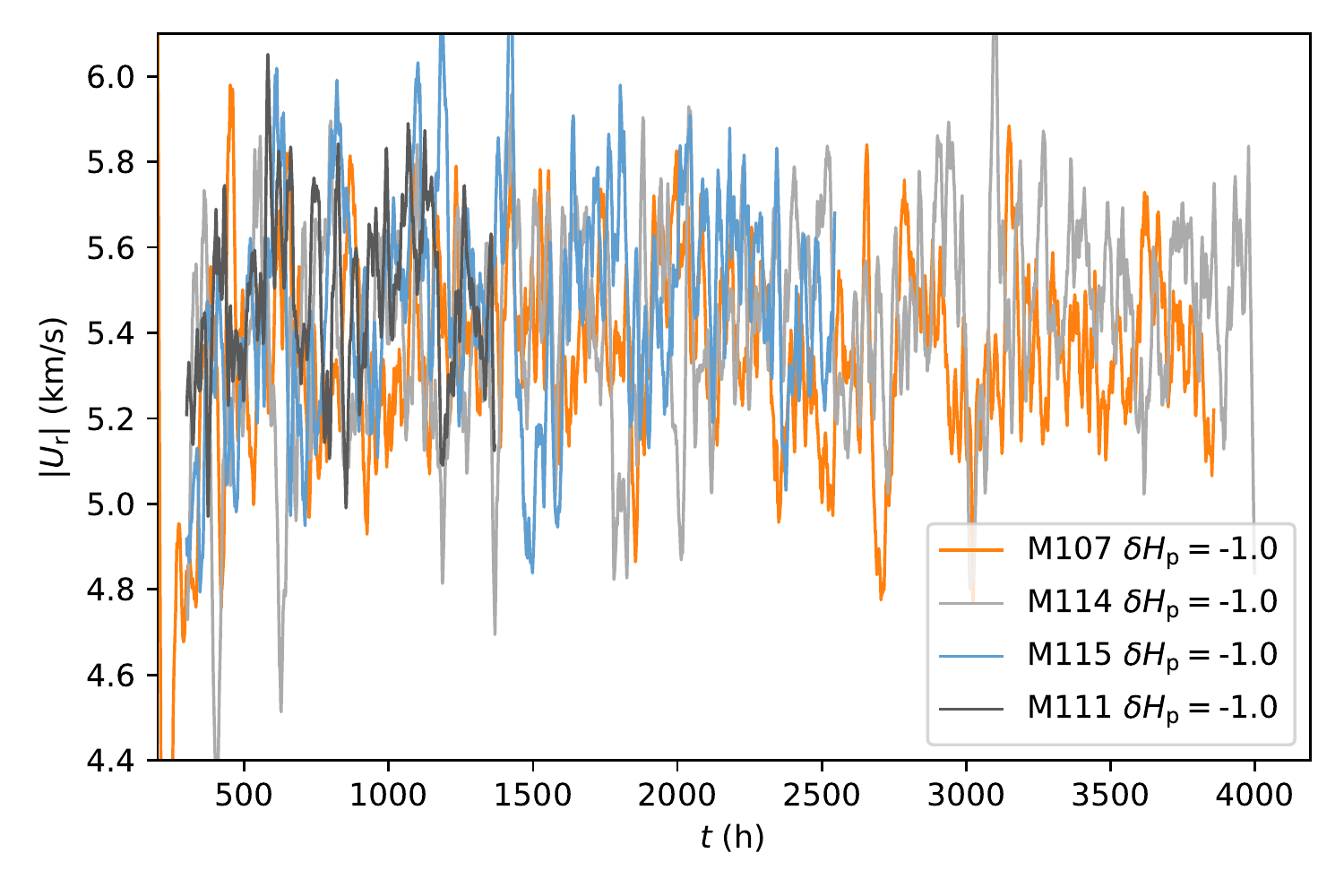}
  \includegraphics[width=\columnwidth]{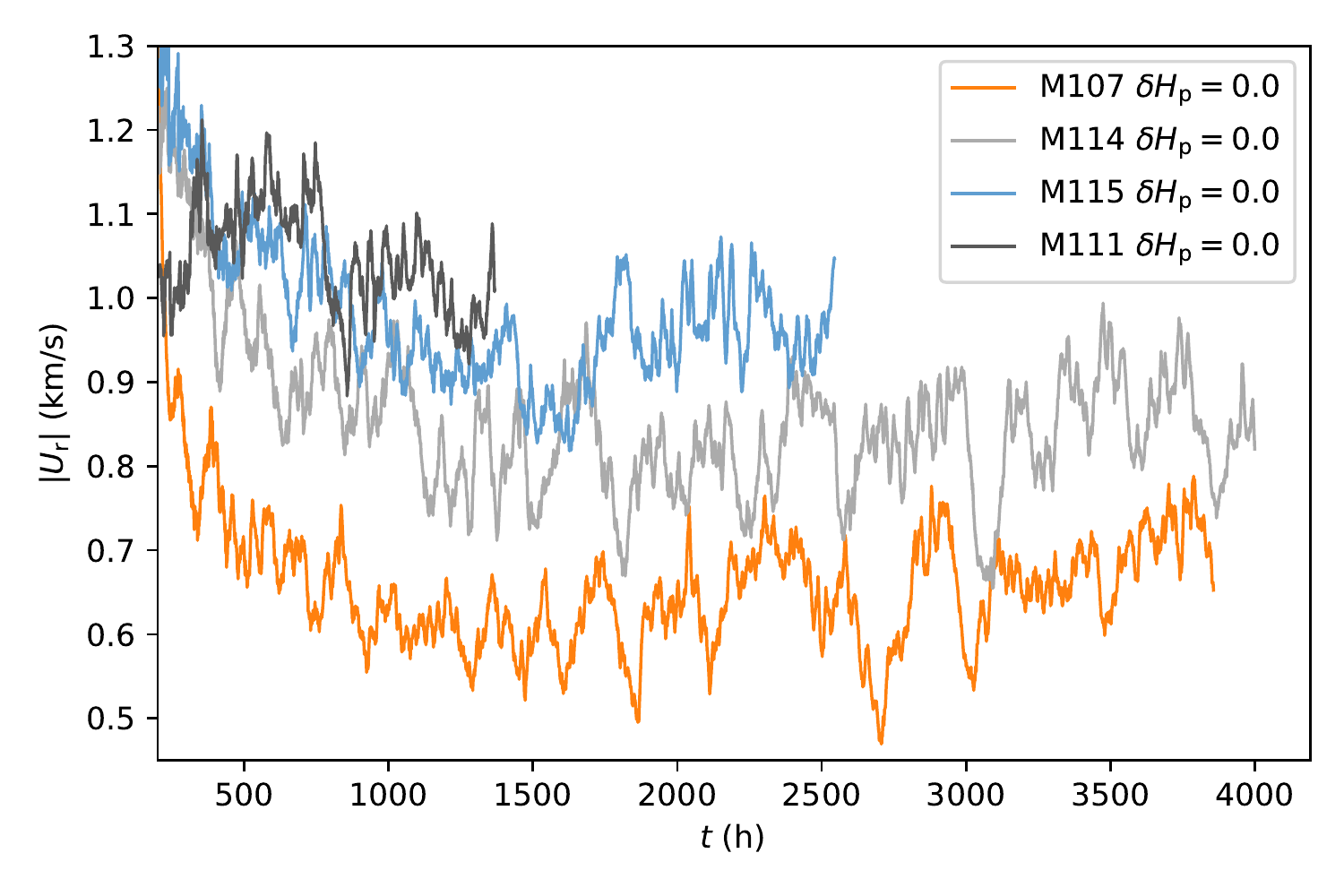}
  \includegraphics[width=\columnwidth]{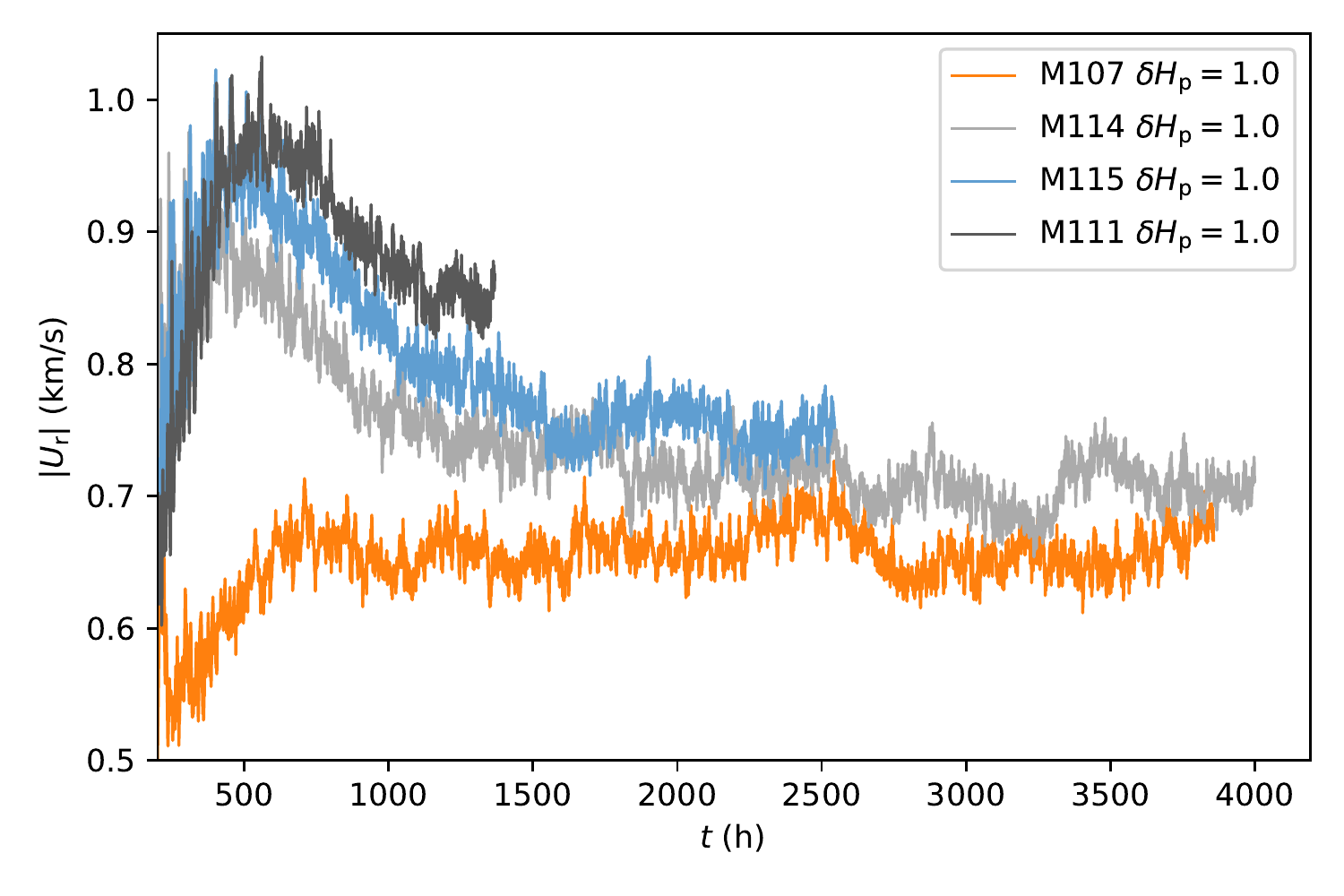}
  \caption{Spherically averaged radial velocity magnitude time
    evolution for grid resolutions $768^3$ (M107), $1152^3$ (M114),
    $1728^3$ (M115), and $2688^3$ (M111, cf.\ \Tab{sims}) at
    $r_\mathrm{N^2-peak} \pm \Hpzero$ (labels \textsf{$\pm \delta
      H_\mathrm{p}$}), where $r_\mathrm{N^2-peak}$ is where $N^2$ has a
    maximum in the boundary region and $\delta H_\mathrm{p}$ is in
    units of \Hpzero. At $r_\mathrm{N^2-peak}$ $\Hpzero = 349\Mm$ or
    $4.6\Msun$ at $t=1662\hour$ of simulation M115.}
  \lFig{fig:M114-velr-evol}
\end{figure}

\subsubsection{Radius dependence of the velocity spectrum in core convection}
The global dipole nature of the flow is also revealed in spherically
averaged radial and tangential velocity component profile plots shown
in \Fig{fig:M114-velocities_U}. The convective boundary is located
approximately at mass coordinate $12.5\Msun$
(cf.\ \Sect{conv_bound_mix} for more details on the determining the
location of the convective boundary more accurately). Near the
convective boundary inside the convection zone ($7.5\Msun$ to
$12.5\Msun$), the tangential velocity component is up to $50\%$ larger
than the radial velocity. This broad peak of the tangential velocity
magnitude represents the sweeping circular boundary layer flow from
where the dipole approaches the boundary toward the antipode where the
downdraft originates. Correspondingly, the radial velocity component
continuously decreases toward the boundary with peak values found in
the central region of the convective core. This asymmetry between
radial and tangential velocity components is similar to what shell
convection shows where the tangential velocity magnitude peaks both
just below the upper boundary and above the lower boundary
\citep{Jones:2017kc,Andrassy:2020,stephens:21}.

\begin{figure}
  \includegraphics[width=\columnwidth]{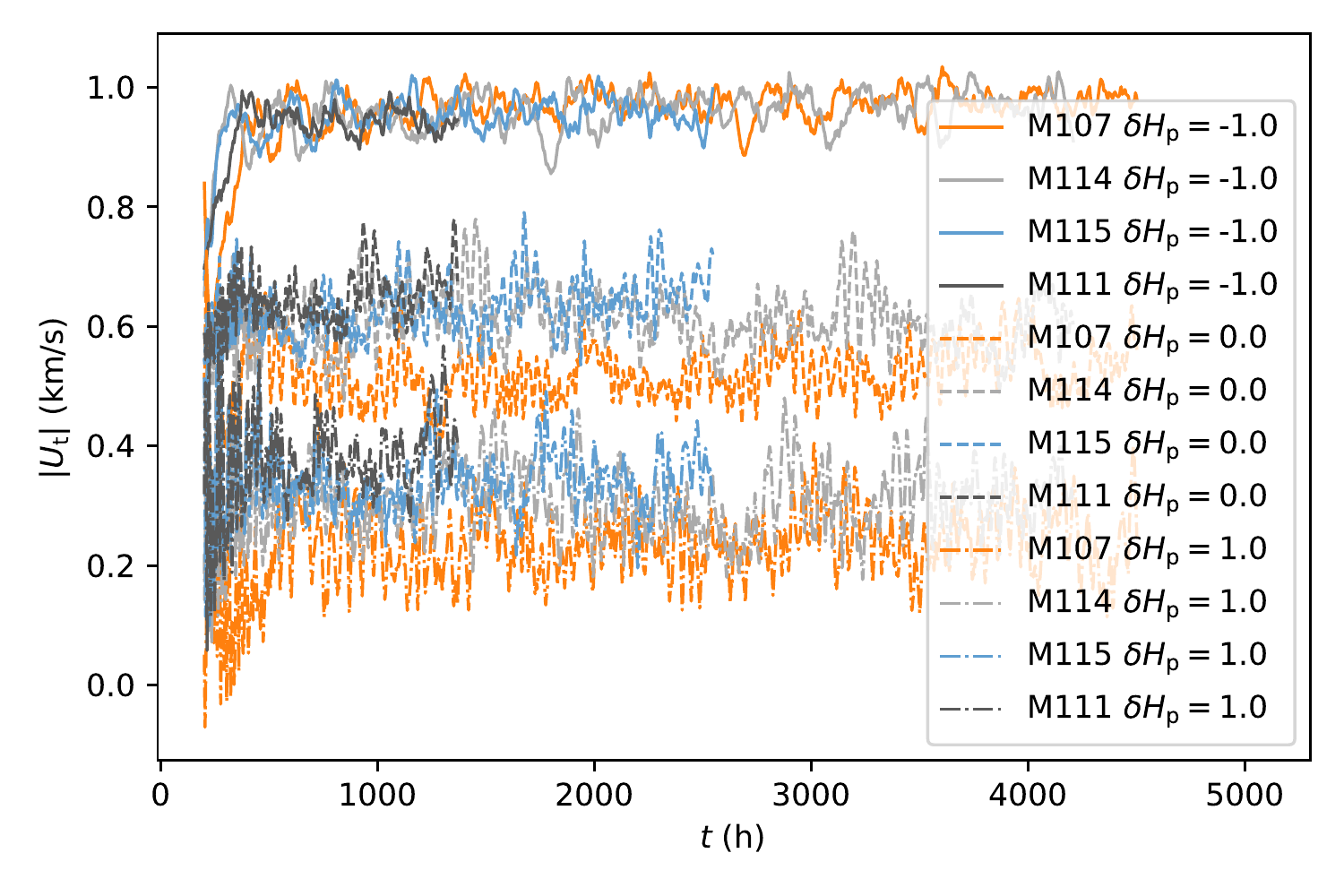}
  \includegraphics[width=\columnwidth]{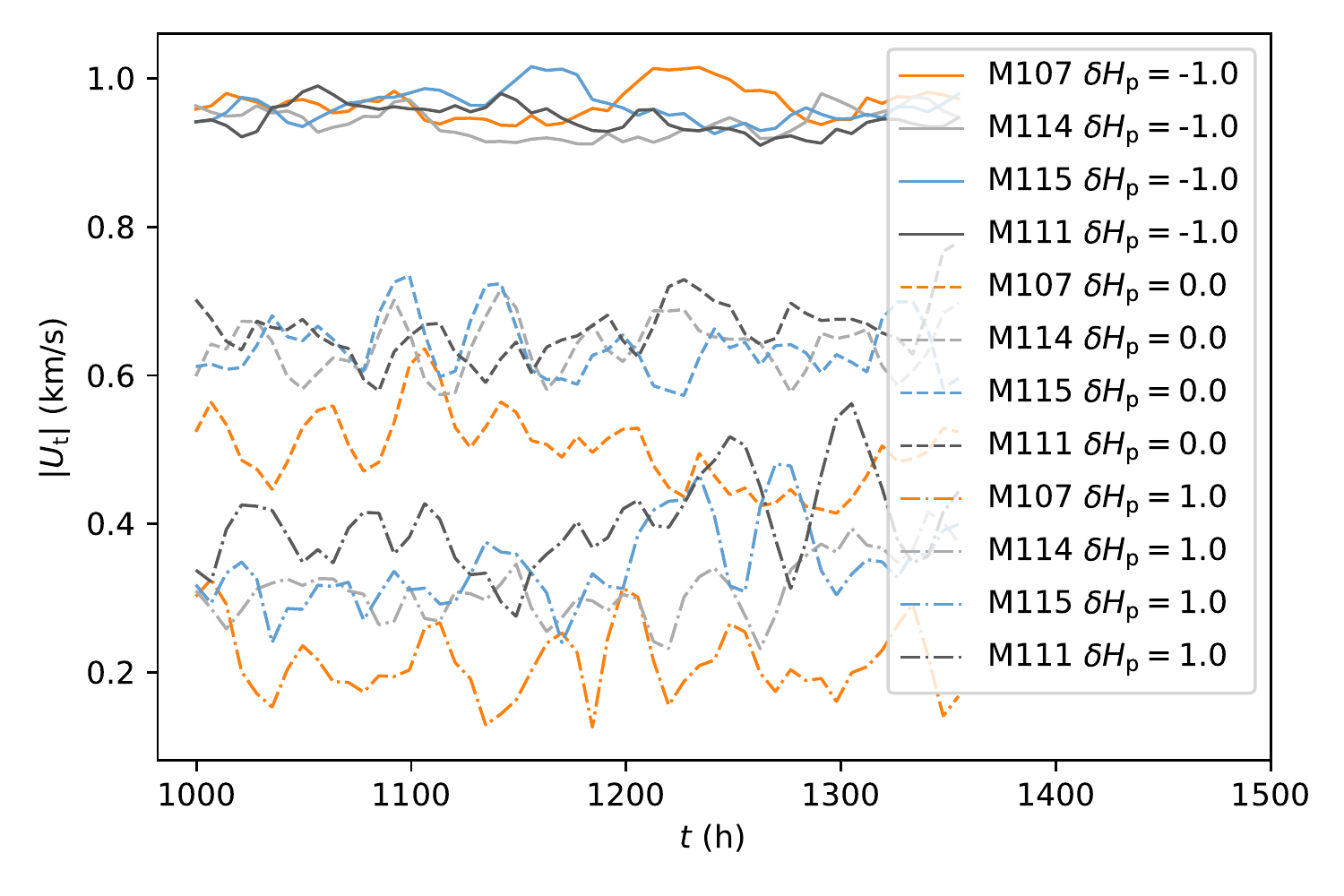}
  \caption{Similar to \Fig{fig:M114-velr-evol} for the tangential
    velocity component, except in decimal logarithms.}
  \lFig{fig:M114-velt-evol}
\end{figure}
The change of the convective flows from radial-component dominated in
the deep core and tangential-component dominated near the boundary has
important consequences for power spectra of convective motions that
break down radial velocity by spherical harmonic degree $l$.  In the
deep core, the scales from the large dipole mode to the the smallest
homogeneous turbulent motions assume a turbulent spectrum in which the
largest scales dominate (\Fig{fig:convective-spectrum}). However,
closer to the boundary, large-scale radial motions are suppressed
compared to the deeper layers, simply because they do not fit into the
smaller remaining vertical distance to the stiff convective
boundary. The spectrum of $U_\mathrm{r}$ becomes flatter and the
relative importance of smaller-scale motions increases. At or just
below the convective boundary the spatial radial velocity spectrum is
indeed flat. Interestingly, the spatial spectrum of the horizontal
velocity component remains $\propto l^{-5/3}$ near the boundary, and
even into the stable layer (cf. \Sect{s.neq1mode}).

\begin{figure*}	
       \includegraphics[width=\columnwidth]{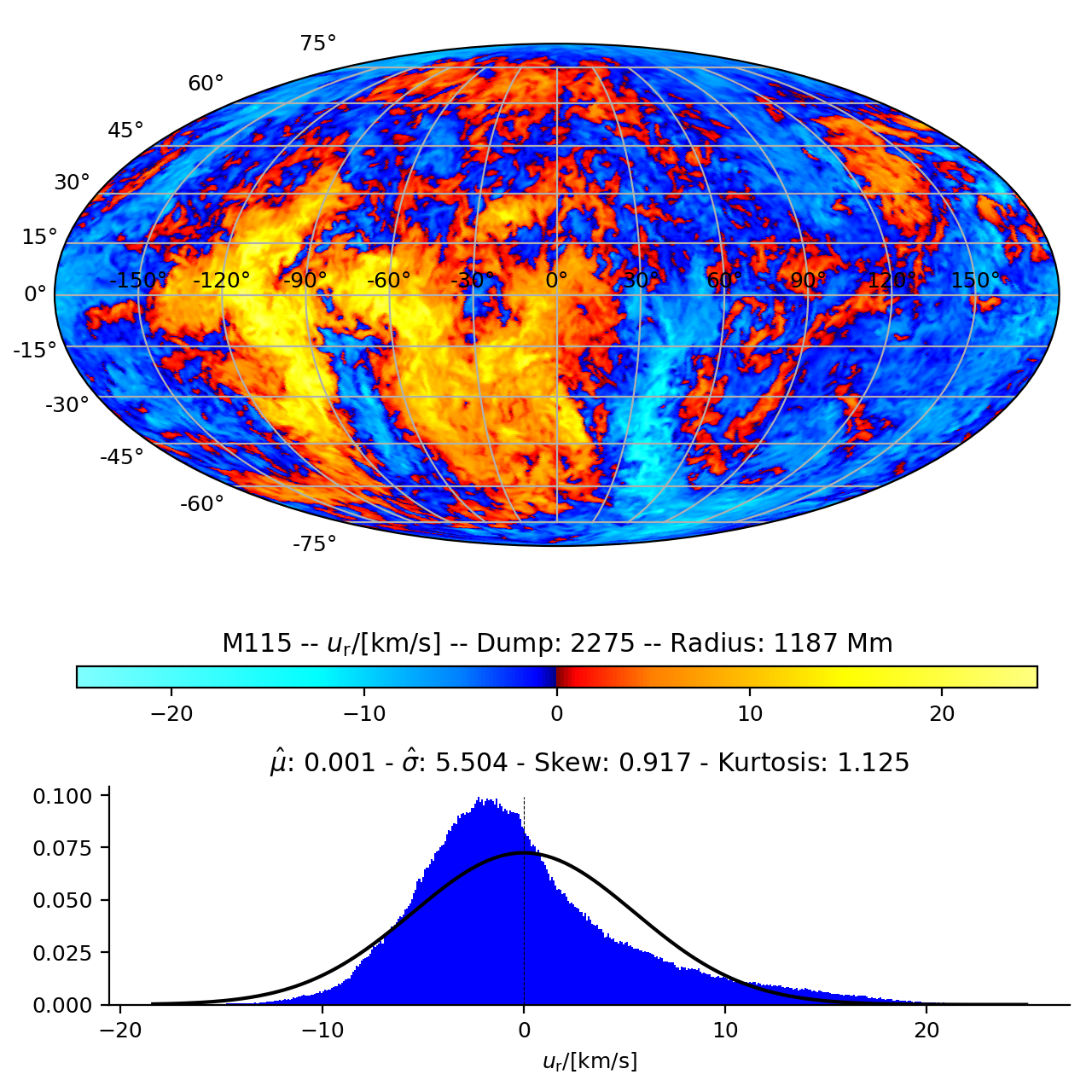}
       \includegraphics[width=\columnwidth]{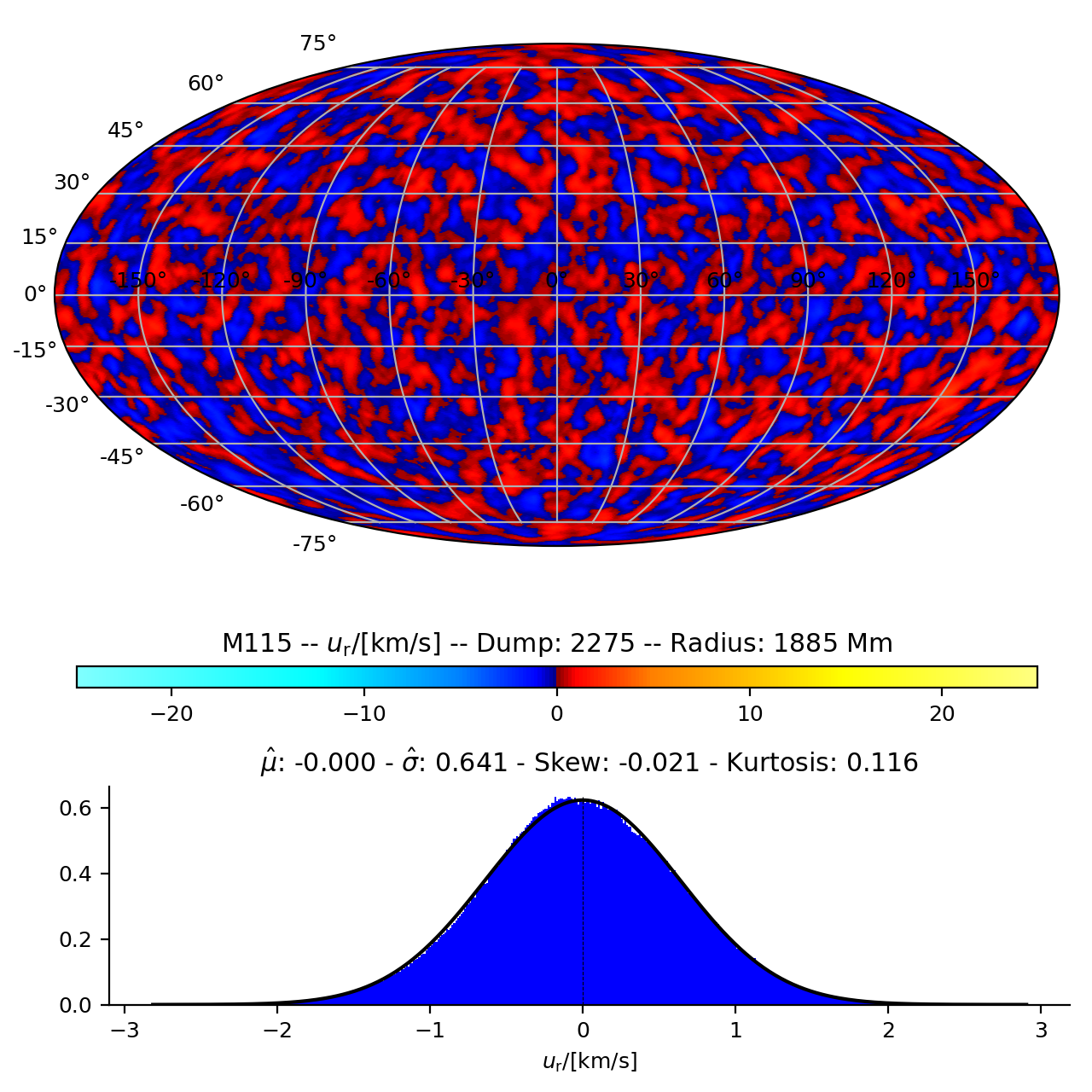}
       \caption{Mollweide projections of the radial velocity component
         and probability distribution function on the $4\pi$ sphere
         for run M115 ($1728^3$-grid). Left: For radius $r =
         r_\mathrm{N^2-peak} - \Hpzero = (1536-349)\Mm=1187\Mm$ in the
         convective core. Right: For $r = r_\mathrm{N^2-peak} +
         \Hpzero = (1536+349)\Mm=1885\Mm$ in the stable envelope.  The
         radius $r_\mathrm{N^2-peak}$ is at or just above the
         convective boundary as discussed in
         \Sect{conv_bound_mix}. The black solid lines are Gaussians
         fitted to the PDFs. Labels provide the values for the moments
         mean, standard deviation, skew, and excess kurtosis.}
       \lFig{fig:M115-Mollweide-conv}
\end{figure*}
This change of the spectrum of scales from the central region of the
convection zone to the boundary has already been noted for the case of
He-shell flash convection in rapidly-accreting white dwarfs by
\citet{stephens:21}, who used this spectral profile information from
the 3D simulations to feed a reduced-dimensionality advective mixing
and nucleosynthesis post-processing scheme. The centre-plane images in
\Fig{fig:M115-images-HcoreM025Z0} reveal that the boundary-layer
separation wedges described above are locations where motions of all
scales including very small scales originate. This spectrum of motions
is also an important ingredient in modeling the excitation of
IGWs. One key result is that the immediate convective region below the
boundary is not dominated by a few low wave numbers, but rather power
of the radial velocity component is almost equally distributed over a
wide range of represented scales.

\subsubsection{Wave motions in the stable envelope}
\lSect{s.waves_envelope} The time evolution over long and short
periods of radial and tangential velocity components in the core,
boundary, and envelope region is shown in
\FigTwo{fig:M114-velr-evol}{fig:M114-velt-evol}. The amplitudes of the
tangential velocity component waves in the envelope are larger by
about a factor $\approx 2.5$ than those of the radial component in
both temporal and spatial dimension (see also
\Fig{fig:M114-velocities_U}), as expected for IGWs.

\begin{figure}     
  \includegraphics[width=\columnwidth]{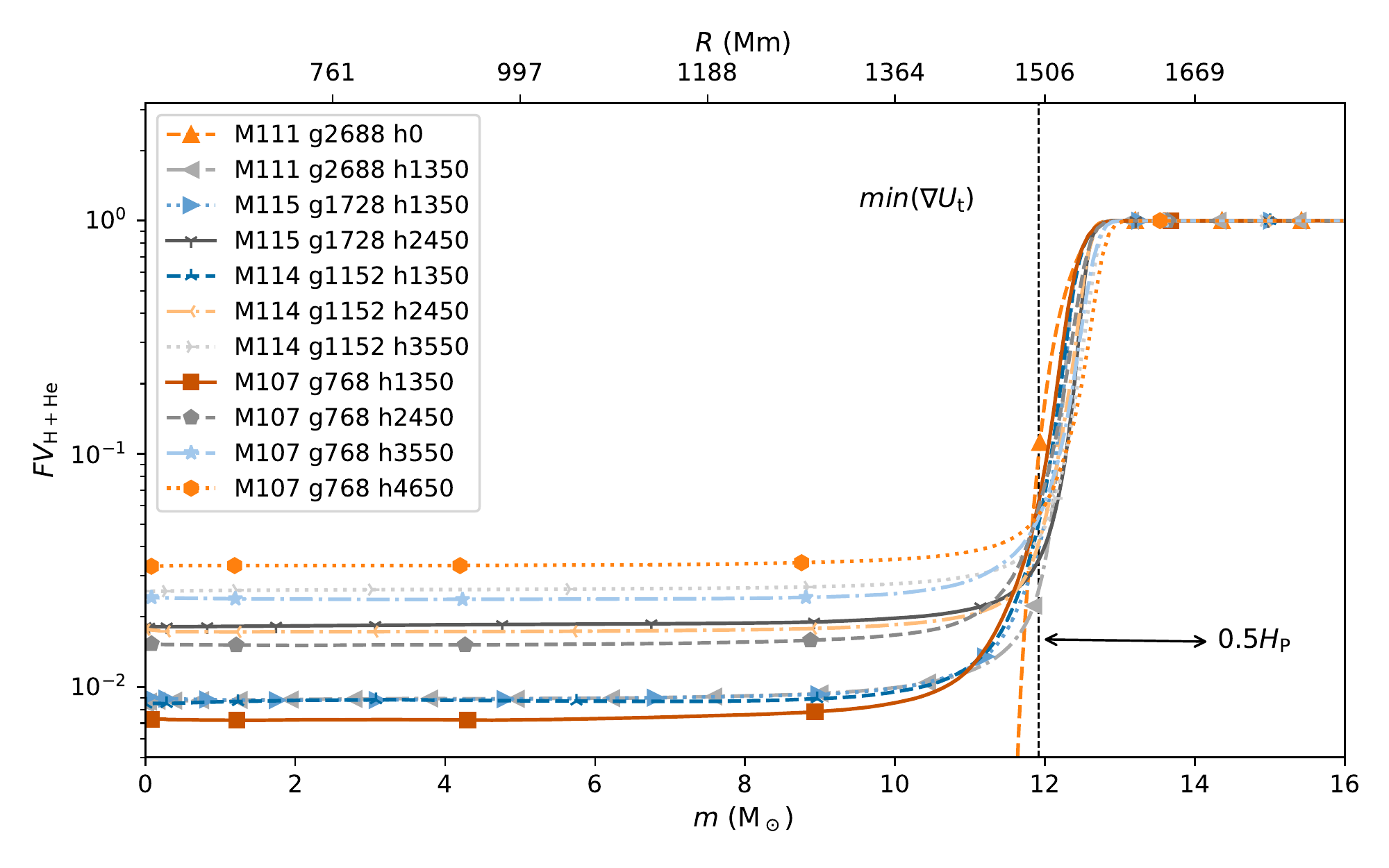}
  \caption{Fractional volume of the fluid in the stable layer
    (\code{H+He}) for three grid resolutions at different times. The
    vertical dashed line shows the location where the radial gradient
    of the tangential velocity has a minimum (\mindudr) in run M107 at
    $t=1600\hour$, which can be taken as the location of the
    convective boundary (see \Sect{conv-bound} for a discussion of
    where the convective boundary is located).}
  \lFig{fig:Resolution-core-FVcld}
\end{figure}   
Both the tangential and the horizontal velocity component ultimately
adopt a steady-state in which neither velocity component changes nor
\emph{drifts} noticeably as a function of time. The $768^3$-grid
simulation M107 has been followed for \unit{7059}{\hour} and does not
show a trend of the velocity magnitude in the envelope beyond the time
shown in \FigTwo{fig:M114-velr-evol}{fig:M114-velt-evol}.

\begin{figure}
  \includegraphics[width=0.93\columnwidth]{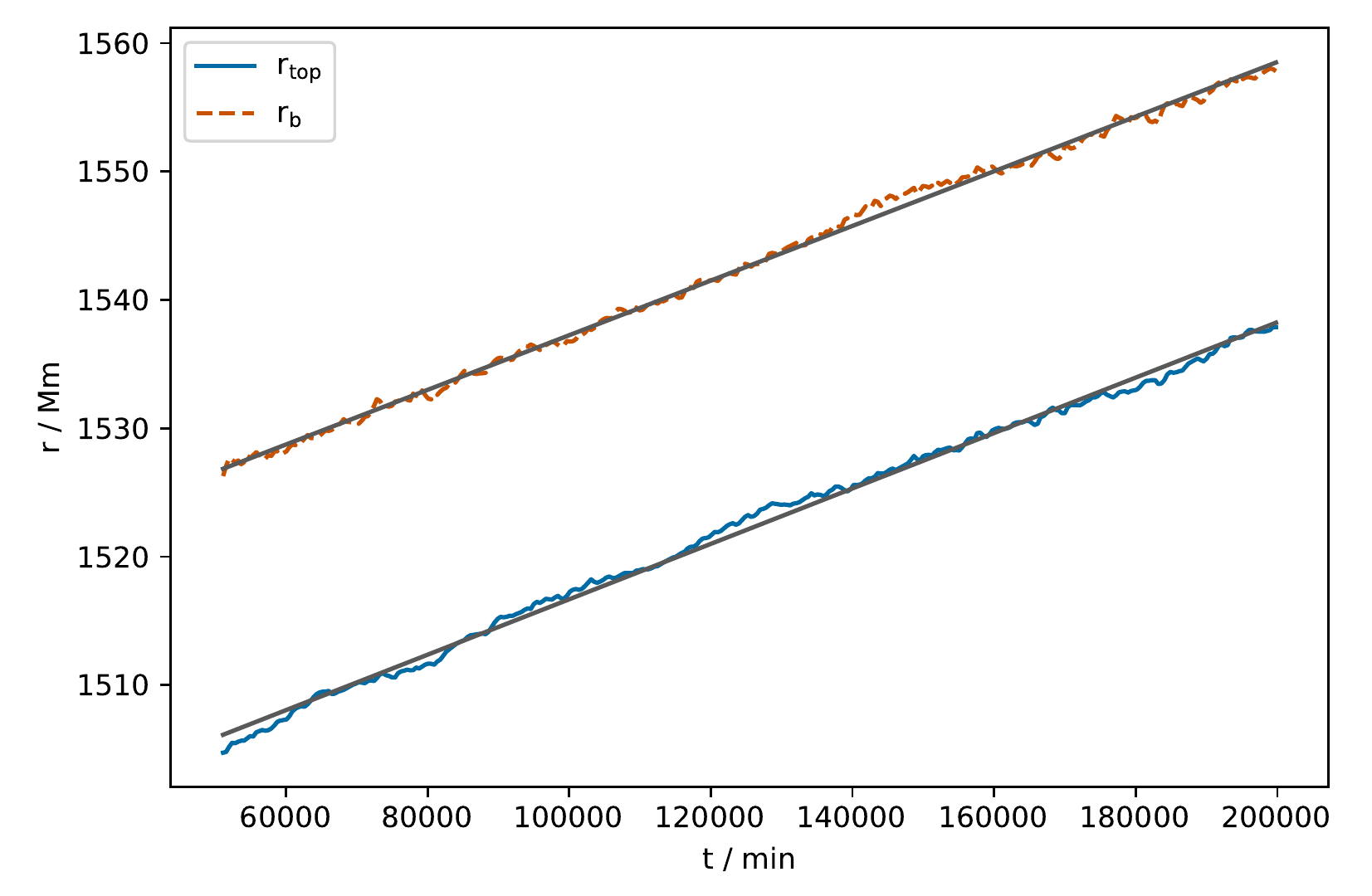}
  \includegraphics[width=0.93\columnwidth]{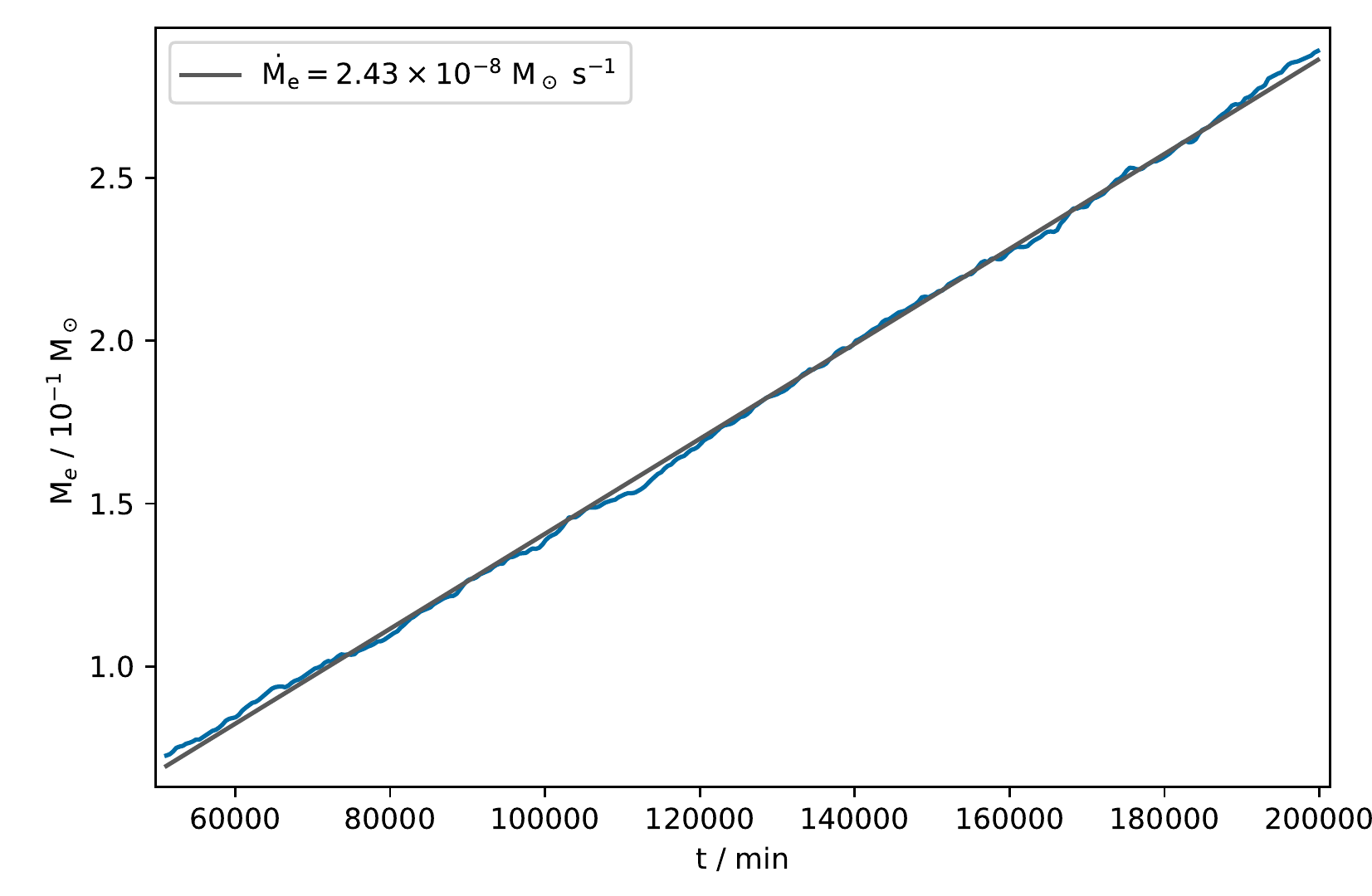}
  \caption{Time evolution of the convective boundary radius (top
    panel). Shown is the radius $r_\mathrm{b}$ where \maxdfvdr\ and
    the radius $r_\mathrm{top}$ which is one $FV$ scale height further
    inward ($r_\mathrm{top} = r_\mathrm{b} - H_{FV}$). The mass
    integration is carried out to $r_\mathrm{top}$. A linear fit of
    the resulting entrained mass evolution (bottom panel) yields the
    entrainment rate.}  \lFig{entrainment-demo}
\end{figure}         
For the tangential velocity component, the initial transient to reach
this steady state is of the order of a convective turn-over time (see
also discussion in \Sect{vorticity}). The radial component appears to
go through a longer transient period of $\approx\unit{1500}{\hour}$
until it settles into the steady state value. This longer time scale
is also the time it takes for the boundary to migrate through the
initial $N^2$ profile due to mass entrainment (cf.\ discussion in
\Sect{s.prop_1D_model} and results in \Sect{s.bnd_sph_ave}). Possibly
the radial velocity component is more sensitive to the exact shape of
the boundary in the \npeak\ region.

The time-evolution comparison of results obtained from simulations on
different grid sizes indicates that velocity component magnitudes and
their oscillation properties are essentially in good agreement among
the different grid resolutions. Only the $768^3$-grid results diverge
by a small amount in predicted envelope velocities.

\subsubsection{Statistics of convective and wave motions}
\lSect{s.statistics_pdfs} Another way to visualize the fundamental
difference between the flow patterns in the core and the envelope is
to use velocity images on spheres of constant
radius. \Fig{fig:M115-Mollweide-conv} shows such images for the radial
velocity component for a radius well inside the convection zone and a
radius one pressure-scale height above the convection zone in the
stable layer.\footnote{Movie versions scanning through radius are
available in the digital supplement at \url{http://www.ppmstar.org}}
These projection images onto a sphere reveal large and coherent upflow
areas and somewhat more narrow downflow lanes. Smaller-scale
structures are distributed throughout the sphere and blur the
distinction between up- and downflow areas. In the stable layer,
smaller regions of upward- and downward-directed flow are ordered in a
semi-regular fashion. This difference can be expressed quantitatively
through the probability distribution function (PDF) shown below each
image and its higher-order moments. The convective PDF has a large
skew, meaning it has significant asymmetry compared to a
Gaussian. There are more units of area on the sphere with downflows
than upflows. However, the largest radial velocities are found in
areas with upflows rather than downflows. The relative strength of the
far tail is measured by the excess kurtosis. Large values of kurtosis
indicate an overabundance of far-tail events. Turbulent convection is
intermittent with gusts of larger-than-average flow speeds occurring
at random times. This is reflected in the larger kurtosis compared to
the PDF of the stable envelope.
\begin{figure}
  \includegraphics[width=\columnwidth]{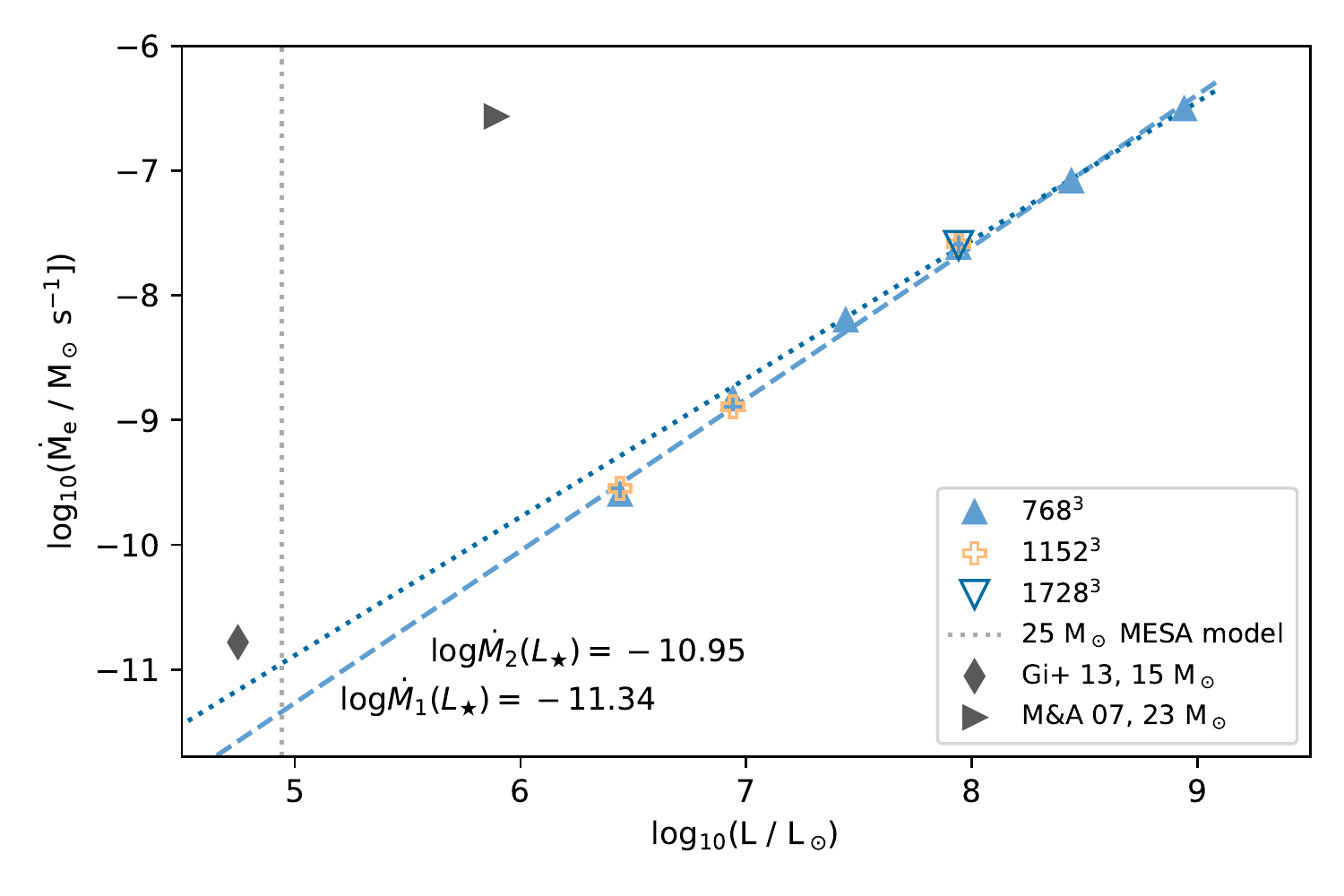}
  \caption{ Entrainment rates as a function of driving luminosity for
    runs with different grid resolutions as indicated in the
    legend. Two linear fits are provided. Fit $\dot M_1$ (dashed line)
    is based on the highest resolution run available for each heating
    rate. The second fit (dotted line) includes only the $768^3$ grid
    runs for the three highest heating rates. Fit parameter
    uncertainties are standard deviations of the fitting only. The
    vertical grey dotted line indicates the nominal luminosity of the
    underlying 1D \mesa\ model. The intercept values of both fits with
    the nominal heating ordinate are indicated as well as the
    entrainment rates reported by \citet{Gilet:2013bj} and
    \citet{Meakin:2007dj} for similar simulations.}
  \lFig{fig:entrainment_rate_vs_luminosity}
\end{figure}

The PDF of radial velocities in the envelope on the other hand is
represented almost perfectly with a Gaussian
distribution. Anticipating the presentation of the spectra of IGWs in
the boundary region in \Sect{s.neq1IGWmode} and in the entire envelope
\cite[][\papertwo]{Thompson:2023a}, we note that the IGWs in the
stable layers of these simulations have power in the radial velocity
component peaking at spherical harmonic degree $l \approx 30$ and that
IGWs with all eigenfrequencies below the \brunt\ frequency are
well-represented. As demonstrated elsewhere in more detail, the
coherent superposition of this wide range of IGW eigenmodes results in
a Gaussian PDF. This feature of the $U_\mathrm{r}$ PDF is thus a
quantitative symptom of a velocity field dominated by IGWs. We use
this statistical difference between convective and wave motions in
\Sect{conv_bound_mix} to characterize the motions in the convective
boundary region.

\subsection{Entrainment} \lSect{entrainment} 
In \Sect{dipole} we discussed the fluid morphology of the giant dipole
modes and the intimately related boundary-separation wedges. The
orientation of the dipole drifts as the simulation proceeds, which can
be best observed by watching the movies\footnote{Movies are available
in the digital supplement \url{http://www.ppmstar.org}}. This drift
time scale is that of several convective turnovers of $\approx
\unit{128}{\hour}$ (\Sect{dipole}). The convective boundary is stiff,
and at the location where the uprising flow impacts the radial
position of the convective boundary it is only minimally displaced. At
the boundary, the flow is redirected from a radial to a dominantly
horizontal component. The large impact zone of the outward flow of the
dipole approaching the boundary at the south pole in
\Fig{fig:M115-images-HcoreM025Z0} would be the closest thing to what
one might consider a plume in these simulations. However, the southern
hemisphere where this impact takes place is not where the entrainment
takes place. The bottom-right panel in
\Fig{fig:M115-images-HcoreM025Z0} shows the concentration of the
material initially only in the stable layer that we refer to as the
\code{H+He} fluid.

The instabilities induced by the boundary-layer separation flows that
we call the \emph{wedge} features are responsible for the entrainment
of material from the stable layer into the convection zone. This is
evident by comparing the velocity centre-plane images
(\Fig{fig:M115-images-HcoreM025Z0}) with the image of the \code{H+He}
fluid, where orange, yellow, and white colors indicate partially-mixed
zones. These regions of effective entrainment are located opposite to
each other on the boundary circle near the east and west equator
coordinates. The dipole axis is tilted a bit to the west in the dump
shown. The dominant hydrodynamic mechanism of entrainment is here, as
it was in He-shell flash convection \citep{Woodward:2013uf}, the
instabilities induced by boundary-layer separation of large-scale flow
sweeping along the convective boundary.

The continuous entrainment process leads to an accumulation of
\code{H+He} fluid in the convective core. Spherically averaging the
\code{H+He} concentration leads to profiles as shown in
\Fig{fig:Resolution-core-FVcld} for the three different grids used in
the $1000\times$ heating simulations. After $4000\hour$, or $31$
convective timescales (\unit{128}{\hour}, \Sect{dipole}) the
\code{H+He} stable-layer fluid has accumulated to a level of $\approx
\natlog{3}{-2}$ in the convective core.

Integrating over the \code{H+He} fluid concentration from the centre
to the convective boundary gives the total entrained mass. With
respect to the upper boundary for the entrained mass integration, we
use a similar approach as \citet{Jones:2017kc}, but instead of the
minimum gradient of the tangential velocity component we adopt here
the maximum gradient of the \code{H+He} fractional volume $FV$ reduced
by one $FV$ scale height. This is essentially equivalent to
integrating to the radius at which $FV = 0.1$. Using $FV$ instead of
$U_\mathrm{t}$ gives a smoother boundary evolution for main-sequence
simulations because the \mindudr\ criterion often finds a location
just outside the dynamic boundary that is dominated by the $n=-1$ IGWs
(\Sect{s.neq1IGWmode}). The entrained \code{H+He} mass evolves
linearly with time, and examples are shown in \Fig{entrainment-demo}
for simulations with different grid resolutions and heating
factors. In each case, the initial transient ($\approx 300$ to
$\approx 1000\hour$) of the simulation was discarded for the purpose
of fitting a linear relation to the entrainment evolution. During this
initial transient, the entrained mass as a function of time would
still show non-linear behaviours that can in part be understood in
terms of the evolution of the convective boundary profile as a
function of time and heating factor as discussed in
\Sect{conv-bound}. This fit of the entrained mass determines an
entrainment rate for each simulation. For simulations of O-shell
convection in massive stars, we have previously found a linear
relation between the heating factor and the entrainment rate
\citep{Jones:2017kc,Andrassy:2020}. \Fig{fig:entrainment_rate_vs_luminosity}
shows the entrainment rates for all heating factors and grid
resolutions included in this paper.

For a few heating factors, we have carried out simulations for
multiple grid resolutions. The entrainment rate does not depend
systematically or significantly on grid resolution for our
simulations. Next, we note that again, as in O-shell simulations, the
entrainment rates follow a linear trend over a wide range of heating
rates. Two linear fits are shown in
\Fig{fig:entrainment_rate_vs_luminosity}. One includes only the three
highest heating rates $1000$, $3162$, and $10000$, while the other fit
includes all heating rates.

Either way the resulting mass entrainment rate at nominal heating is
unrealistically large, \unit{\approx \natlog{5}{-12}}{\Msun/\second}
for the second fit. This is $992\Msun$ applied over the H-core burning
lifetime of \unit{\natlog{6.91}{6}}{\yr}. This interpretation of the
derived entrainment rate obviously does not make sense. Our
simulations are not alone in predicting very large entrainment rates,
but in good agreement with those of \citet{Gilet:2013bj}, who used a
low-\Mach\ number solution scheme. They include radiation pressure but
like us ignore radiative diffusion. \citet{Meakin:2007dj} report an
entrainment rate about three orders of magnitude higher than our
simulations and included both radiation pressure and radiation
diffusion. Preliminary tests that we will describe in detail in a
forthcoming publication indicate that neither the addition of
radiation pressure, radiative diffusion or the addition of rotation
resolves the unrealistically high entrainment rate. Instead, the
strong time-dependence of the response of the 3D hydrodynamic
simulation to the adopted MESA base state signals that the initial
MESA base state is out of thermal-dynamic
 equilibrium in the 3D hydro
framework. The large entrainment rate leads to an increasing
nearly-adiabatic layer outside the initial convective boundary. Thus
the large entrainment rate phase of these 3D simulations represents
the approach toward a thermal-dynamic equilibrium state with a larger
nearly-adiabatic core. Our own preliminary tests and simulations by
\cite{anders:21} show that indeed such simulations reach a
quasi-equilibrium state.

\section{The convective boundary}
\lSect{conv-bound}
\begin{figure}
  \includegraphics[width=\columnwidth]{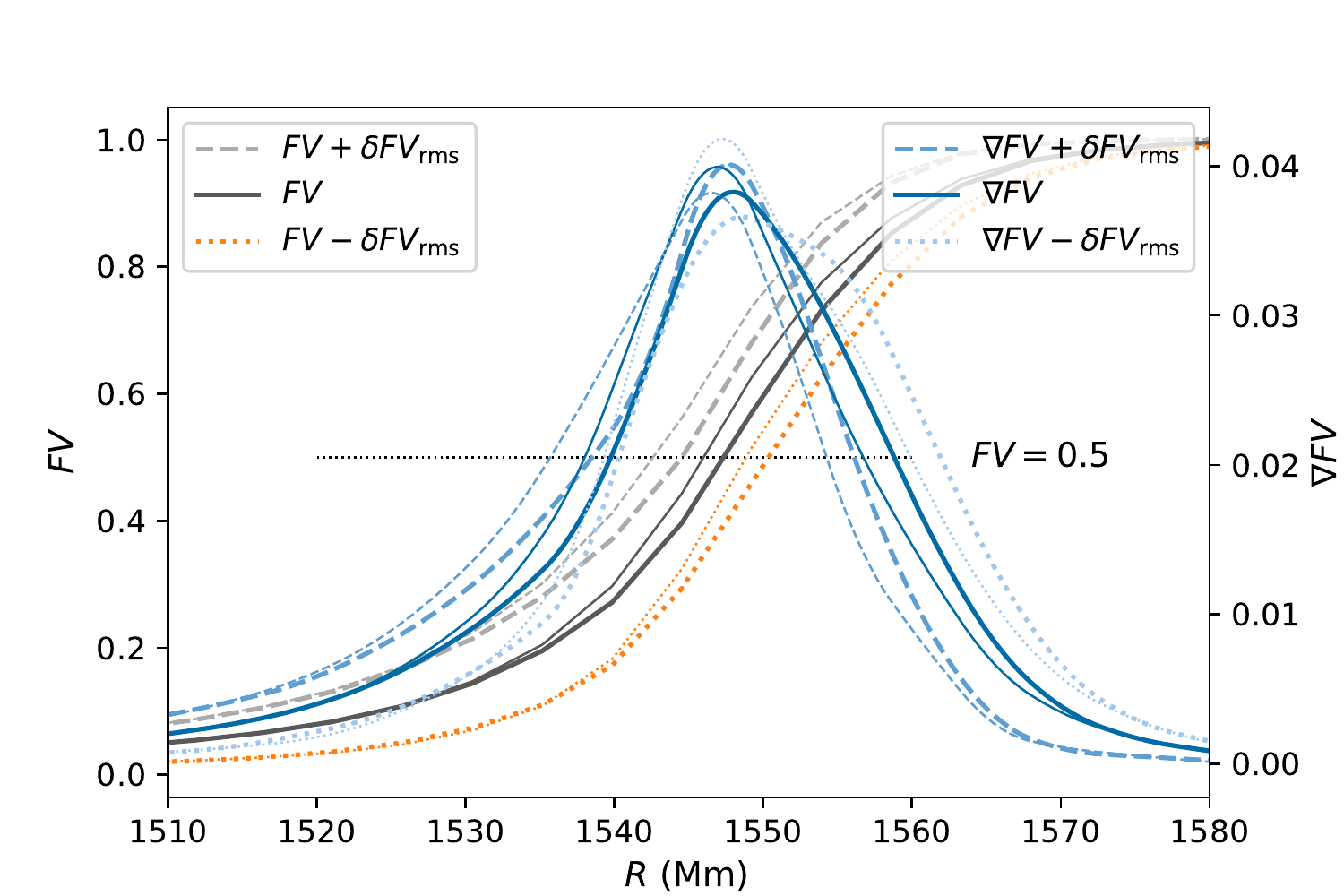}
    \caption{FV profiles (left ordinate) and FV gradients from FV
      spline interpolations of FV (right ordinate) at the boundary for
      \unit{2450}{\hour} (thick lines) and \unit{2350}{\hour} (thin
      lines), run M115 ($1000\times$ heating factor, $1728^3$ grid)}.
    \lFig{fig:FV-boundary-M115-dump3000}
\end{figure}  
\begin{figure}
  \includegraphics[width=\columnwidth]{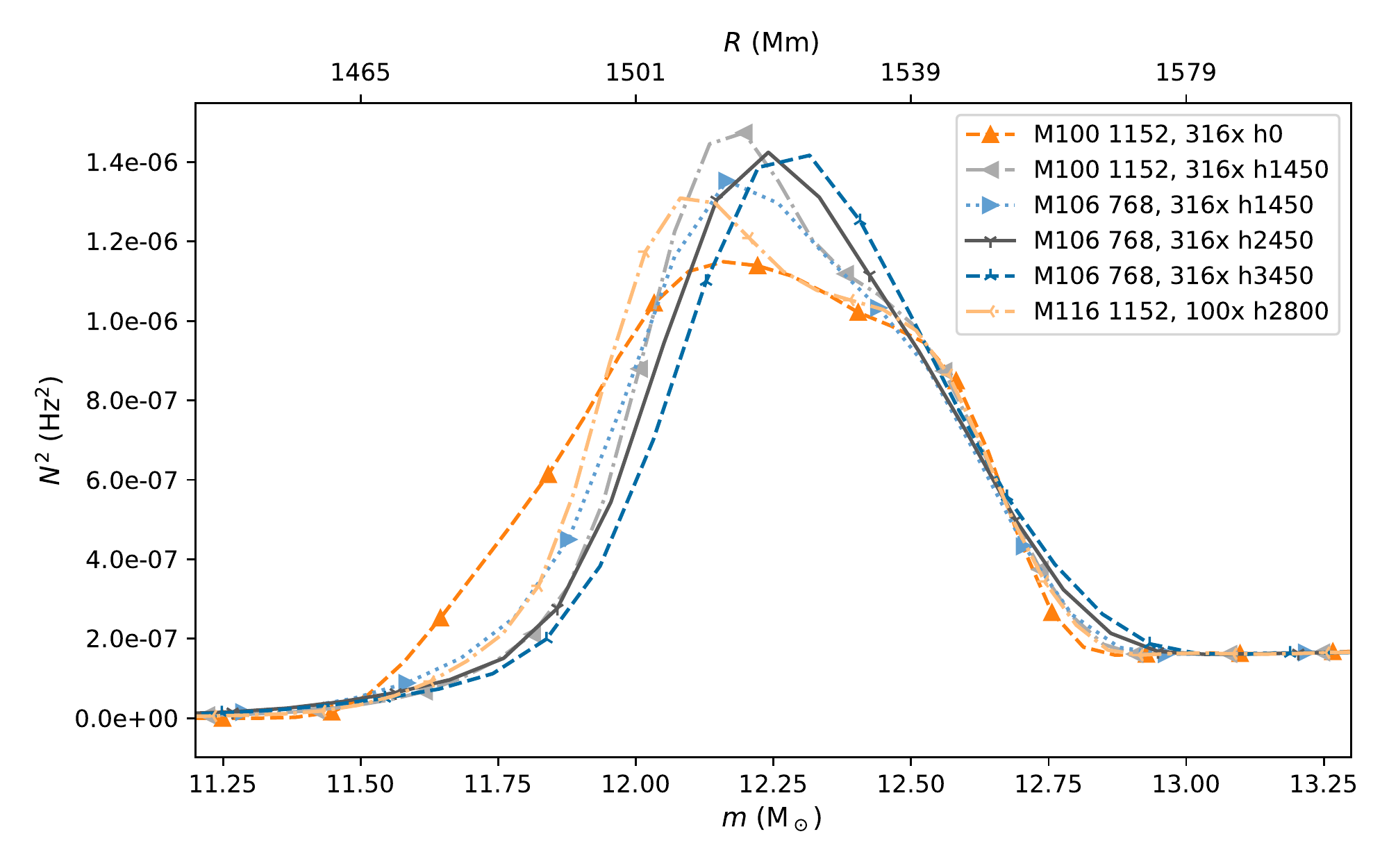}
    \caption{Radial profile of \brunt\ frequency for low heating
      factor runs (\Tab{tab:runs-summary}).}  \lFig{fig:Low-L-N2}
\end{figure} 
An important goal of this paper is
to improve our understanding of the hydrodynamic processes and
properties of the convective boundary. In this section, we describe the
properties of the boundary, how to determine its location, and the
different types of motions in the boundary region. Again, an important
aspect is to demonstrate how the results depend on grid resolution. We
will focus the discussion on the $1000\times$ simulations, for which
simulations with four grid resolutions have been used (M107 --
$768^3$, M114 -- $1152^3$,  M115 -- $1728^3$ and M111 -- $2688^3$).
\begin{figure}
  \includegraphics[width=0.99\columnwidth]{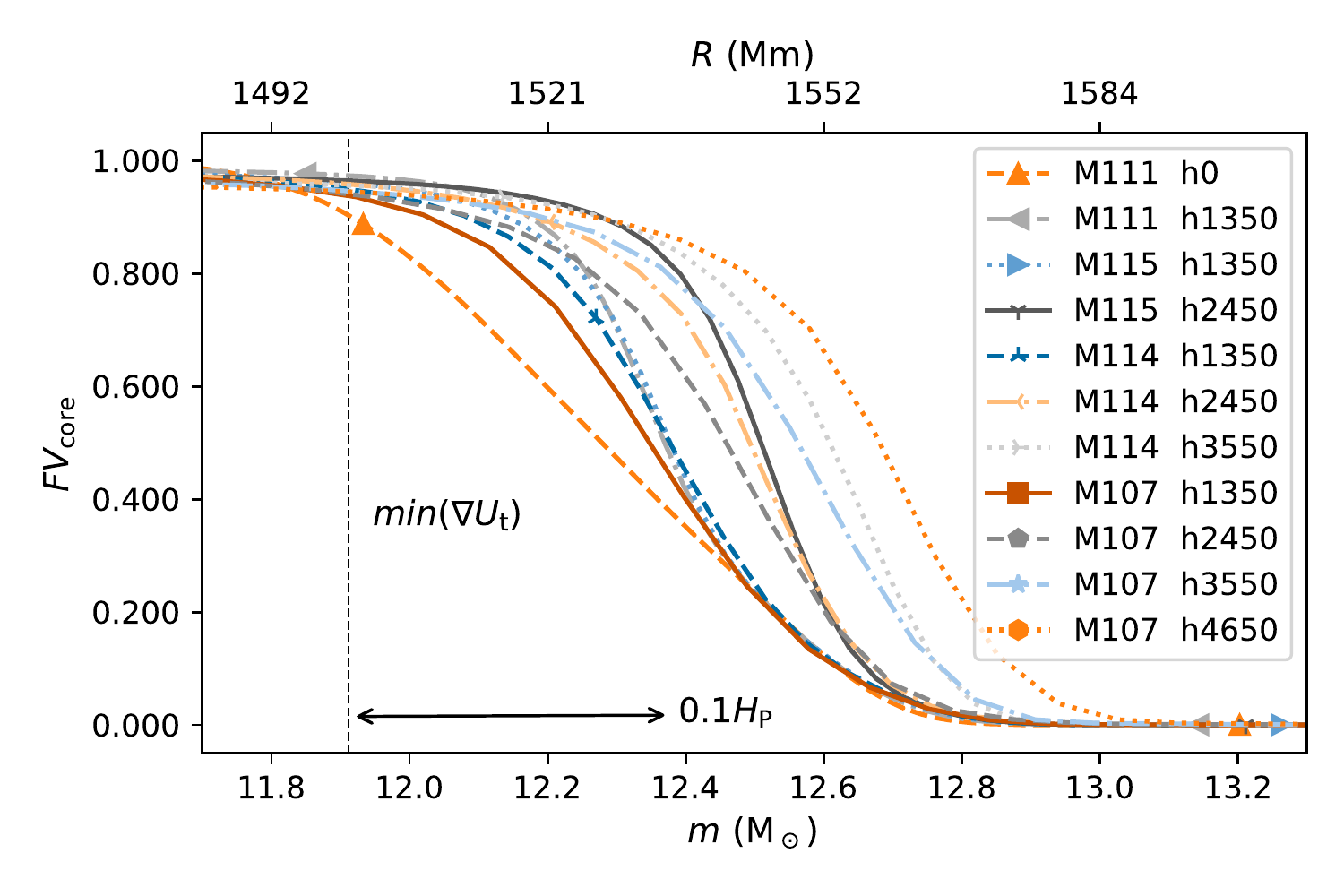}
  \includegraphics[width=0.99\columnwidth]{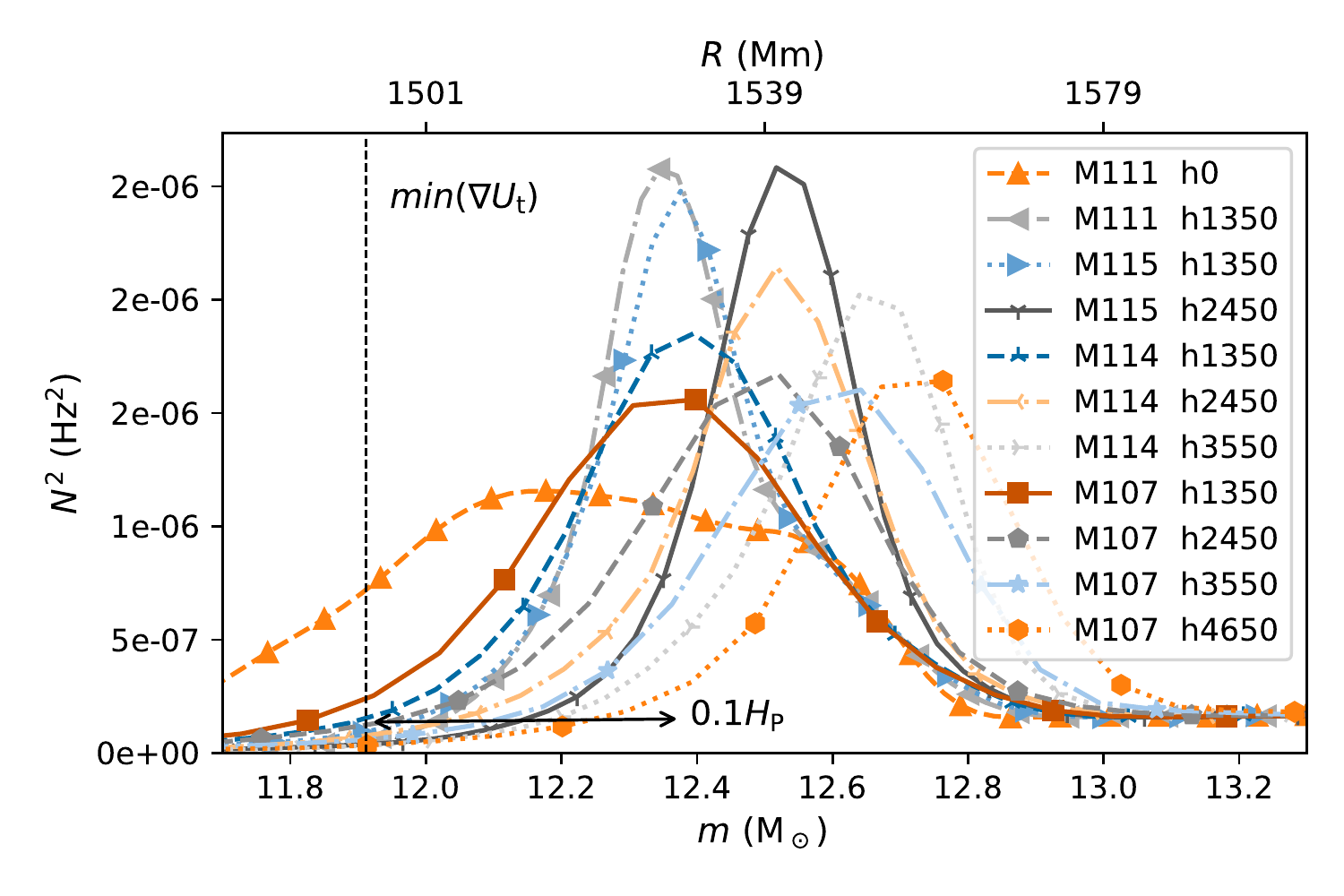}
  \includegraphics[width=0.99\columnwidth]{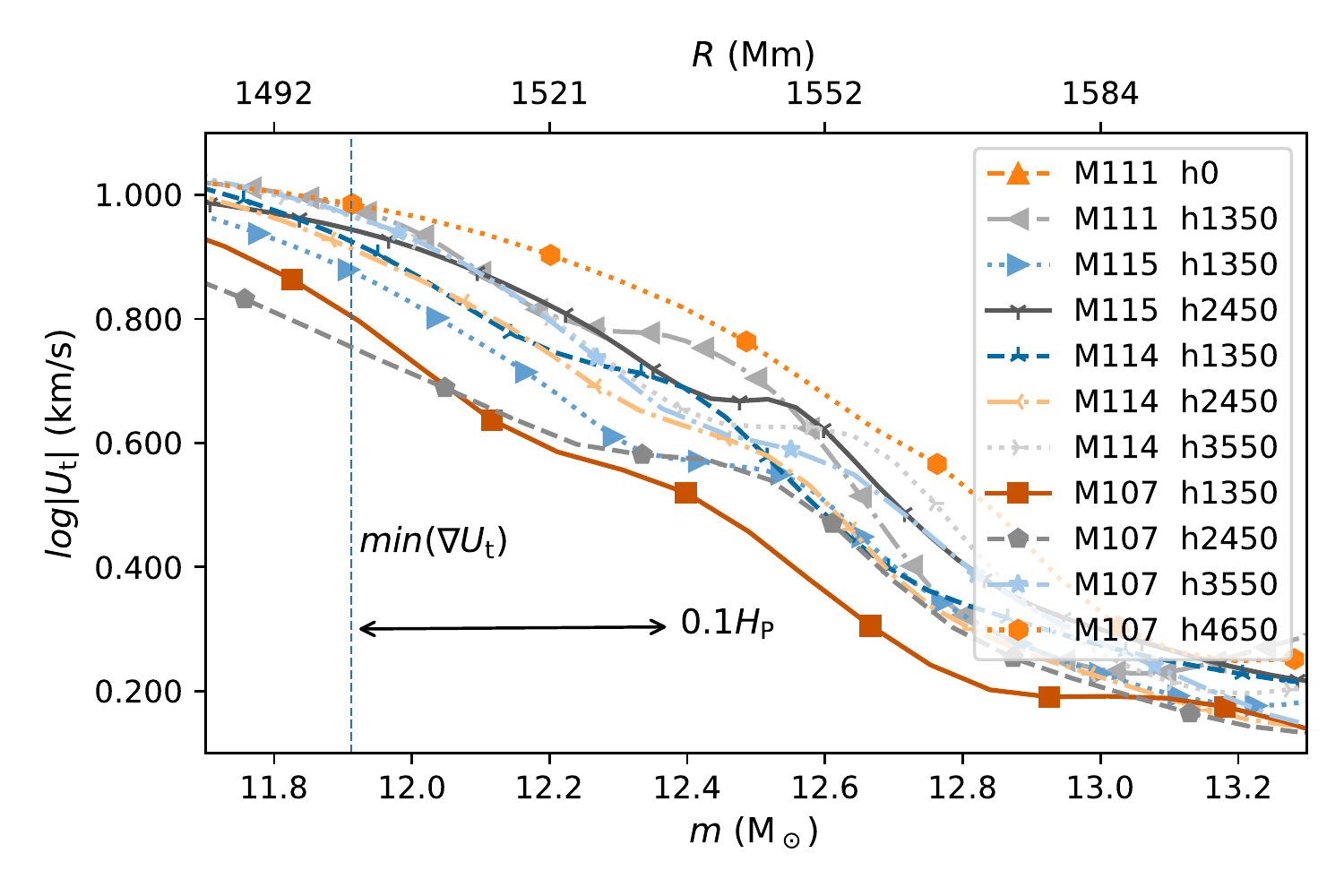}
    \caption{Radial profiles of the \code{core} fluid, $N^2$, and the
      magnitude of the tangential velocity component. The legend gives
      the run ID (\Tab{sims}), the grid size, and the time in
      hours. Markers are shown every three data points. The vertical
      line shows for run M107 and \unit{1350}{\hour} the location of
      the minimum (steepest) gradient of the tangential velocity
      component, which is one of the criteria for the boundary of the
      convection zone (cf.\ \Sect{conv_bound_mix}). For reference the
      size of $0.1\Hp$ is shown with a horizontal arrow. The abscissa
      at the top gives the radius according to $r(m)$ for
      \unit{1350}{\hour} of run M107, see text for details.}
    \lFig{fig:RProf-N2FVUt-boundary}
\end{figure}
\afterpage{\clearpage}

\subsection{Evolution of the boundary in terms of spherical averages}
\lSect{s.bnd_sph_ave} Radial profiles of 3D simulation quantities
averaged on spheres are an obviously useful dimensional reduction when
the goal is to develop models for applications in 1D stellar evolution
codes. We are mentioning two complications. The first is that because
we keep heating the core at rates that are much larger than the
nominal heating rate, and we do not include radiation diffusion
\cite[simulations with radiative diffusion will be presented in
  \paperthree][]{Mao:2023a}, the core is expanding and thereby the
radial coordinate of the boundary is moving slightly. This effect is
easily taken care of by working with radial profiles in terms of the
mass coordinate.

\begin{figure}   
  \includegraphics[width=\columnwidth]{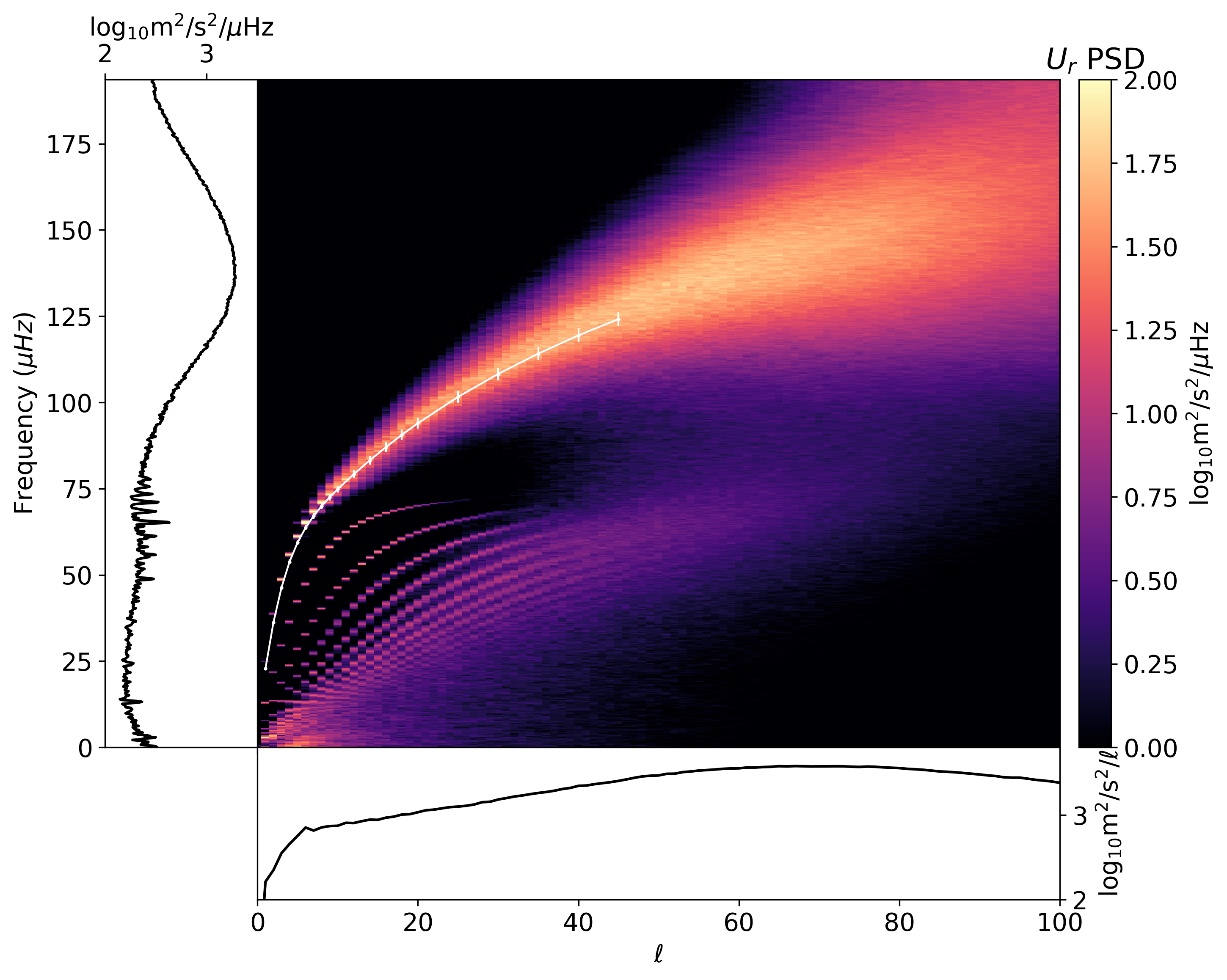}
  \includegraphics[width=\columnwidth]{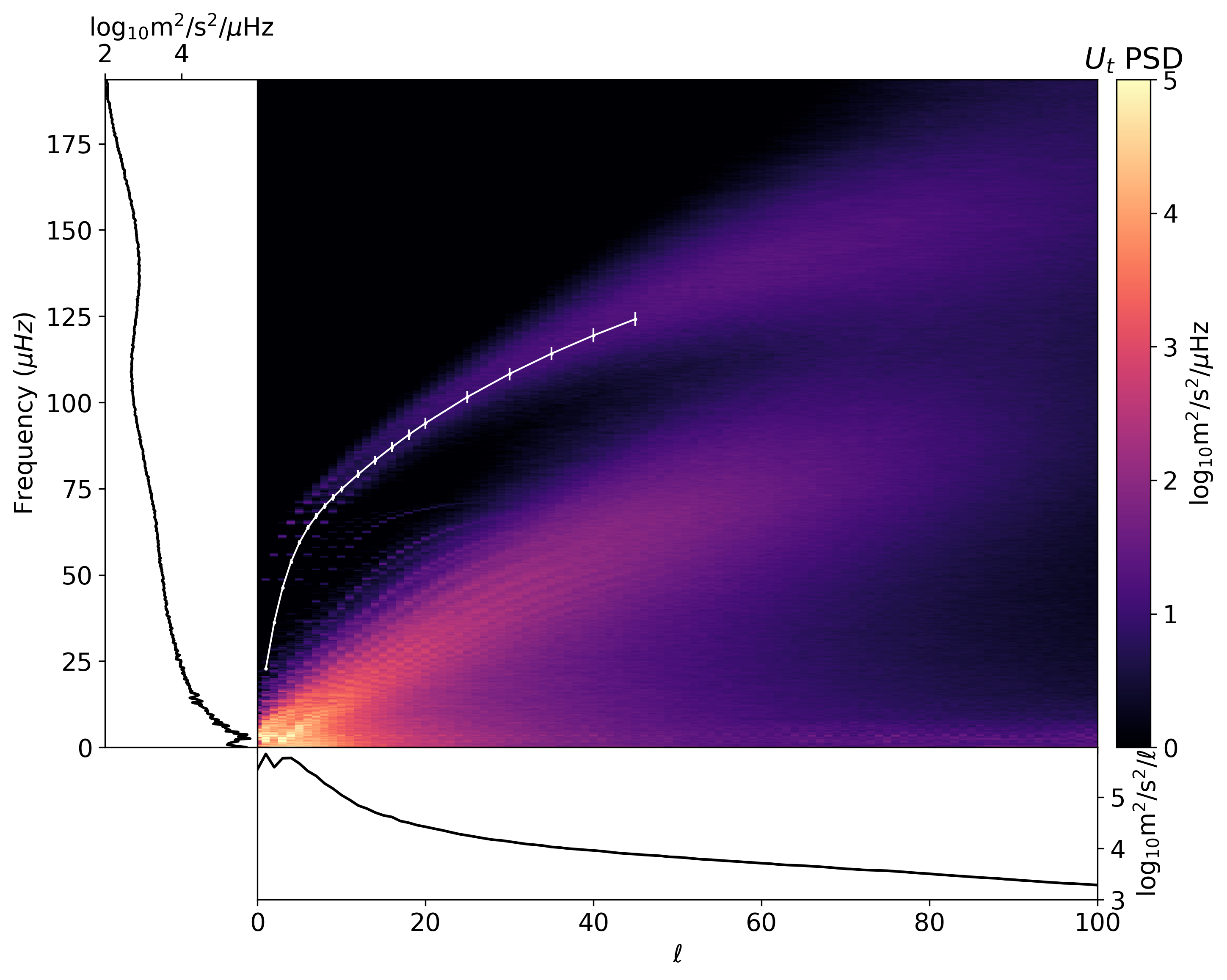}
 \caption{Power spectral density as a function of spherical
   harmonic angular degree $l$ and cyclic frequency (\komega) for the
   radial (top) and horizontal (middle) velocity components in
   simulation M114 ($1000\times$, $1152^3$ grid) at $r = 12.6\Msun$,
   where the peak of $N^2$ is located. Dumps 3925 to 5924 were
   used. The white line with dots are $n=-1$ modes determined
   from the \code{GYRE} calculations for the spherically-averaged
   stratification of dump 3925. The abscissa and
   ordinate include projected spectra in terms of spherical harmonic
   $l$ and frequency, respectively.}
 \lFig{fig:kw-boundary}
\end{figure} 
The second point is a bit more subtle. The boundary according to the
adopted \mesa\ base state is very stiff, which implies that the
boundary layer is narrow in the radial direction. The largest-scale
modes of the convection may lead to a non-spherical deformation of the
position of the convective boundaries, or of certain features. When
taking an average over a deformed surface, for example of the
concentration, one obtains a smoothly varying profile. However, when
averaging over an undeformed surface, which has for each boundary
surface element a turbulently mixed interface where each vertical
volume element truly consists of a mix of the two materials separated
by the boundary, then the resulting averaged vertical profile is
likewise a smoothly varying profile. Only in the latter case does the
radial concentration profile represent partial mixing. The situation
is like that of ocean swell from a distant storm on a calm day. Taking
horizontal averages will yield a smoothly varying vertical profile,
but nowhere except on the molecular level can there be found a volume
element that contains water and air. This second aspect of radial
profiles from spherical averages of 3D data is much more difficult to
take into account.

An estimate for the magnitude of this effect can be obtained from the
3D spherical rms-deviations of the FV
profile. \Fig{fig:FV-boundary-M115-dump3000} shows these for two times
approximately one convective turnover apart, as well as their
derivatives, which would correspond to the $N^2$ profiles. The maximum
of the gradient differs by $\loa \unit{3}{\Mm}$ between $FV \pm \delta
FV_\mathrm{rms}$ profiles, whereas the radius at which $FV = 0.5$
differs at both times by $\approx \unit{5}{\Mm}$. Further insight into
how much the dominant spherical boundary features are subject to
deformation and what the internal structure of the boundary is will be
explored in \Sect{conv_bound_mix}.

The key quantities are shown for the narrow $\approx 100\Mm$-wide
boundary layer region in \Fig{fig:Low-L-N2} for lower heating runs and
for $1000\times$ heating factor simulations in
\Fig{fig:RProf-N2FVUt-boundary}.  The concentration of the core
material traces mixing while $N^2$ (which is proportional to the
entropy gradient) represents the stability of the stratification, and
the magnitude of the tangential velocity represents the average of the
actual fluid flow. For low heating factors the initial boundary
stratification (cf.\ \Sect{stellar_evolution}) is somewhat deformed
but the entrainment rate overall is too small to migrate the boundary
signifcantly over the durartion of the simulation. At $1000\times$
heating rate the initial boundary structure is entirely erased after
about $\unit{1000}{\hour}$, as discussed already in
\Sect{entrainment}. In these simulations, the entrainment rate is so
high that the boundary migrates through the mass region of the
original $N^2$ peak region and establishes a new $N^2$ profile that
has no memory of the initial stratification and is only due to
hydrodynamic processes. The $N^2$-peak becomes higher and narrower,
and this trend is more pronounced for higher-resolution grids.  Visual
inspection of the peak $N^2$ values as a function of grid resolution
shows that the maximum steepness of the boundary is increasing with
grid resolution. This can be understood in terms of IGW mixing in the
\npeak\ region decreasing with grid resolution, as explained below.
The rate of boundary progression is constant and the same for each of
the three grids, and it is equal to the mass entrainment rate reported
in \Fig{fig:entrainment_rate_vs_luminosity}.
\begin{figure}   
  \includegraphics[width=\columnwidth]{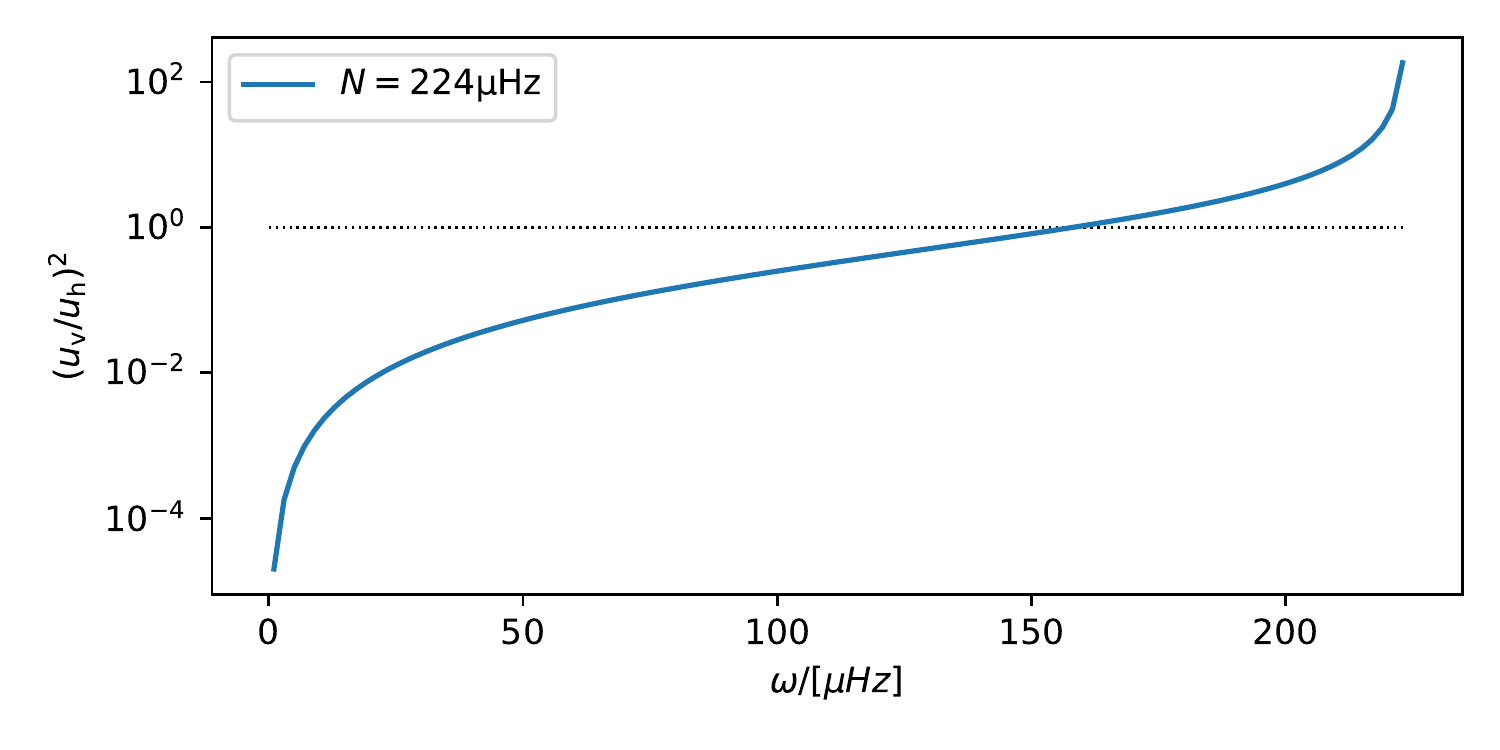}
  \includegraphics[width=\columnwidth]{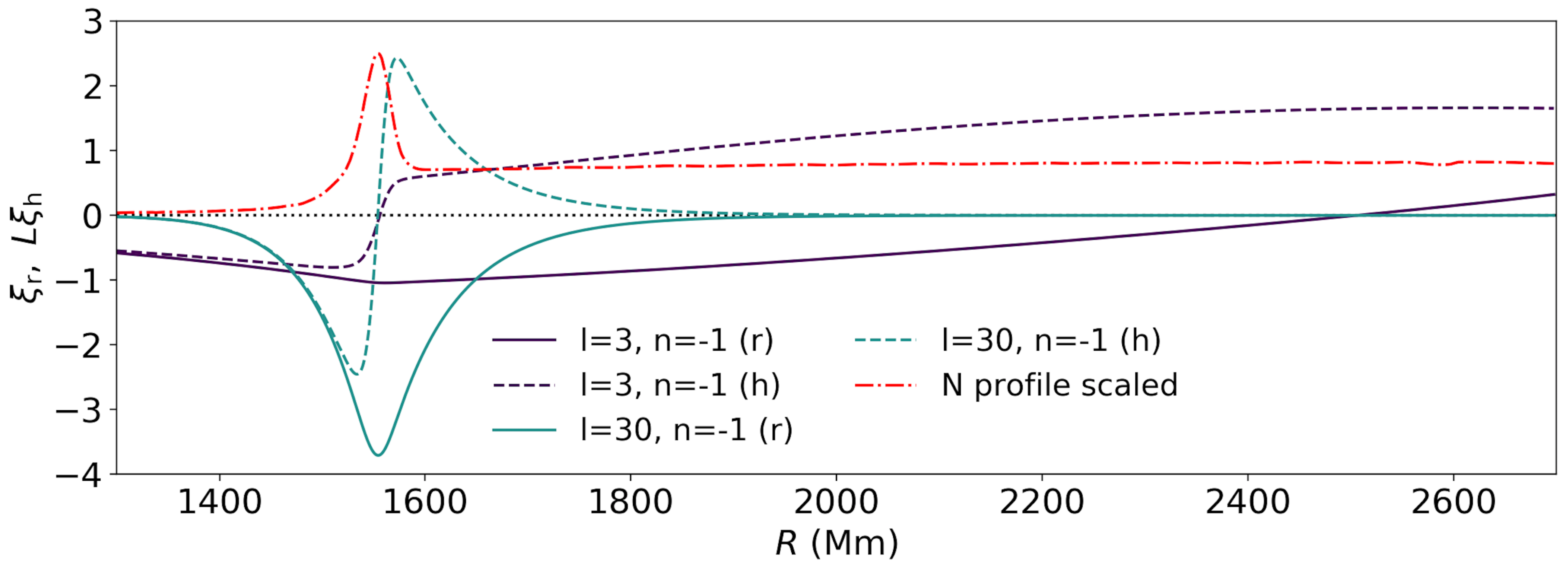}
 \caption{Top panel: Ratio of vertical to horizontal velocity
   component according to \Eq{e.uvuh_ratio} for N at the
   \npeak\ location of run M114 (\Fig{fig:RProf-N2FVUt-boundary}).
   Bottom: Radial $\xi_r$ and horizontal $L\xi_\mathrm{h} =
   \sqrt{l(l+1)}\xi_\mathrm{h}$ oscillation displacement amplitudes of
   two $n=-1$ modes with $l=3$ and $l=30$ calculated with \code{GYRE}
   for dump $3925$ of simulation M114 (cf.\ \Sect{s.wave_analysis}). }
 \lFig{fig:igw-properties}
\end{figure}  
\begin{figure*}
    \includegraphics[width=\textwidth]{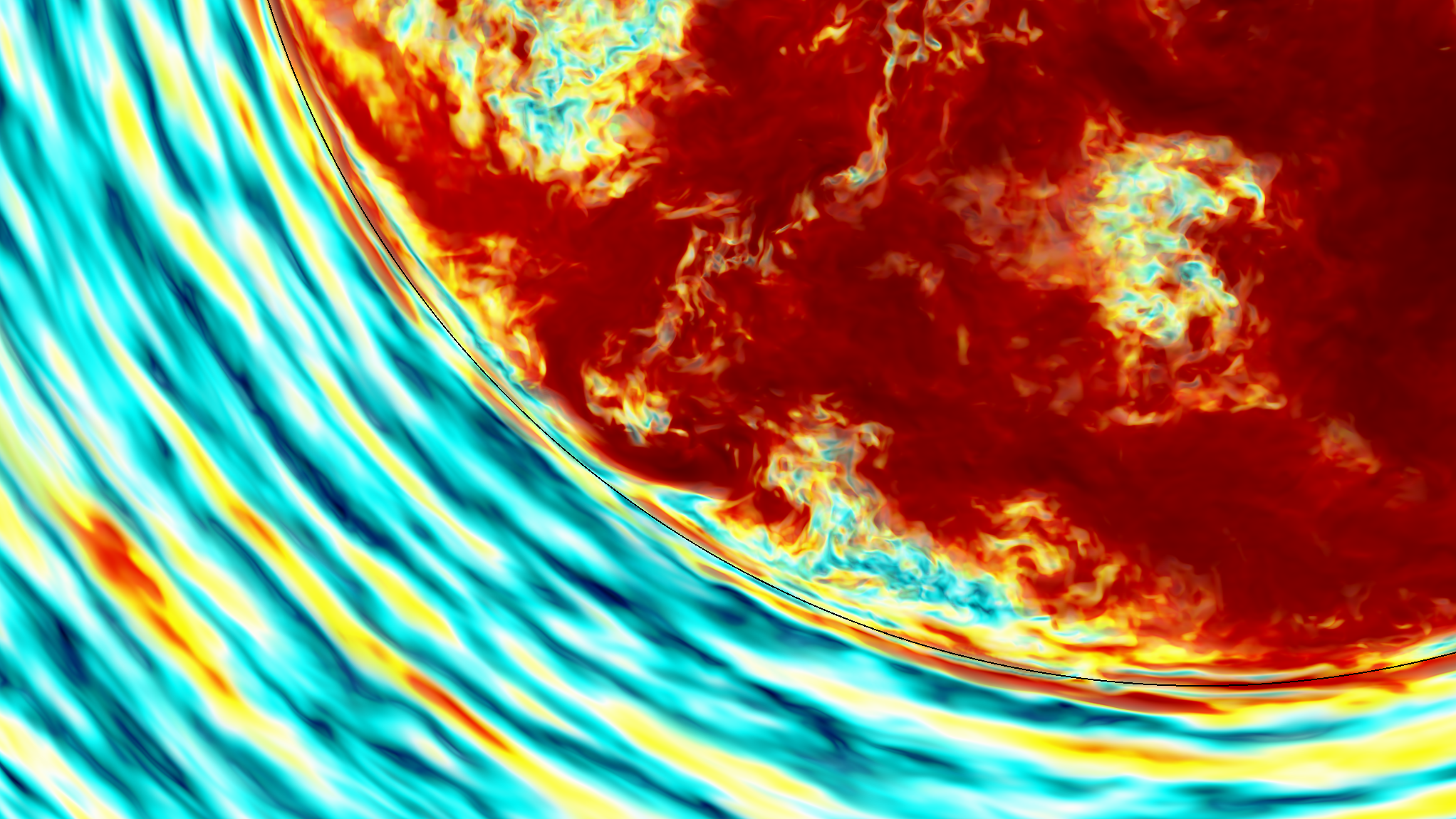}
    \includegraphics[width=\textwidth]{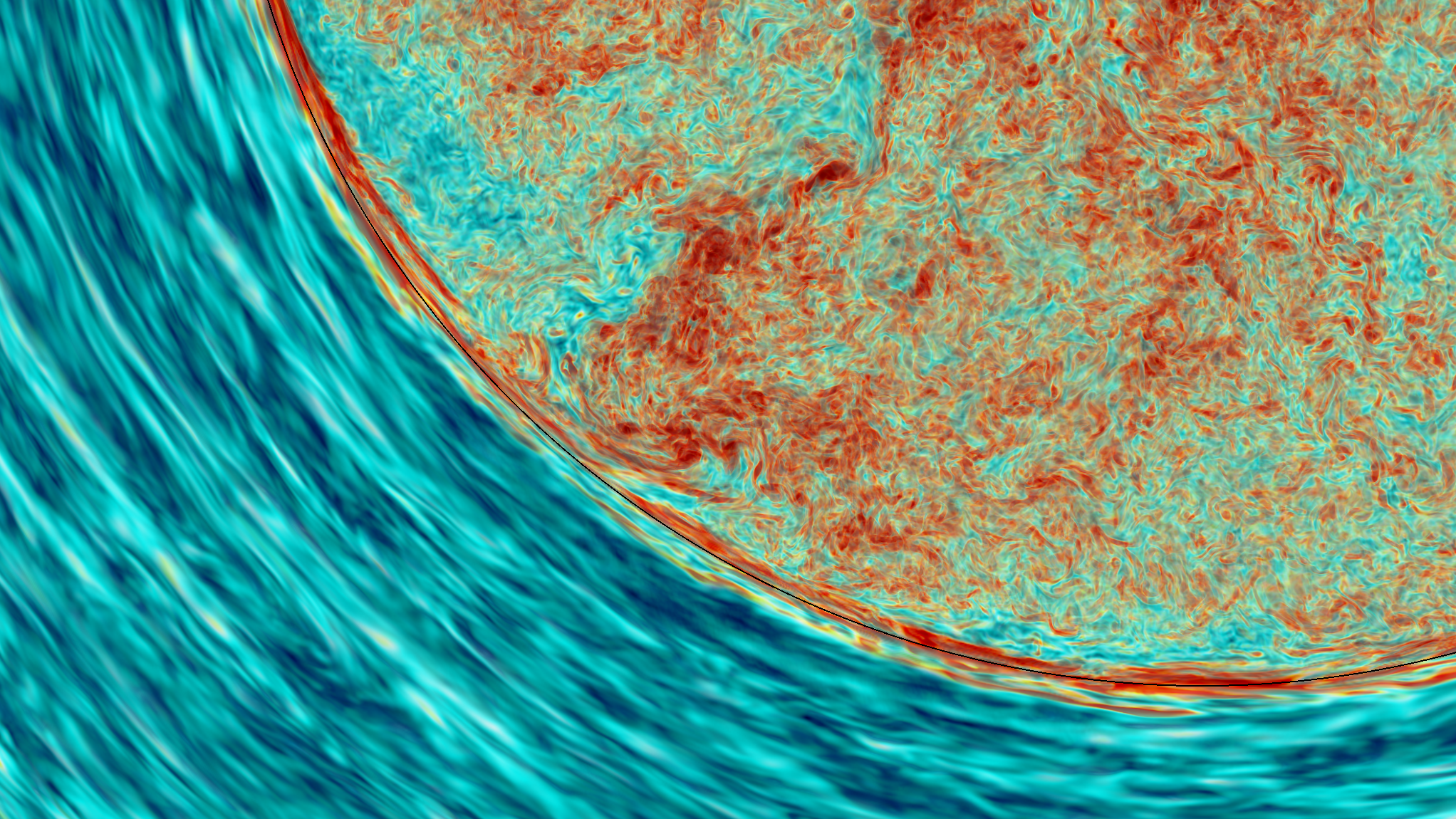}
        \caption{Zoomed-in images of the tangential velocity component
          (top) and vorticity magnitude in the convective boundary
          region from run M111 ($2688^3$ grid) dump 1225. The black
          circular lines show the radial location of the \npeak.}
        \lFig{fig:neq1-mode}
\end{figure*}  

$N^2$ (which is $\propto d S/dr$) follows $d FV/dr$ in this transition
region except where these gradients transition to different envelope
values at the top of the $N^2$ profile. Once the $N^2$ peak has passed
through the initial \npeak\ region given by the initial
stratification, it migrates outwardly in a self-similar form. The same
is true for the concentration profile which, like the $N^2$-peak
region, has an approximate width of $0.2\Hp$, as shown in the top
panel of \Fig{fig:RProf-N2FVUt-boundary}. This means that mixing
processes must occur on both sides of the peak of $N^2$ and across the
entire peak region. We establish in \Sect{mixing-1D} that the
\npeak\ region experiences mixing due to IGWs. It then follows that
the shape of the \npeak\ profile is a convolution of its migration in
mass coordinate and the IGW mixing, similar to how it works for the
convective boundary in a 1D stellar evolution model (second-last
paragraph in \Sect{s.prop_1D_model}). IGW mixing in the \npeak\ region
is inversely proportional to grid resolution (\Sect{s.therm-cond}),
while the entrainment rate is essentially independent of grid
resolution (\Fig{fig:entrainment_rate_vs_luminosity}). Therefore, the
\npeak\ profile for higher-resolution runs is narrower as the
simulation evolves toward quasi-equilibrium.

\citet{Jones:2017kc} adopted the criterion \mindudr\ to locate the
convective boundary. In
\FigTwo{fig:Resolution-core-FVcld}{fig:RProf-N2FVUt-boundary} that
location is shown by a vertical line for run M107 at
$t=\unit{1000}{\hour}$. It is also clear that the decrease of
$|U_\mathrm{t}|$ is not monotone nor steady in many of the cases
shown, as we would expect from an exponential decay of convective
velocities assumed for the 1D exponential diffusive CBM model. At
times, the tangential velocity component can even increase with
radius, indicative of wave motions. For this reason, as discussed in
\Sect{entrainment}, we did not adopt the \mindudr\ criterion as the
entrained mass integration boundary but instead the
\maxdfvdr\ criterion.

\subsection{The $n=-1$ IGW mode in the convective boundary region}
\lSect{s.neq1mode}
\lSect{s.neq1IGWmode} In the \unit{1000}{\hour} profile of M107
(\Fig{fig:RProf-N2FVUt-boundary}), the dashed vertical line indicates
the location where the dominant convective flow velocities are
dropping off rapidly. In this section, we demonstrate that the
velocity field transitions rapidly above the vertical dashed line from
convection-dominated to wave-dominated, and that the layers at and
above \npeak, according to our diagnostics, are exclusively populated
by wave motions.
\begin{figure*}  
  \includegraphics[width=\columnwidth]{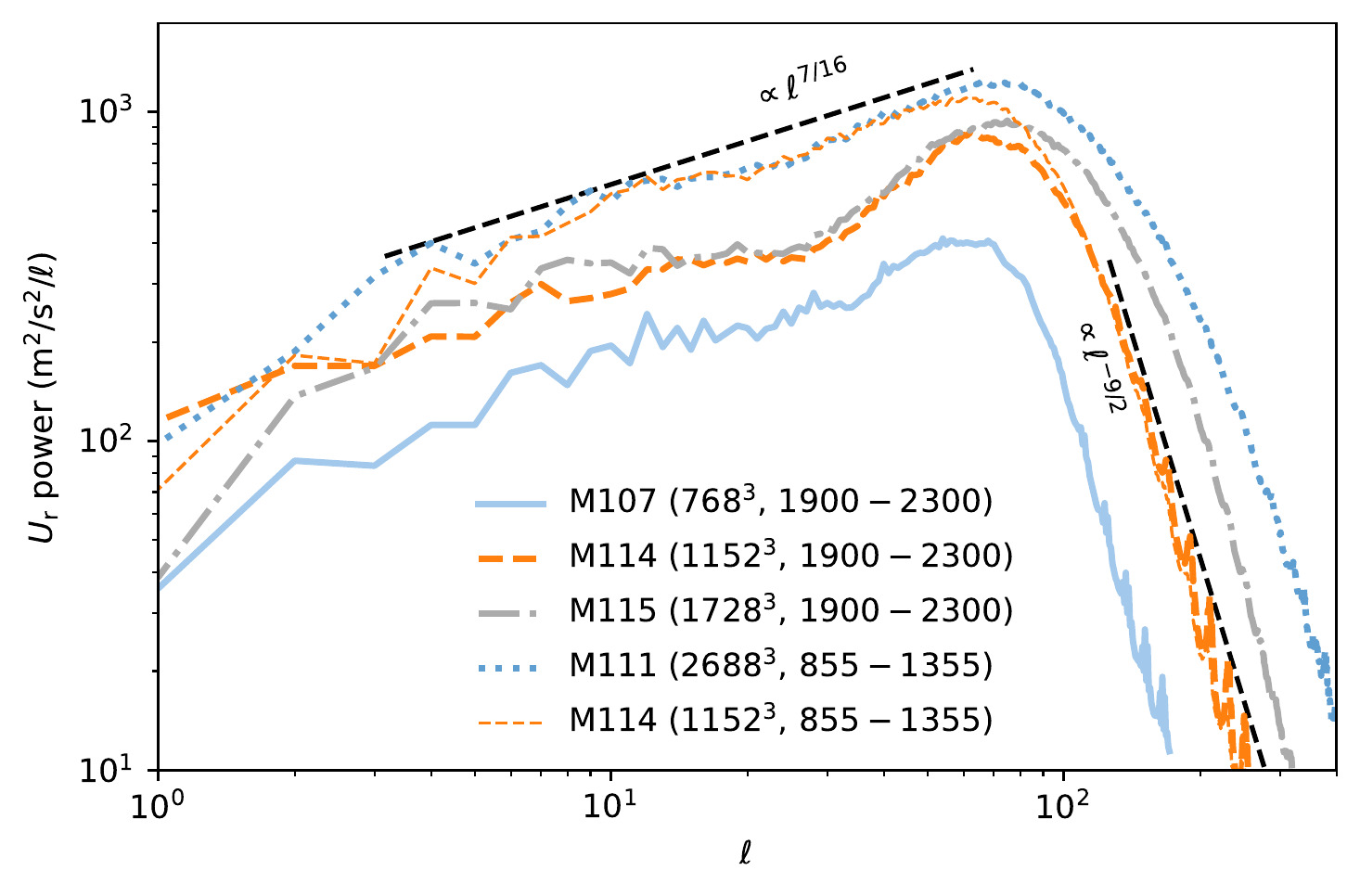}
  \includegraphics[width=\columnwidth]{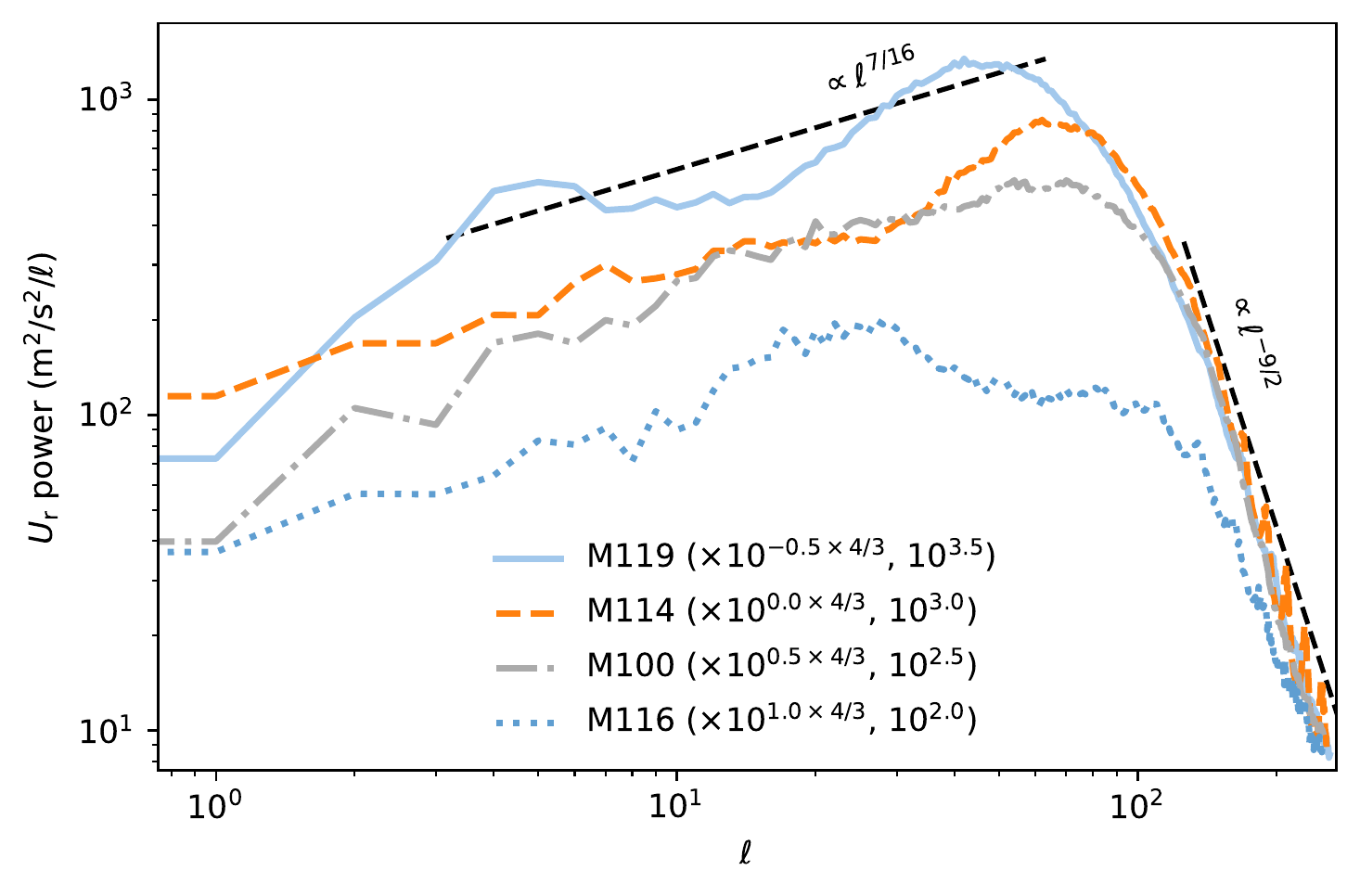}
  \includegraphics[width=\columnwidth]{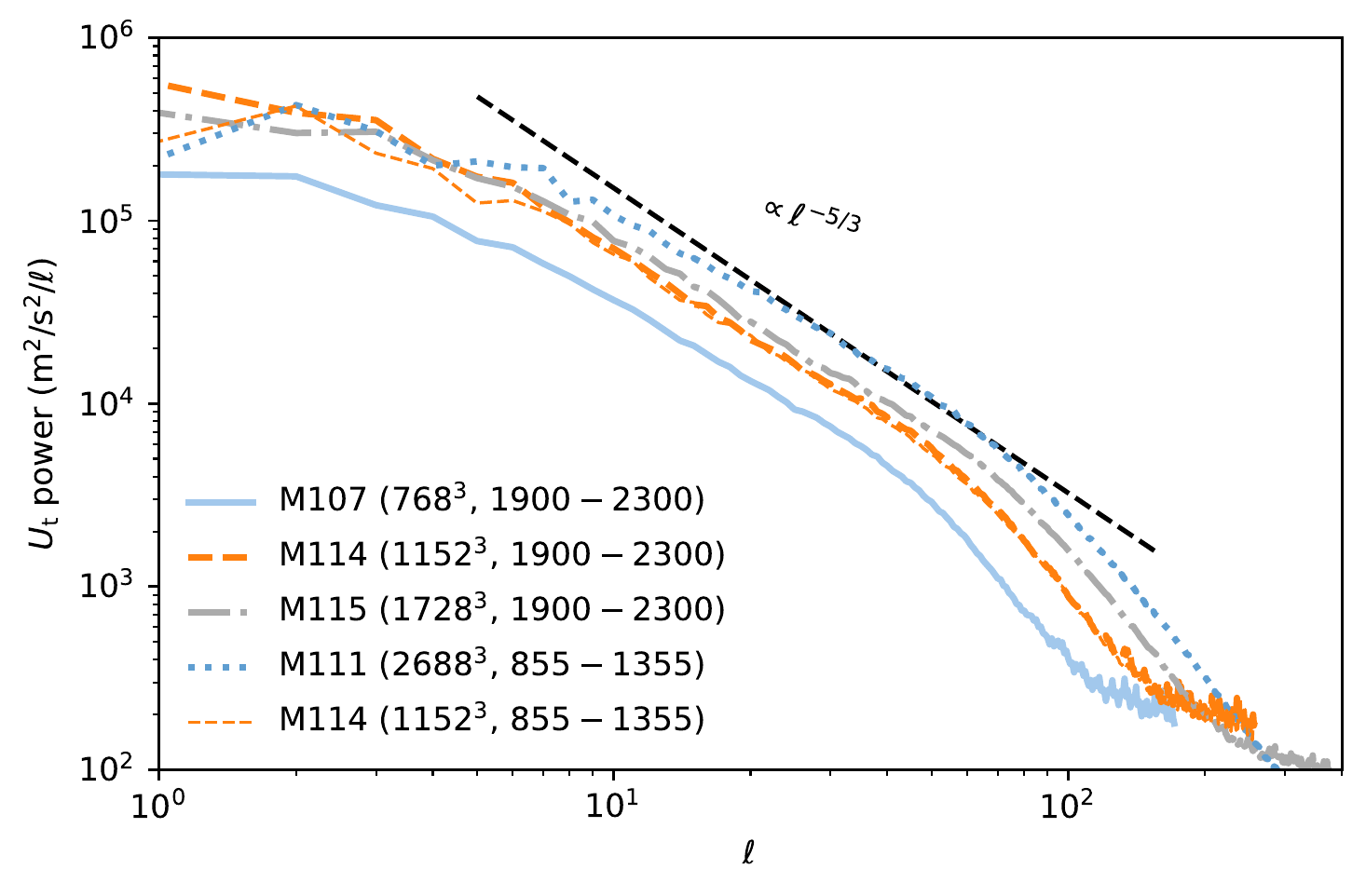}
  \includegraphics[width=\columnwidth]{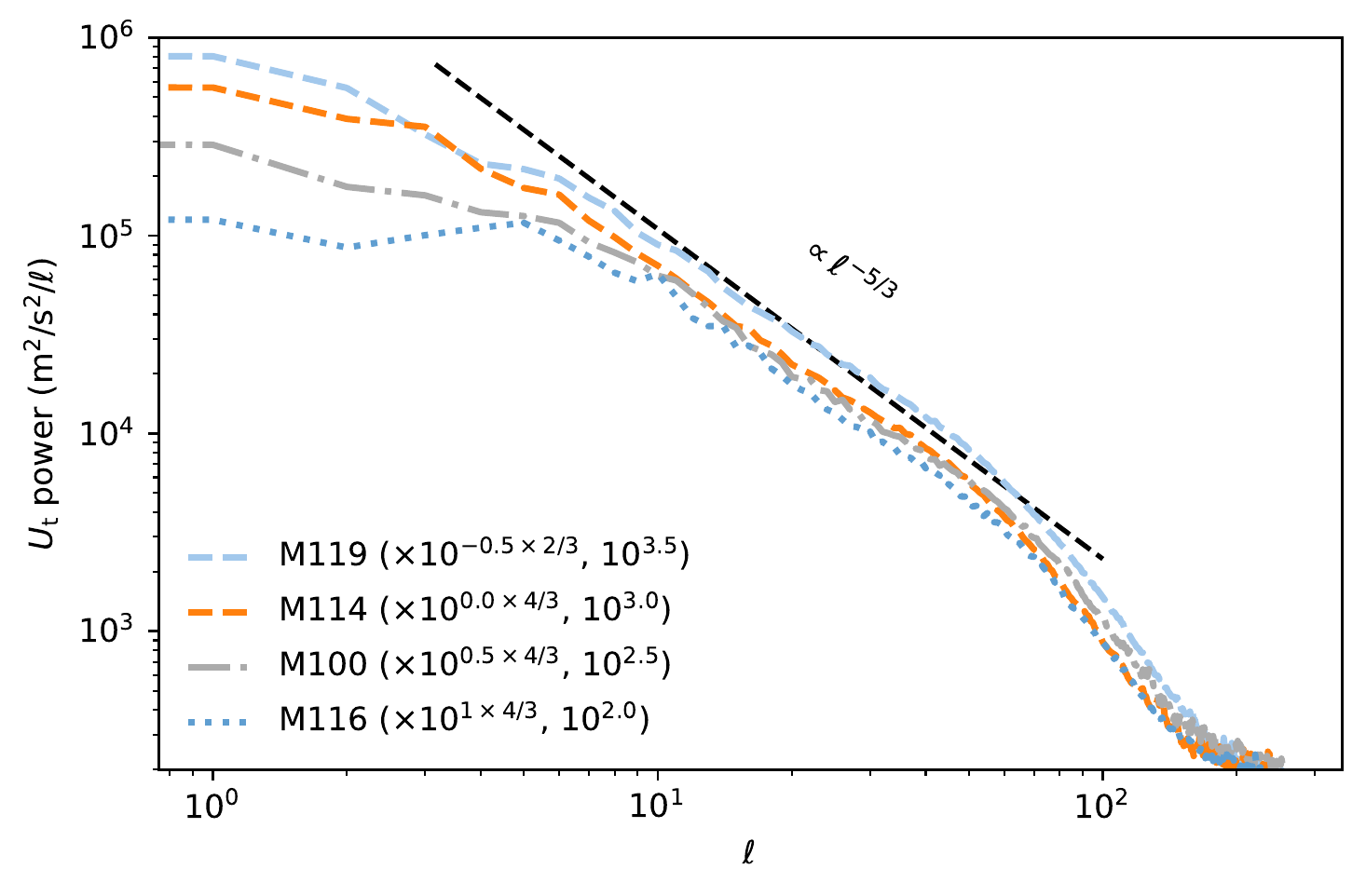}
  \caption{Spatial power spectral distribution of radial (top row) and
    horizontal (bottom row) velocity component for different
    resolutions at $1000\times$ heating factor (left column) and for
    different heating factors (right column, $1152^3$ grids)
    at the location of the maximum of the \npeak. Spectra for
    different heating factors are scaled according to the velocity
    scaling relations established in \Sect{vorticity} as indicated in
    the legend. Various power laws are shown to guide the
    eye. Spectra with different boost factors shown in the
      right panels are also averaged over $300$ to $400$ dumps.}
  \lFig{fig:N2-spectrum}
\end{figure*} 
\begin{figure}  
  \includegraphics[width=\columnwidth]{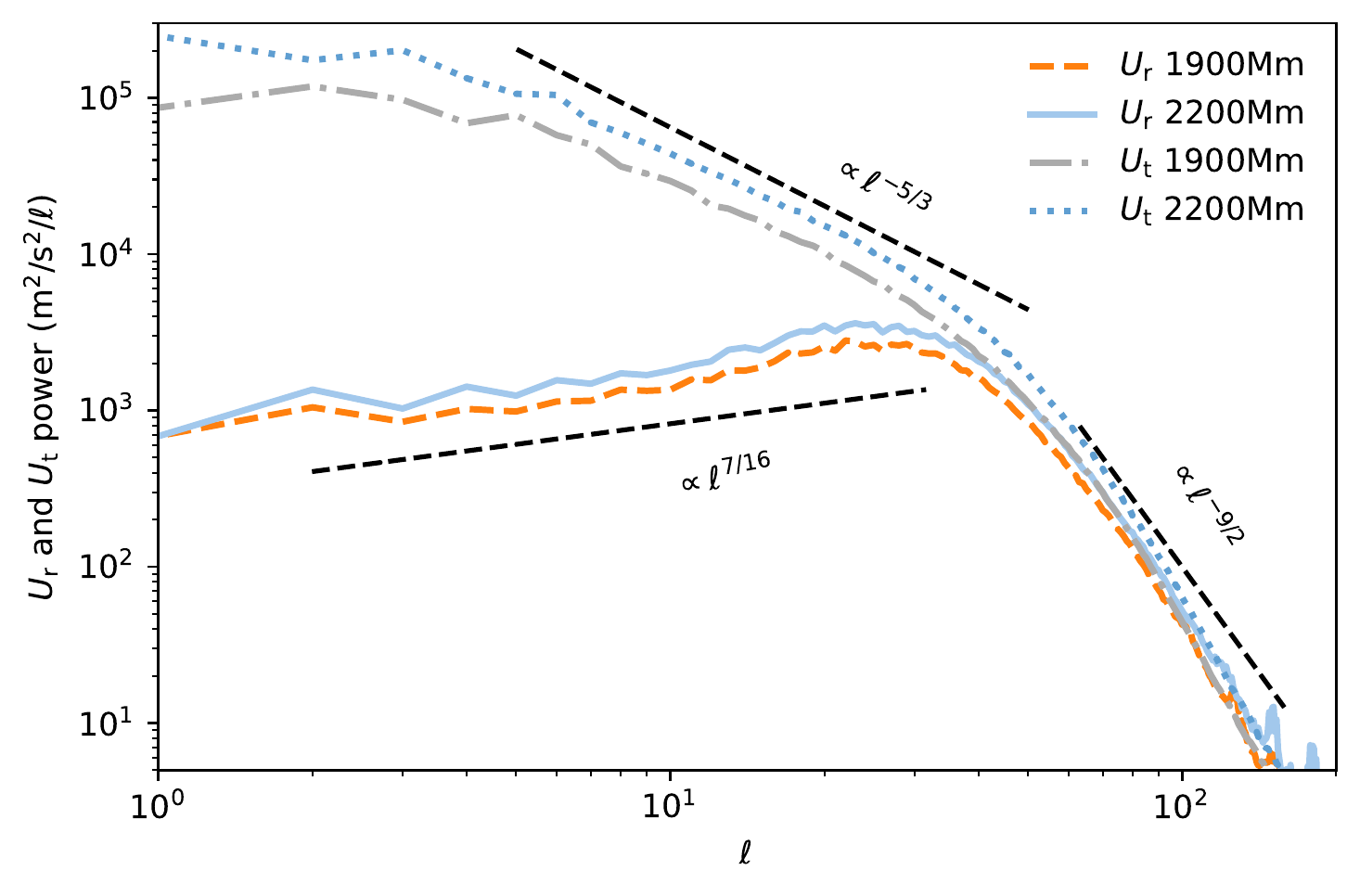}
  \caption{Horizontal and radial velocity component power
      spectra as in \Fig{fig:N2-spectrum} for run M114 (1152-grid)
      averaged over dump range $1900$ to 2300 at two radii in the
      envelope one to two pressure scale heights above the
      \npeak\ location.}  \lFig{fig:N2-spectrum-env}
\end{figure}

This is demonstrated by performing a spatio-temporal wave analysis
(see \Sect{s.wave_analysis} for details). \komega s for core and
envelope radii are be presented by
\citet[][\papertwo]{Thompson:2023a}. Here we focus just on the wave
analysis of the immediate convective boundary
layer. \Fig{fig:kw-boundary} shows the \komega\ derived from the 3D
simulations for the \npeak\ radius along with the $n=-1$ modes
predicted by \code{GYRE} for the M114 stratification.

As expected for IGWs, low-frequency modes have overall a large ratio
of horizontal to vertical velocity component
($u_\mathrm{h}/u_\mathrm{v}\gg 1$). Given the kinetic energy flux of
IGWs \citep[e.g.\ Eq.\,39][]{press:81}
\begin{equation}
  F_\mathrm{IGW} = \rho u_\mathrm{v}^2
  \frac{N}{k_\mathrm{h}}\sqrt{1-\frac{\omega^2}{N^2}} = \rho
  u_\mathrm{h}^2\frac{\omega^2}{Nk_\mathrm{h}}(1-\frac{\omega^2}{N^2})^{-1/2}
  \lEq{e.Figw}
\end{equation}
the frequency dependence of the ratio of the velocity components is 
\begin{equation}
\frac{u_v^2}{u_h^2} = \frac{\omega^2}{N^2 (1-\omega^2/N^2)}  
\lEq{e.uvuh_ratio}
\end{equation}
and shown in \Fig{fig:igw-properties}. The velocity ratio is
$u_\mathrm{v}/u_\mathrm{h} >1$ for $\omega / N >0.7$ but for $\omega
\ll N$ the power of the horizontal velocity component exceeds the
power in the radial component by two orders of magnitude\footnote{We
use both $r$, $v$ and $t$, $h$ indices synonymously for the
vertical/radial and tangential/horizontal components.}.

\begin{figure}     
   \includegraphics[width=\columnwidth]{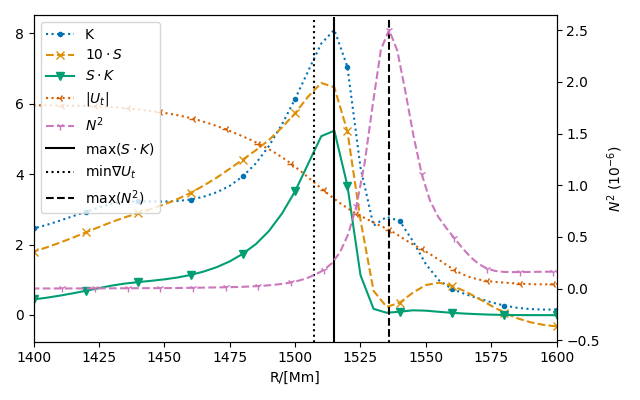}
   \includegraphics[width=\columnwidth]{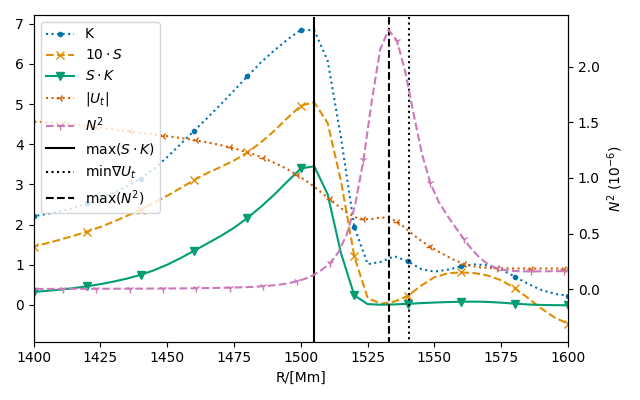}
   \includegraphics[width=\columnwidth]{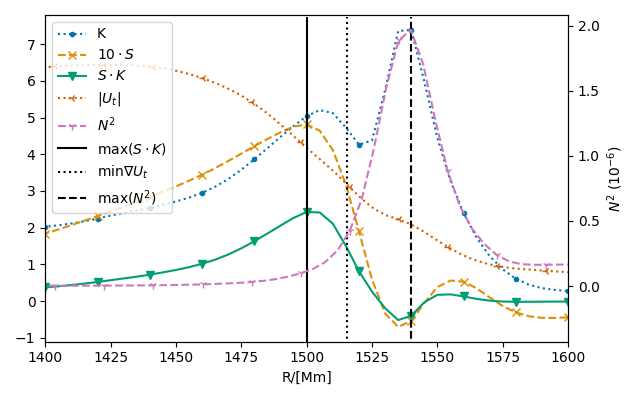}
   \caption{Radial profiles of kurtosis $K$, weighted skewness $S$
     (scaled a factor of $10$ for visibility), their product $S \cdot
     K$, the norm of the tangential velocity $|U_\mathrm{t}|$, and the
     \brunt\ frequency $N^2$ in units of $\mathrm{rad/s}$. Top: M115
     at $1615.25$hr (dump 2275 shown in
     \FigTwo{fig:M115-images-HcoreM025Z0}{fig:M115-Mollweide-bound});
     middle: M115 at $1444.15$hr ; bottom: M114 at $1619.5$hr.  }
   \lFig{fig:MomsProf-KSN2}
\end{figure}
\begin{figure*}
       \includegraphics[width=\columnwidth]{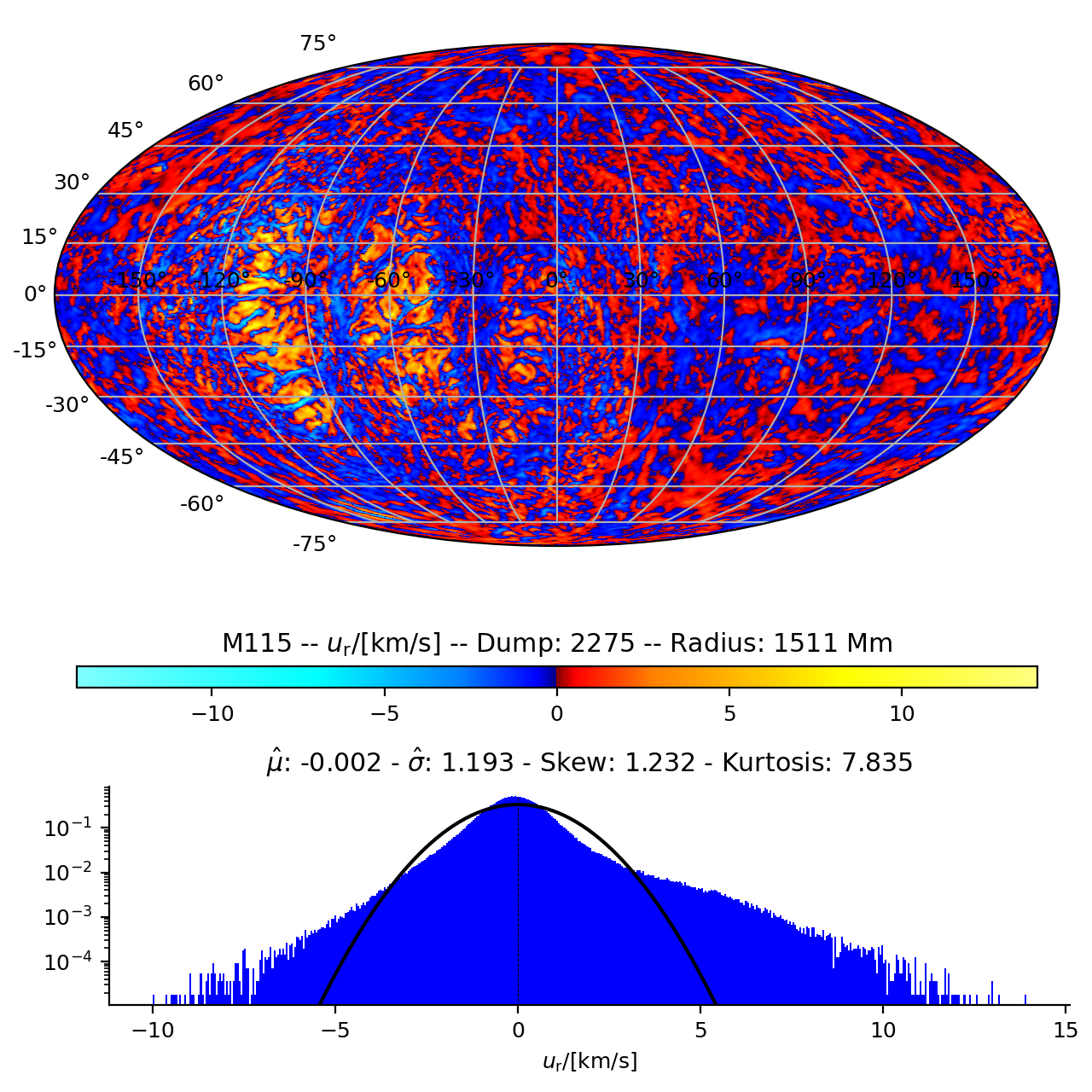}
       \includegraphics[width=\columnwidth]{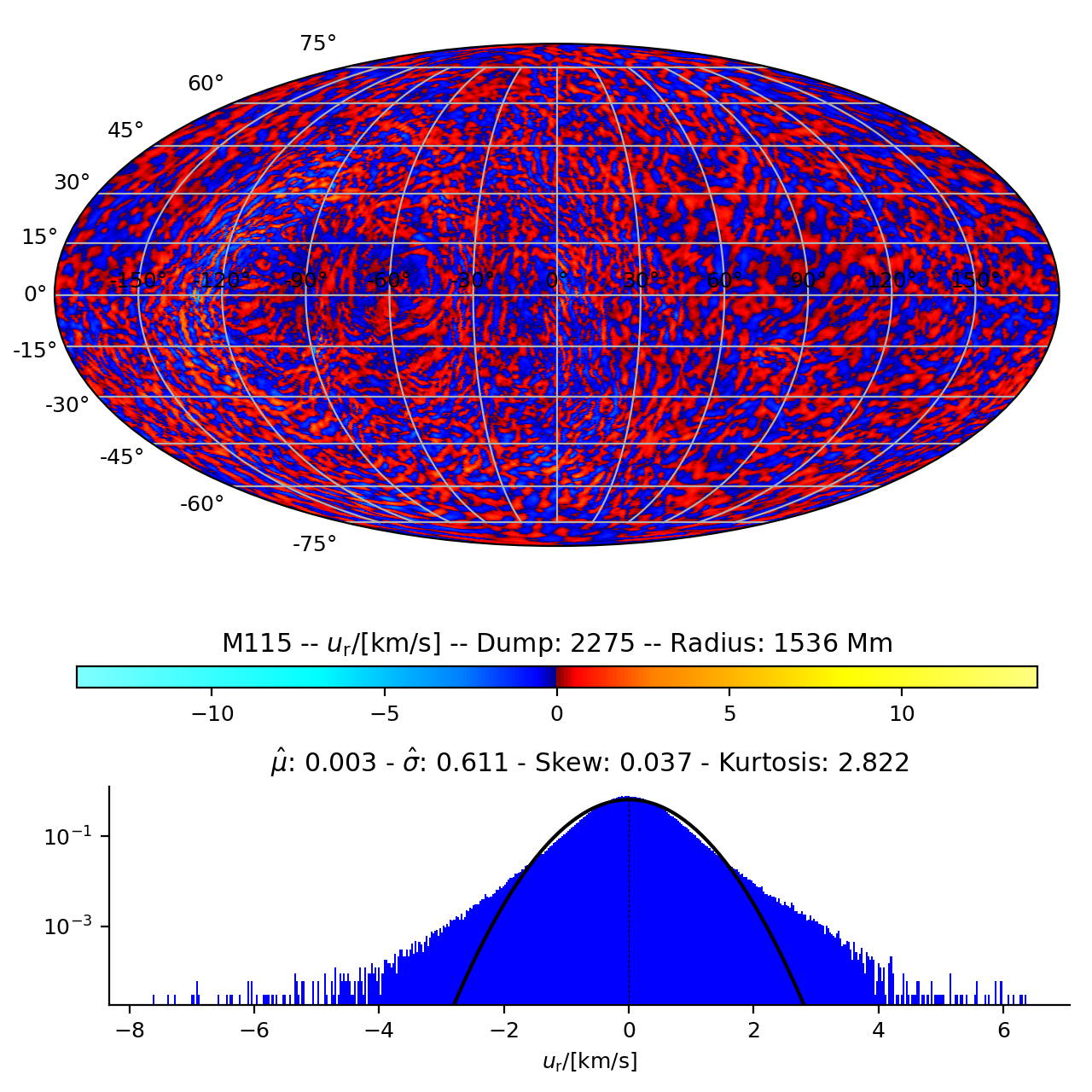}
       \caption{Mollweide projections of the radial velocity component
         and PDF on the $4\pi$ sphere. Left: At the location of the
         convective boundary according to the K-S-product-peak
         criterion at $r_\mathrm{KS-peak}=1511\Mm$.  Right: At the
         location of the peak of $N^2$ at $r_\mathrm{N^2-peak} =
         1536\Mm$.}  \lFig{fig:M115-Mollweide-bound}
\end{figure*}
The power spectral distribution shown in \komega s for the radial and
horizontal velocity components (\Fig{fig:kw-boundary}) reflect this
expectation that $U_\mathrm{t}$ is overall larger than $U_\mathrm{r}$
power. The power in the radial component $U_\mathrm{r}$ is dominantly
associated with the $n=-1$ mode at high $l$. At low frequencies, a
much smaller amount of power is associated with higher $n$ modes. The
$U_\mathrm{r}$ power is largest at high $l \geq 50$ and high
frequencies $f \approx \unit{140}{\mu\Hz}$. This frequency is much
higher than the \emph{convective frequency}, which is $\approx
\unit{2.5}{\mu\Hz}$ at $1000\times$ heating.  Power associated with
convective motions is found in the lower-left corner at $f\leq
\unit{50}{\mu\Hz}$ and $l\leq 60$
\citep[Fig.\ 13][\papertwo]{Thompson:2023a}.  The $U_\mathrm{r}$
\komega\ shows essentially no power that could be associated with
those frequencies. The $U_\mathrm{t}$ power on the other hand is
dominantly concentrated in eigenmodes with low frequencies and
correspondingly low wave numbers.

\begin{figure*}
  \includegraphics[width=\textwidth]{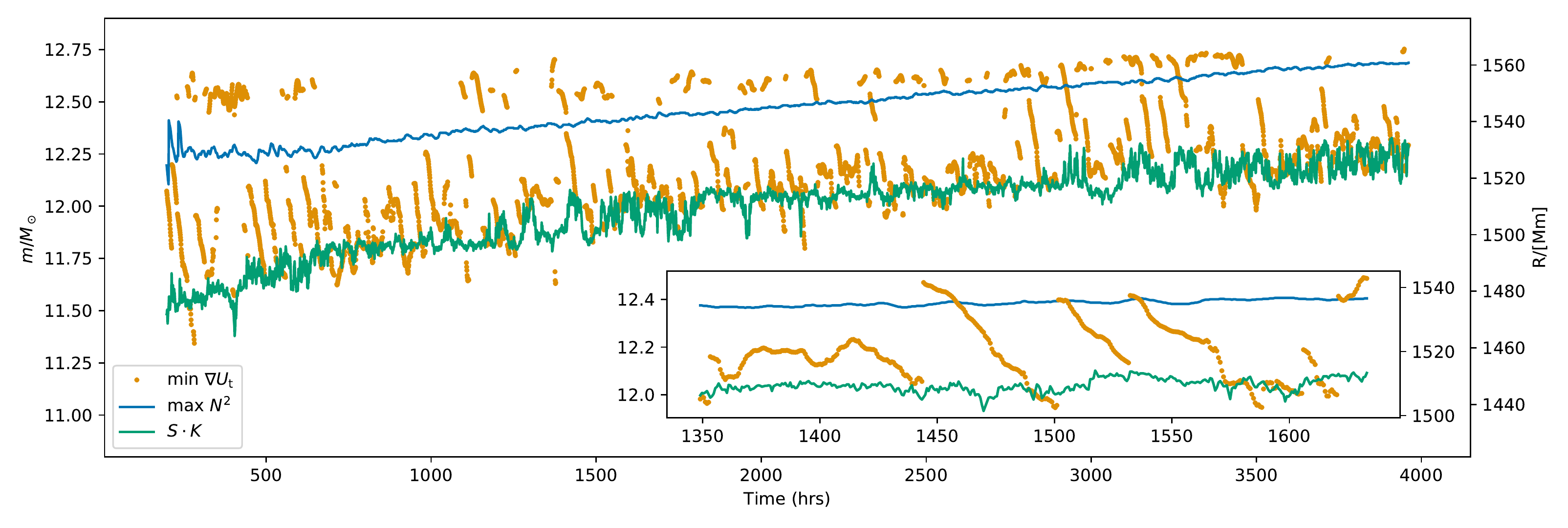}
  \caption{Time evolution of the mass coordinate
    of the maximum of the skew-kurtosis product $S K$, the steepest radial gradient
    of the tangential velocity component \mindudr\ and
    the peak of the squared Brunt-V\"ais\"al\"a frequency $\max(N^2)$
    for simulation M114 ($1152^3$) and, in the inset, more detail from
    the M115 simulation ($1728^3$). These three quantities mark the
    location of the convective boundary in different ways, see text
    for discussion.} \lFig{fig:Time-ConvBound}
\end{figure*}         
\Fig{fig:igw-properties} (bottom panel) shows the radial and
horizontal components of the displacement amplitude of two $n=-1$
modes from the \code{GYRE} calculation, for $l=3$ and $l=30$. The
high-$l$ modes have peaks of opposing direction bracketing a node at
the $N^2$ peak location. The $l=3$ mode as an example for a low-$l$
mode also has a sign change in the horizontal velocity component at
the \npeak\ radius. This means that these IGW modes have horizontal
components exactly opposite right above and below the
\npeak\ location, where the radial component has a single maximum. For
high $l$ values, the mode amplitude is sharply peaked in and around
the narrow $N^2$-peak region and falls off quickly both outwardly in
the stable layer and inwardly in the convectively unstable layer.

Inspection of centre-plane horizontal-velocity component slices
(\Fig{fig:neq1-mode}) immediately reveal these $n=-1$ modes and
specifically the nodal location that separates opposite directions of
horizontal flow. The location of the \npeak\ radius is shown as a thin
black line and coincides for most of the boundary arc shown with a
minimum in $|U_\mathrm{t}|$. Along the boundary, the IGW fluid motion
is detached from the convective horizontal flow and its independent
and distinct nature becomes apparent. The vorticity image also reveals
the layered nature of the flow in the \npeak\ region, which is
distinctly different from the irregular vorticity distribution
characteristic of convection as seen in the core.

The discussion of IGW mixing in terms of the shear-mixing model
(\Eq{eq:digwtherm}, \Sect{digw-thermal}) involves scaling relations of
vorticity (\Sect{vorticity}). The spatial spectra represent the scale
distribution of velocity power and therefore determines the
denominator of the velocity derivative vorticity. It is therefore
useful to establish how the spatial spectra at \npeak\ depend on
heating factor and on grid resolution. This is shown in
\Fig{fig:N2-spectrum}. For both velocity components the spectra are
truncated at high $l$ according the grid resolution. The spectra are
extracted from the filtered \emph{briquette} data outputs
(\Sect{hydro_sims}) with grid resolution reduced by a factor 4 in each
dimension. If $N_\mathrm{r}$ is the radius of \npeak\ in units of
simulation grid cells then the maximum resolvable $l$ is
$l_\mathrm{max} = \pi N_\mathrm{r}/4$ which is $174$ for a $768^3$
grid, and accordingly higher for finer grids.  Of course, the largest
$l$ that can be captured on the simulation grid that has $4x$ higher
resolution than the \emph{briquette} data is accordingly higher. In
simulations with radiation diffusion high-$l$ modes would be truncated
due to radiative damping. Thus, while in these simulations the
downturn or truncation at high $l$ is impacted by the given resolving
power the overall shape of the spectrum is expected to be similar to
that in simualtions with radiative damping in which the high-$l$
truncation is not caused by the limited grid resolution.

The radial and horizontal velocity components have very different
spatial spectra.  $U_\mathrm{r}$ spectra are peaked around $l \approx
60$ with a steep decline at higher wave numbers. $U_\mathrm{t}$
appears to resemble the power law characteristic for
turbulence. However, carefull inspection of the bottom-left
  region of low $l$ and $\nu$ in \Fig{fig:kw-boundary} shows that in
the \komega\ power is associated with discrete IGW
eigenmodes, and not chaotically distributed at lower wave number and
frequency as is typical for convection (see \papertwo\ for
  examples of \komega s for the proper convection zone).  In addition,
  the spectra further away from \npeak\ (see
  \Fig{fig:N2-spectrum-env}) show the same spectral shapes and power
  laws for both radial and tangential velocity components at a
  location where the velocity is undoubtedly purely of IGW nature. In
  particular, also in the envelope does the horizontal velocity
  component power spectrum follow up to the $l$ where the peak in
  power for $U_\mathrm{r}$ is located, the $-5/3$ power law. The only
  difference is that the peak in power for $U_\mathrm{r}$ is at $l
  \approx 30$ in the envelope rather than near $\approx 60$ at
  \npeak. The envelope IGW spectrum is further discussed in
  \papertwo. It therefore appears that the familiar $-5/3$ power law
  that $U_\mathrm{t}$ IGW spectra show at lower $l$ is not a symptom
  of the flow actually being turbulent. However, IGW eigenmodes have a
  radial displacement amplitude profile (\Fig{fig:igw-properties}) and
  spectra are global. Since there is little resistance to flow in the
  horizontal direction it is maybe reasonable to expect that the
  turbulent excitation spectrum manifest at least in the horizontal
  velocity component does imprint itself onto the IGW
  spectrum. However, as shown here, the radial velocity component does
  not follow this pattern.

According to our analysis the $U_\mathrm{r}$ power is independently of
heating distributed along the $n=-1$ and other IGW eigenmodes
(\Fig{fig:kw-boundary}). Run M119 should have more power at larger $l$
than the lower heating runs which is not the case. Irrespective of
heating rate the $U_\mathrm{r}$ spectra drop off steeply at high $l$
in a similar manner. The spectrum of neither velocity component
depends much on grid resolution for $l \loa 60$, but more power
appears for higher wavenumbers for finer grids. For all cases shown in
the top-left panel the peak of the spectrum falls in the range $l \in
[60 ... 74]$. The details of the spectra depends on the exact shape of
the \npeak\ feature. In the left panels the M114 case is shown for
both the later dump range when the boundary has migrated through the
initial profile, and the earlier dump range that is also available for
the highest resolution run M111 (cf.\ \Sect{s.bnd_sph_ave}). For the
$U_\mathrm{r}$ power the M114 run for the early dump range and M111
(also for the early dump range) agree very well on the left up-sloping
part of the spectrum. At the peak these two lines depart from each
other and at the down-sloping high-$l$ part to the right of the peak
instead both M114 spectra for the different dump ranges agree very
well. This indicates that for the left part of the spectrum
corresponding to larger-scale modes the shape of the \npeak\ dominates
over grid resolution, whereas for high wavenumbers the resolving power
of small scales corresponding to grid resolution becomes important.

For the tangential velocity component (bottom row in
\Fig{fig:N2-spectrum}) the spectrum does not depend significantly on
heating rate, nor on grid resolution, except that again for more
refined grids power extends to higher wave numbers. If anything, it
appears that lower heating rates have less power at the lowest wave
numbers which generally for IGWs correspond to lower frequencies,
despite having lower convective frequencies. However, this difference
is probably rather attributed to the systematic difference in
\npeak\ shape considering the discussion above concerning the two dump
ranges shown for M114.

The conclusion of this section is that the various diagnostics of the
velocity field support the finding that at the radius of the \npeak\,
the flow is dominantly due to IGWs, and that the radial velocity power
is dominantly in the $n=-1$ mode.

\subsection{Where is the convective boundary?}
\lSect{conv_bound_mix} 
As shown in \Sect{s.statistics_pdfs}, convective and wave fluid
motions have very different statistical properties. Convective flow
has an asymmetric (high skew) and fat-tailed (high excess kurtosis)
radial velocity distribution function. Wave motions have a Gaussian
PDF. In addition to the wave analysis presented in the previous
section, we can use this statistical property to characterize the
boundary layer and determine quantitatively how convective motions
transition into wave motions.

Using the 3D briquette data output, we determine the higher-order
moments skew ($S$) and excess kurtosis ($K$) as a function of
radius. \Fig{fig:MomsProf-KSN2} shows the profiles of these quantities
for three times in the M115 simulation. The times were selected to
demonstrate properties of different quantities to track the location
of the convective boundary, as explained below. The general behavior
of the higher-order moments is to increase substantially outward
toward the convective boundary. Both skew and kurtosis have a
prominent peak approximately $20$ to \unit{40}{\Mm} below the location
of the $N^2$ peak. However, for the kurtosis this may be a local
maximum, with the global maximum at times aligning with the peak of
$N^2$ as in the example shown in the bottom panel of
\Fig{fig:MomsProf-KSN2}. The skew may have a second local maximum just
outside of the $N^2$ peak. However, the product $S\cdot K$ has one
easily detectable maximum at the top of the convection zone, close to
the location where the gradient of the tangential velocity has a
minimum most of the time.

In \Fig{fig:M115-Mollweide-bound}, we show radial velocity projections
using the same colour maps as in \Fig{fig:M115-Mollweide-conv}. Now,
the PDFs are on a logarithmic scale to better show the far-tail
distributions. Shown are the projected $U_\mathrm{r}$ image and PDF
for the radius of $\max S\cdot K$ and $\max N^2$ for dump 2275, which
is also shown in the top panel of \Fig{fig:MomsProf-KSN2}. In both
distributions, maximum and mean are now nearly identical, reflecting
the symmetry of up- and downflows for most fluid elements. Comparing
the left panels of \Fig{fig:M115-Mollweide-conv} with both panels in
\Fig{fig:M115-Mollweide-bound} shows that two side-by-side upflow
regions at longitude $-60\deg$ and $-100\deg$ and latitudes ranging
from $+15\deg$ to $-45\deg$ leave clear imprints at the location $\max
(S\cdot K)$, where they represent the largest radial velocities. These
convective motions are still identifiable in the right panel of
\Fig{fig:M115-Mollweide-bound} at the radius of $\max N^2$, although
at velocity magnitudes that represent less of an outlier to the
general distribution. While most surface areas at the location of
$\max (S\cdot K)$ approach the Gaussian distribution characteristic of
wave motions with generally lower and symmetric radial velocities,
substantial convective incursions take place, especially where the
dipole impacts the convective boundary. These populate the far tail of
the distribution, leading to very large kurtosis values. These
far-tail velocity elements are predominantly contributing positive
radial velocities as shown in the PDF in the left panel of
\Fig{fig:M115-Mollweide-bound}, which causes the asymmetry of the
distribution reflected in the large skew. However, only about
\unit{0.1}{\Hp} further out, at the location of $\max N^2$, the skew
has a minimum close to values of $0.0$ in all cases. The $n=-1$ IGW
mode (cf.\ \Sect{s.neq1IGWmode}) enforces an almost perfectly
symmetric radial velocity distribution.

At this location, the kurtosis has smaller values than where $\max
S\cdot K$, but not always. The bottom panel of \Fig{fig:MomsProf-KSN2}
shows an example where the global maximum of $K$ coincides with $\max
N^2$. However, the skew is nearly zero at this location, which
excludes the possibility that far-tail events indicated by high $K$
are due to a convective intrusion of the dipole impacting the
convective boundary, as that would be a far-tail event with only
positive velocity. Fluctuating kurtosis and nearly-zero skew may
rather be the signature of a time-variable spectrum of $n=-1$ modes.

$S$ has at all times a clear minimum of nearly-zero values at the
location of $\max N^2$, where the radial oscillation amplitude of the
$n=-1$ mode has a maximum (bottom panel \Fig{fig:igw-properties}). The
$n=-1$ mode enforces the symmetry of the flow pattern at this
location. Just above $S$ sees a low relative maximum. This is where
oscillation power shifts from the $n=-1$ to more-negative modes, and a
mix of distributions with different mean values causes asymmetry.

Although the eye is able to recognize the convective flow pattern in
the radial velocity projection at $\max N^2$ (right panel
\Fig{fig:M115-Mollweide-bound}), the PDF does not show the
characteristics of convection (large $K$ and $S$). This suggests that
coherent convective motions are not able to penetrate past the $\max
N^2$ radius.

The increasing stability of the stratification from the convection
zone to the radius of $\max S\cdot K$ and $\max N^2$ is reflected by
the decrease of the variance (given along with each PDF plot in
\FigTwo{fig:M115-Mollweide-conv}{fig:M115-Mollweide-bound}) as the
average of the convective radial velocity magnitude decreases. At
$\max N^2$, the variance is the same as it is further above in the
stable layer. This is consistent with the notion that at and above the
radius $\max N^2$, convective motions play a minor role, and that
$\max N^2$ is above the convective boundary.

The maximum of kurtosis and skew at the convective boundary can then
be interpreted as the result of a radial velocity PDF generally
contracting in terms of variance across the boundary, supplemented
however with occasional massive incursions of the large-scale
convective system, most prominently the large dipole mode. We
therefore propose the condition $\max S\cdot K$ as a dynamic criterion
for the convective boundary, above which fluid motions are dominated
by waves and below which fluid motions are predominantly
convective. This criterion is more reliable than the
\mindudr\ criterion used in \citet{Jones:2017kc}. As shown in
\Fig{fig:MomsProf-KSN2}, the \mindudr\ location is not well-defined in
these main-sequence simulations due to the strong IGW velocity
component in the region just above the convective boundary, and it can
also be located above $\max N^2$, as in the case shown in the middle
panel. This effect only becomes noticeable in simulations with high
grid resolution in which the radial morphology of the IGWs is
sufficiently resolved.

The long- and short-term evolution of the different convective
boundary criteria candidates are shown in
\Fig{fig:Time-ConvBound}. The derivative of this boundary mass
migration gives the same entrainment rate as in
\Sect{entrainment}. The difference in the variability of the three
locations is noteworthy. As explained above, the \mindudr\ location is
highly variable, as high-resolution simulations resolve IGWs and place
it on the edge of individual oscillations or at the edge of the
convection zone in some erratic and alternating fashion. The $\max
S\cdot K$ criterion, on the other hand, is well-defined, and the
radial fluctuation of the convective boundary according to this
criterion is $\approx \unit{15}{\Mm}$ or $\lessapprox
\unit{0.05}{\Hp}$. However, the radial variability of the location of
$\max N^2$ is ten times smaller, corresponding to only $1/2$ grid cell
size of run M115 with a $1728^3$ grid. This small variability over
long time-scales corresponds to the estimate of the magnitude of
spherical deformation's effects discussed in \Sect{s.bnd_sph_ave}
(cf.\ \Fig{fig:FV-boundary-M115-dump3000}).

\section{Mixing due to internal gravity waves}
\lSect{mixing-1D}
In this section, we determine the mixing efficiency in the convective
core and at the \npeak\ location using the technique outlined in
\Sect{mixing-analysis}. We present scaling relations with heating and
interpret the simulation results in the framework of shear-induced
mixing outlined in \Sect{mixing-analysis}.

\subsection{Mixing in terms of diffusion due to convection and IGWs}
\lSect{digw-dynamic} In the previous section we have demonstrated how
the flow transitions from convective advection-dominated to wave
motion-dominated in the region between the $S \cdot K$ peak and the
$N^2$ peak (\Fig{fig:MomsProf-KSN2}, \Sect{conv_bound_mix}). Around
the \npeak\ radial fluid motions are dominated by the $n=-1$ IGW mode
(\Fig{fig:kw-boundary}), and to the left of the \npeak\, mixing is
mostly due to the decaying convective boundary flow.

\Fig{fig:f-bound-mix} shows the determination of the $D$ profile from
the diffusion equation inversion method (as described in
\Sect{mixing-analysis}). For this method to work well, it is required
that the FV gradient be not almost zero. For this reason, we measure
the convective mixing well inside the convective boundary but not too
deep inside the core where the FV gradient is very small
(\Fig{fig:Resolution-core-FVcld}). The coefficient \dconvhydro\ in the
convection zone is taken at the radius $0.75 \Hpzero$ below the radius
of the \npeak. The diffusion coefficient \digwhydro\ is recorded at
the radius where the $N^2$-peak is located and where IGWs dominate
mixing (\Sect{s.neq1IGWmode}). These two mixing coefficients are
measured in the same way for all runs listed in \Tab{tab:runs-summary}
and shown as a function of the heating factor in
\Fig{fig:Scaling-D-core-and-env}.
\begin{figure}
  \includegraphics[width=\columnwidth]{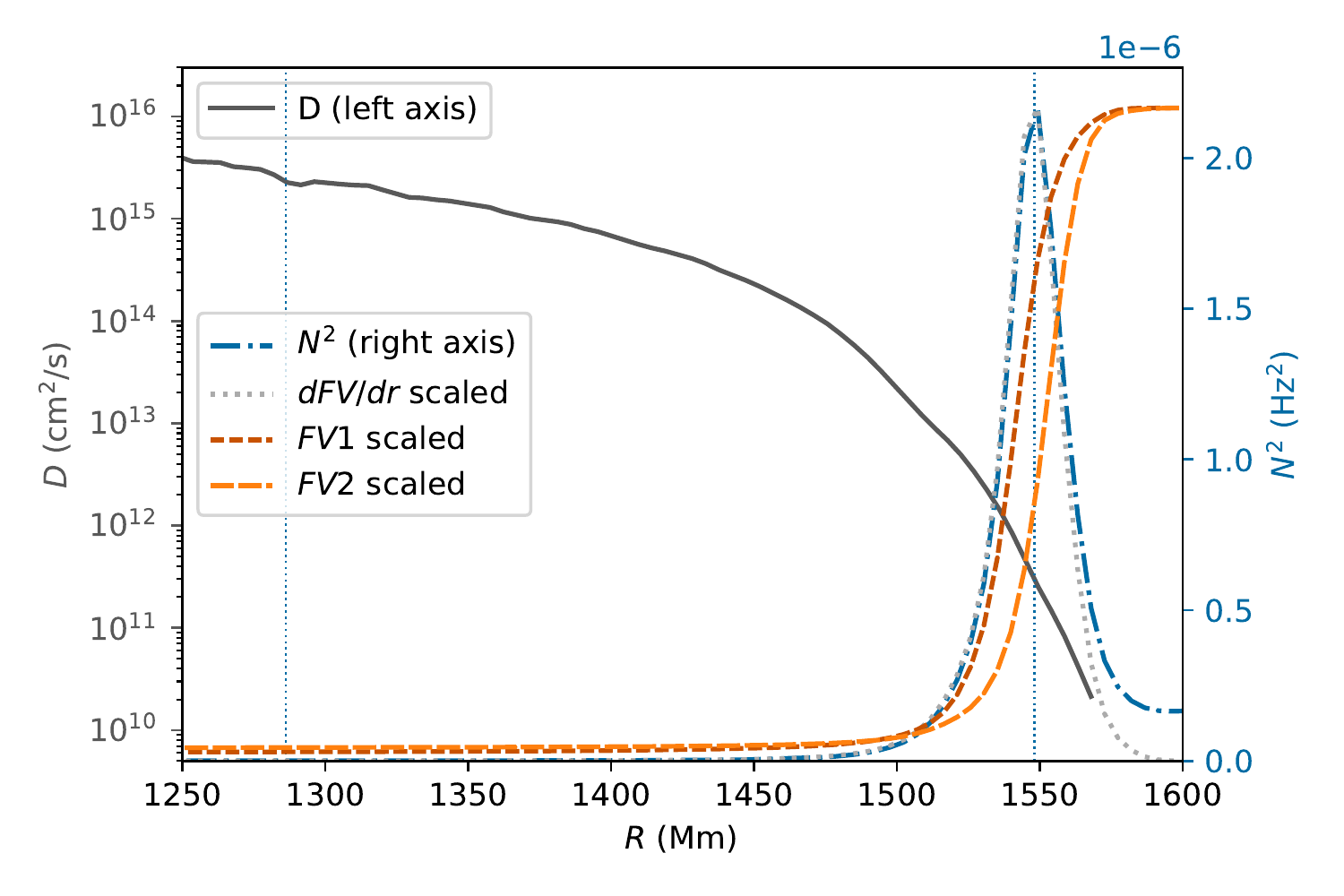}
  \includegraphics[width=\columnwidth]{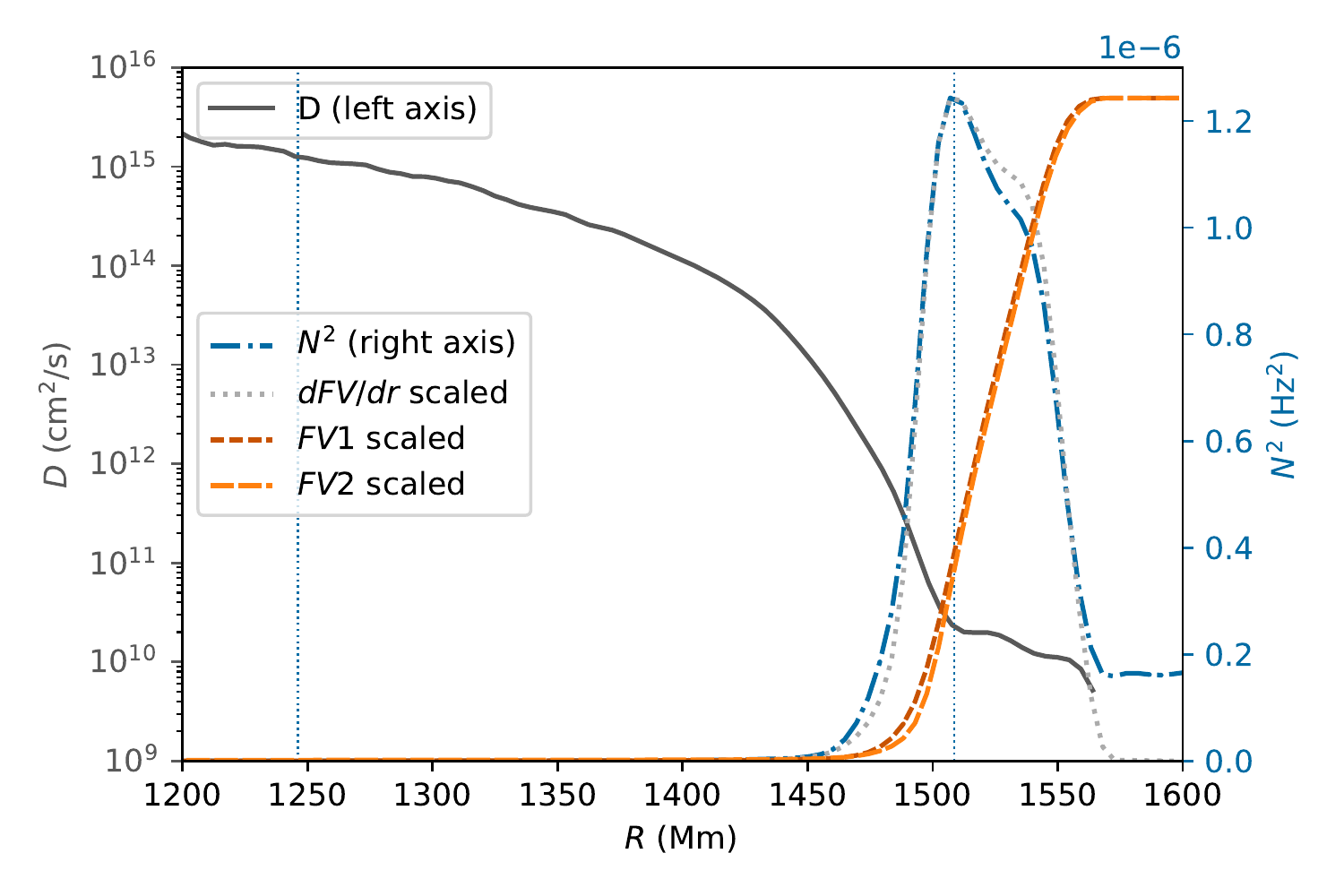}
  \caption{Diffusion coefficient $D$ from evolution of spherically
    averaged abundance profiles for heating factors $1000\times$
    (M114, top) and $100\times$ (M116, bottom). $D$ is
      calculated according to the diffusion equation inversion method
      (\Sect{mixing-analysis}). For M116 the two time average dump
      ranges are $[1050, 2050]$ and $[3050, 4050]$ respectively, and
      for M114 the averaging dump ranges are $[2725, 3200]$ and
      $[3450, 3925]$ respectively. Shown also is the
    \brunt\ frequency squared (blue dash-dotted) and the gradient of
    the fractional volume concentration (dotted grey lines,
      taken of the FV profile at the dump in the middle of the total
      range used for this analysis, which is dump $2550$ for M116 and
      3325 for M114, and scaled so that the FV gradient and $N^2$
    curve peaks match). The thin, blue-dotted vertical lines to the
    right indicate the radius of \npeak\ for which a power spectrum is
    shown in \Fig{fig:kw-boundary} and where \digwhydro\ is
    measured. The left vertical lines are at the location
    $0.75\Hpzero$ further inward, where the convective diffusion
    coefficient \dconvhydro\ is measured. Also shown are the two
    fractional volume concentration profiles FV1 and FV2 (scaled to
    match the range on the right axis, $FV \in [0,1]$) at the
      first and last dump of the overall dump ranges used for the $D$
      analysis.}  \lFig{fig:f-bound-mix}
\end{figure}

Just as found previously
\citep{Muller:2016kta,Jones:2017kc,Andrassy:2020}, both convective
velocity components follow the scaling $U_\mathrm{conv} \propto
L^{1/3}$ (\Fig{fig:hcorem25-heating-Ucomponents}). The convective
mixing coefficient follows the same scaling, consistent with the usual
expression $D_\mathrm{conv} =
\frac{1}{3}l_\mathrm{mix}v_\mathrm{conv}$ where $l_\mathrm{mix}$ is
the mixing length. Since both \dconvhydro\ and the convective
velocities scale with the heating factor with the same power, then,
assuming the above expression for $D_\mathrm{conv}$, the mixing length
$l_\mathrm{mix}$ is independent of the heating factor.
\Fig{fig:hcorem25-M116-alpha_mix} shows
\begin{equation}
  \alpha_\mathrm{mix} = \frac{1}{\Hp}l_\mathrm{mix} =
  \frac{3}{\Hp}\frac{\dconvhydro}{u}  \lEq{e.alphamix}     
\end{equation}
for $u = U_\mathrm{r}$, the
radial velocity component, and $u = U_\mathrm{total} =
\sqrt{U_\mathrm{r}^2+U_\mathrm{h}^2}$, the total velocity magnitude.

The mixing-length parameter determined in this way\footnote{This
mixing-length parameter is just a quantity as defined and measured
from our simulations. It may or may not be related to the
mixing-length parameter from MLT.} increases from the convective
boundary, where it is essentially zero, to order unity at almost one
pressure scale height into the convective core, i.e.\ at about
  one pressure scale height into the convection from the Schwarzschild
  boundary $\alpha_\mathrm{mix} \approx 1$. A gradual increase of the
mixing length from the boundary to well inside the convective core has
previously been observed in hydrodynamic simulations of O-shell
convection in a massive star by \citet[][Eq.\, (4)]{Jones:2017kc}, who
suggested that the mixing length parameter should be
\begin{equation}
  \alpha_1 = \min(\alpha_\mathrm{mix},\frac{|r-r_\mathrm{SB}|}{\Hp}) \, .
  \lEq{e.alpha1}
\end{equation}
adopting $\alpha_\mathrm{mix} = 1$. $\alpha_1$ calculated in
this way is shown in \Fig{fig:hcorem25-M116-alpha_mix}, and the slight
bend reflects the radius dependence of \Hp. In these core-convection
simulations, the mixing-length parameter is better modeled with an
exponential
\begin{equation}
  \delta r = \frac{|r-r_\mathrm{SB}|}{\Hp} \, ,\ \alpha_2 =
  \min(\alpha_\mathrm{mix},4 e^{7(\delta r -1)}) \lEq{e.convbound}
\end{equation}
again adopting $\alpha_\mathrm{mix} = 1$. As shown in
\Fig{fig:hcorem25-M116-alpha_mix} this matches the mixing-length
parameter profile determined from the simulations using the total
velocity magnitude better.

\begin{figure}
  \includegraphics[width=\columnwidth]{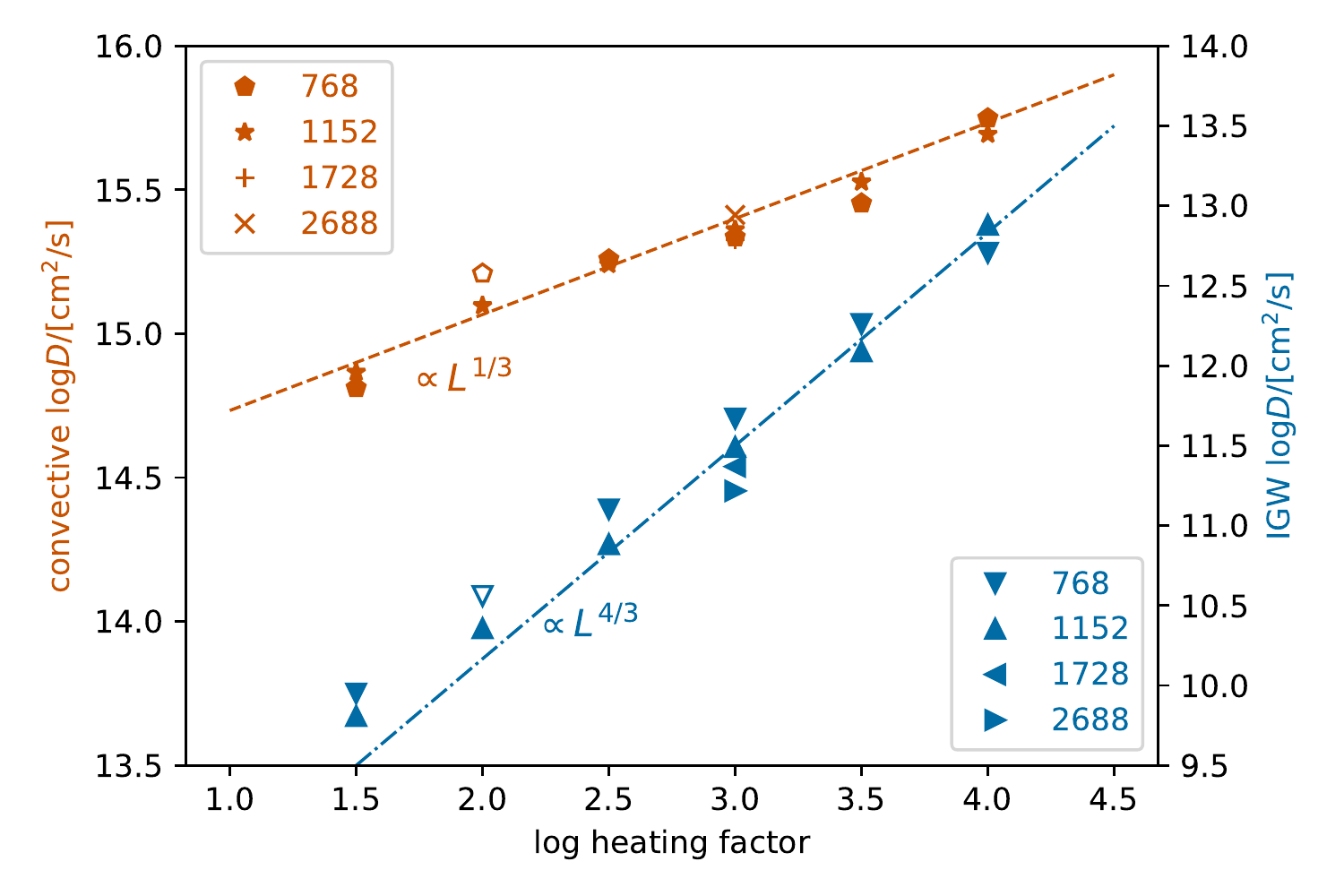}
  \caption{Scaling of diffusion coefficients determined by the
    diffusion equation inversion method (\Sect{mixing-analysis}) as a
    function of heating factor. Hexagons, stars, plusses, and crosses
    represent convective diffusion coefficients \dconvhydro\ taken at
    $0.75\Hpzero$ below the \npeak\ radius (left axis). Triangles show
    the diffusion coefficients \digwhydro\ at the \npeak\ radius as
    shown in \Fig{fig:f-bound-mix}. Two power laws shifted to be
    nearer to where the points are located are shown with dashed and
    dash-dotted lines.}  \lFig{fig:Scaling-D-core-and-env}
\end{figure}
The diffusion coefficient determined at the radius of the
\npeak\ follows a scaling with heating factor $\propto L^{4/3}$ for
heating factors $\geq 10^{2.5}$. This is distinctly different from the
mixing scaling found for \dconvhydro\ measured in the convective core
and implies that the physical mixing process is fundamentally
different from turbulent convection. This adds evidence to the
expectation that mixing at the \npeak\ is caused by IGWs. In
\Sect{digw-thermal} we compare this mixing with predictions in terms
of IGW-induced shear according to \Eq{eq:digwtherm}, and we turn
therefore now to exploring the properties of vorticity in our
simulations.

\subsection{Vorticity scaling relations}
\lSect{vorticity}
Expression \Eq{eq:digwtherm} relies on the assumption that the
horizontal vorticity component is much larger than the radial
vorticity component, so that $\vort \approx \vort_\mathrm{h}$. This is
indeed expected for IGWs and borne out by our simulations, as shown in
\Fig{fig:hcorem25-R-vortN2Ri}. The bottom panel shows the exact
location of the \npeak\ in relation to the vorticity profile, as well
as the Richardson number calculated from the vorticity as outlined in
\Sect{mixing-analysis}, using spline representations to find the
precise vorticity at the radius of the \npeak. This shows that the
local vorticity magnitude peak is $\approx \unit{25}{\Mm}$ further
inward relative to the \npeak, and that at that radius and above, the
ratio of horizontal to vertical vorticity components exceeds $\approx
10$.
\begin{figure}
  \includegraphics[width=\columnwidth]{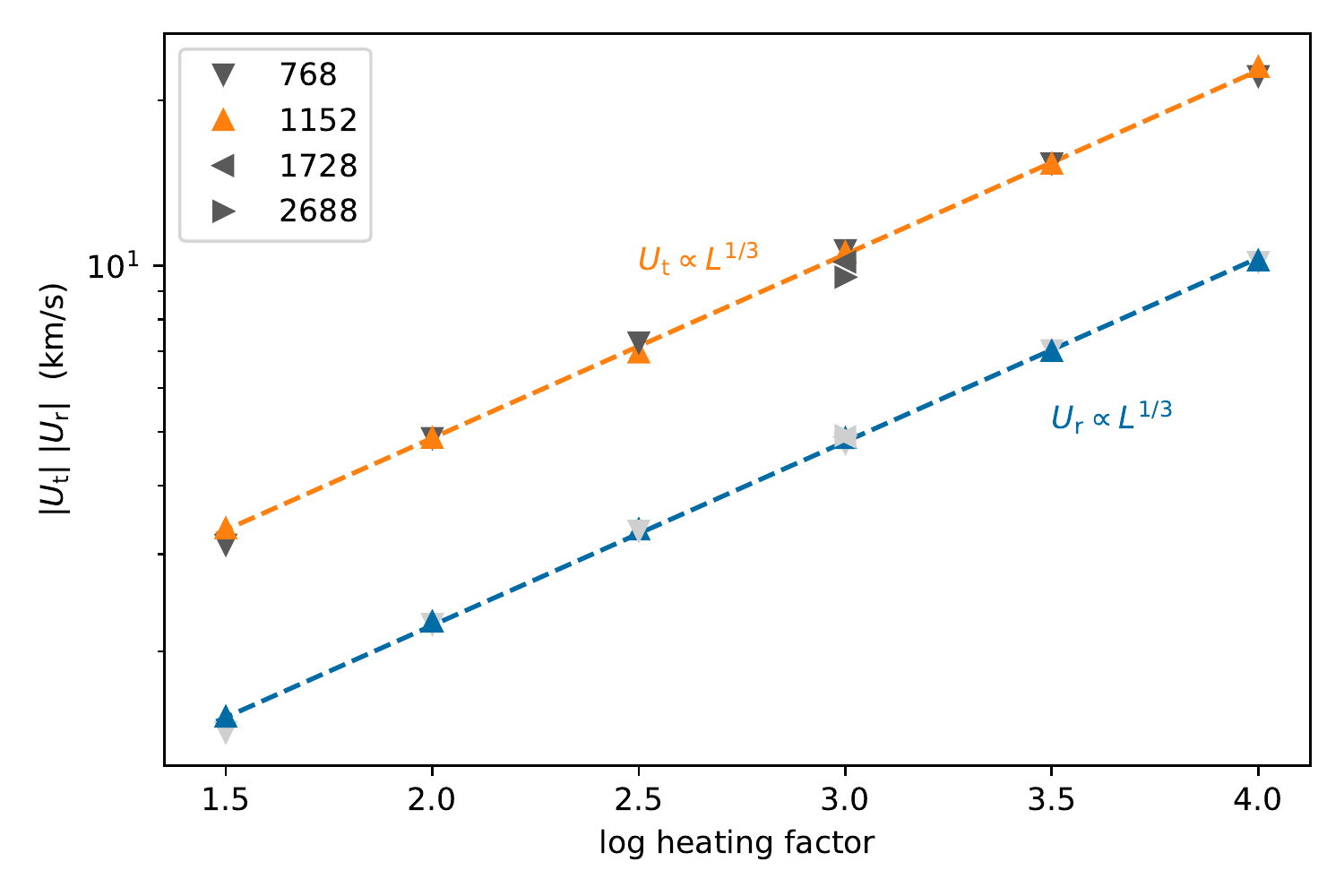}
  \caption{Horizontal and radial velocity magnitude in the convection
    zone at the same radius, $0.75\Hpzero$ inward from the radius of
    the \npeak. }  \lFig{fig:hcorem25-heating-Ucomponents}
\end{figure}
\begin{figure}
  \includegraphics[width=\columnwidth]{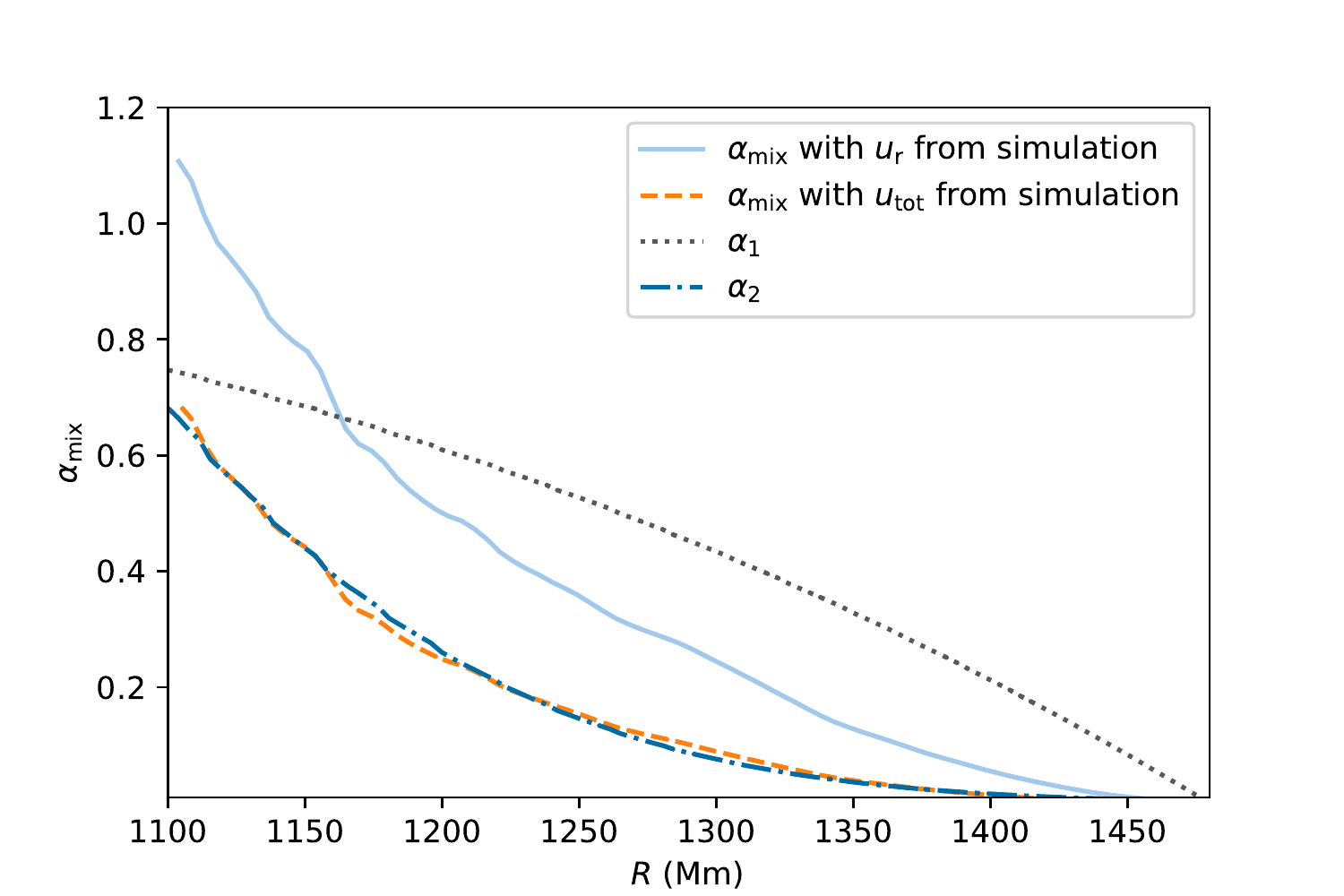}
  \caption{Mixing-length parameter $\alpha_\mathrm{mix}$ as a function
    of radius in the outer third of the convective core for simulation
    M116 ($100\times$ heating factor, $1152^3$-grid) using
      \Eq{e.alphamix} with radial velocity component (blue solid) and
      total velocity magnitude (dashed orange) as well $\alpha_1$
      according to \Eq{e.alpha1} and $\alpha_2$ according to
      \Eq{e.convbound}. See text for details.}
  \lFig{fig:hcorem25-M116-alpha_mix}
\end{figure}

The vorticity profiles for four different grid resolutions are shown
in \Fig{fig:vort_conv_calc}. The lower three grid resolutions are
shown averaged over a time range after the convective boundary has
migrated through the initial \npeak\ profile
(\Fig{fig:RProf-N2FVUt-boundary}). The idea was that we may avoid in
this way a possible dependence of the vorticity profile on the shape
of the \npeak\ profile. The highest resolution run was not followed to
those late times. For this reason, the time ranges over which we
average the three lower resolution runs and the highest resolution
runs are not the same. In any case, in these simulations vorticity
magnitude of IGWs in the stable layer shows no sign of convergence. At
\npeak\ \vortsq\ scales $\propto N_\mathrm{x}^2$
(\Fig{fig:grid-vort-N2peak}).  The question of whether or not the IGW
vorticity converges in the simulations will depend on the effect of
radiative diffusion, which could dampen small-scale fluctuations. This
question will therefore be revisited with simulations that include
radiative diffusion and have reached a quasi-equilibrium.

\begin{figure}
  \includegraphics[width=\columnwidth]{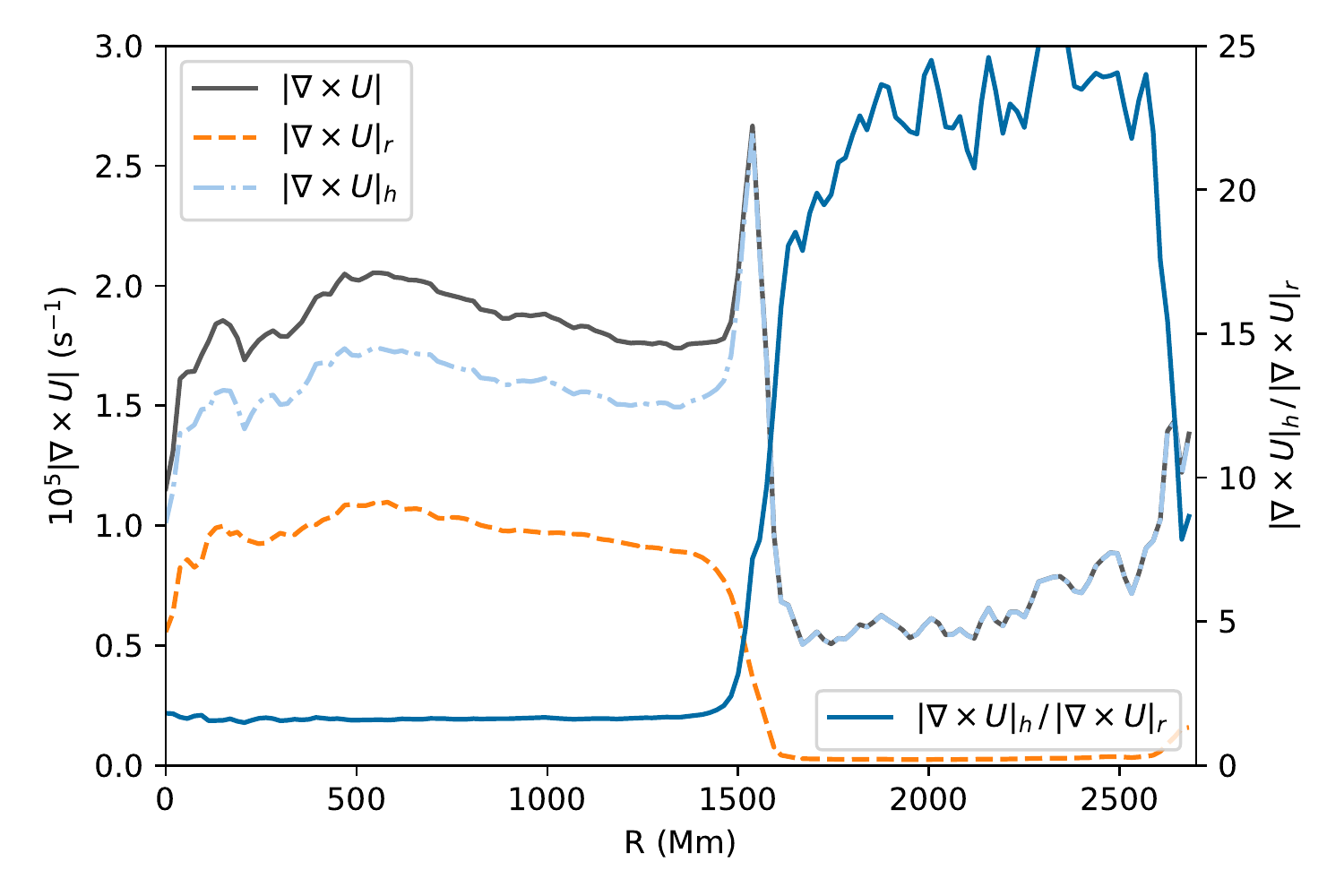}
  \includegraphics[width=\columnwidth]{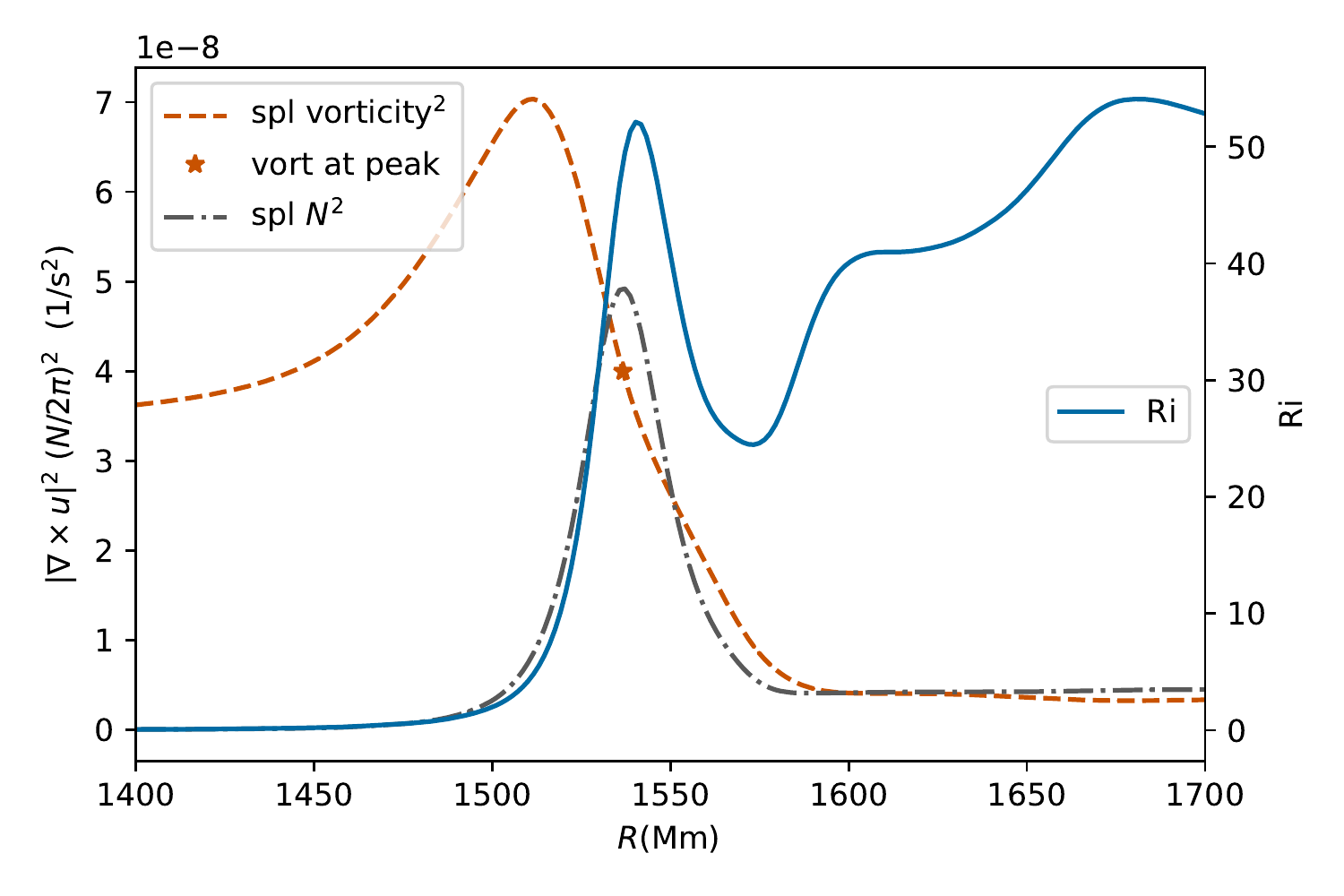}
  \caption{Top: Vorticity profile as well as the radial and horizontal
    components. Bottom: Profiles of \vortsq, $N^2$, and the Richardson
    number $Ri \approx \frac{N^2}{\vortsq}$
    (cf.\ \Sect{mixing-analysis}). The orange star indicates
      the radius of the \npeak. Both plots are for run M114 ($1152^3$
    grid, $1000\times$ heating factor) at dump 4425 corresponding to
    time \unit{3142}{\hour}.}  \lFig{fig:hcorem25-R-vortN2Ri}
\end{figure} 
\begin{figure}
  \includegraphics[width=\columnwidth]{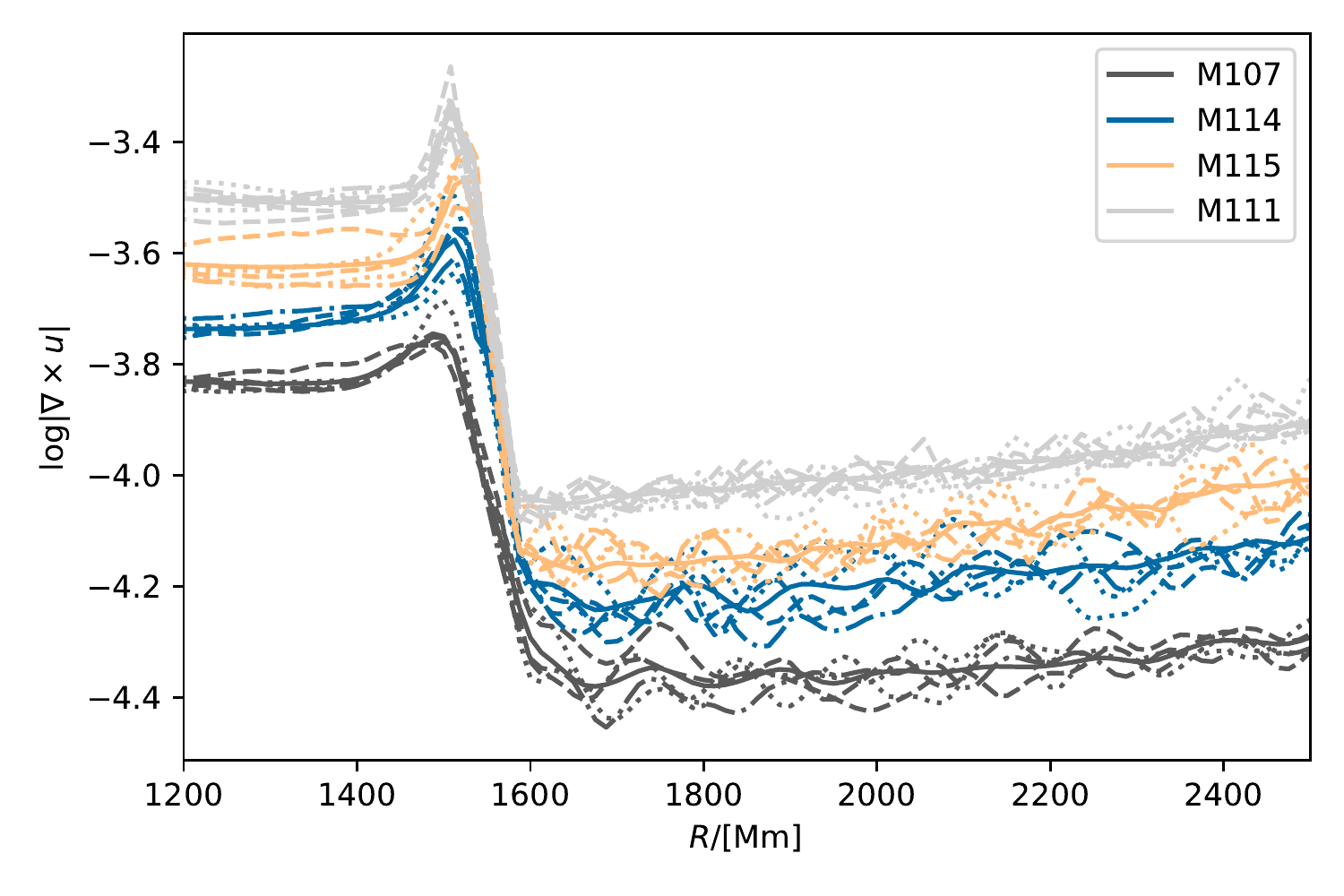}
  \caption{Spherically averaged vorticity profiles for $1000\times$
    heating runs in M107 ($768^3$), M114 ($1152^3$), M115 ($1728^3$),
    and M111 ($2688^3$). Solid lines show the average of 442 dumps
    between $t = 1349$ and \unit{1662}{\hour}, except for M111, where
    817 dumps in the time range $569$ to \unit{1316}{\hour} are
    averaged. The dashed, dotted, and dash-dotted lines show
    spherically averaged profiles for individual dumps in this time
    range every 100 dumps. The 3D vorticity data is calculated from
    the briquette data.
    }  \lFig{fig:vort_conv_calc}
\end{figure}
The analysis of the time evolution of vorticity shows fluctuations
over long periods. The M107 time evolution of the average in the range
$1900 < R/\Mm < 2100$ shows long-term $\approx 700\hour$ variations
with amplitudes of $\approx 10\%$.

Next we establish how vorticity scales with heating rate. The
simulation vorticity in the convection zone displays the scaling
relation $|\nabla \times U| \propto L^{1/3}$ (top panel
\Fig{fig:vort-scaling}). The spectrum represents the spatial scale
distribution and therefore the denominator of the velocity
gradient. The spectrum (\Fig{fig:convective-spectrum}) remains
approximately the same as a function of heating, and thus the
vorticity scaling follows the velocity scaling (\Sect{digw-dynamic}).

The expected scaling of IGW vorticity with heating factor is based on
the assumption that the kinetic energy flux of IGWs is proportional to
the convective flux
$$
F_\mathrm{IGW} = \Mach F_\mathrm{conv},
$$ where $\Mach\propto v_\mathrm{conv}$ is the Mach number \citep[e.g.][]{Rogers:2013fl}. 
Given \Eq{e.Figw}    
and $v_\mathrm{conv}\propto L^{1/3}\propto F_\mathrm{conv}^{1/3}$, it follows that 
$$
U_\mathrm{r}\propto L^{2/3}
$$
and
$$
U_\mathrm{t}\propto L^{2/3}
$$ for fixed values of $n$ and $l$ and for the same stellar model. Our
simulations are mostly consistent with this
scaling. \Fig{fig:hcorem25-heating-UtUr-N2peak} shows the tangential
and radial velocity component magnitudes at the radius of the
\npeak. For the simulations with the lowest heating factors
  the grid resolution becomes insufficient and the radial velocity
  component values drop below the scaling relation. Note how the
  $768^3$-grid simulations drop more than those with $1152^3$ grid.
The scaling relation for these velocities one pressure scale height
\Hpzero\ further out in the envelope looks the same, except with
tangential velocities roughly a factor of two smaller
(\FigTwo{fig:M114-velt-evol}{fig:M114-velr-evol}). However, at both
locations, the tangential velocity component deviates from the scaling
relation for IGWs derived above to follow $\propto L^{1/3}$ for
heating factors $> 1000$.

One may think that this break from the scaling relation by the
tangential velocity component is due to lower heating factor runs
retaining most of their original \npeak\ profiles, while velocities
for higher heating factor runs have been measured at later times, when
the boundary has migrated through the initial \npeak\ profile as
discussed previously in \Sect{s.bnd_sph_ave}. If this were the case,
we would see variation in the tangential velocity component during the
initial \unit{1500}{\hour} as the convective boundary migrates through
the initial \npeak\ profile in a $1000\times$ heating run.
\Fig{fig:M115-N2urut-time} shows that this is not the
case. Immediately after the initial $\approx \unit{100}{\hour}$ (about
a convective turn-over time), when the first convective plumes reach
the boundary, the horizontal velocity magnitude reaches its
steady-state value. During the initial $\approx \unit{1500}{\hour}$,
the maximum $N^2$ value gradually increases, signaling the continued
change of the profile. After this time, the $N^2$ max remains
constant, corresponding to the phase of self-similar migration of the
boundary. During this entire time, the tangential velocity magnitude
remains constant, demonstrating that it does not depend on the shape
of the \npeak.

Instead, a different explanation of the break from the tangential
velocity scaling relation is more plausible. The velocities of IGWs
scale with twice the power the convective velocities scale with. At
some heating factor the IGW velocities would exceed the convective
velocities. For heating factor $10^4\times$, the tangential velocity
according to the scaling law would be \unit{18.6}{\km \sec^{-1}} at
the \npeak\, whereas the convective tangential velocity component at
$0.75\Hpzero$ further inward is \unit{22.6}{\km \sec^{-1}}. Since IGW
motions are excited by convective motions in the core it makes sense
that at some point the IGW velocities cannot continue to follow their
steep power law, but instead will follow the shallower power law of
the convective motions.

For heating rates $\leq 100\times$ for an $1152^3$ grid and $\leq
316\times$ for a $752^3$ grid, \Mach\ numbers become too low for the
given grid resolution to resolve the flow velocities accurately
\citep[see][for same effect in O-shell convection simulations, Fig.\,
  15]{Andrassy:2020} which shows first in the smaller $U_\mathrm{r}$
component.

Again, since the IGW spectrum does not depend much on the heating
factor, especially at high wave numbers (\Fig{fig:N2-spectrum}) the
IGW vorticity follows the same scaling $|\nabla \times U| \propto
L^{2/3}$ as the velocity.  The IGW vorticity data in the radiative
envelope from our simulations is consistent with this scaling for
heating factors in the range $100$ to $1000$ (\Fig{fig:vort-scaling})
for both the envelope and the \npeak\ radius. Extra care has been
taken in view of the steep vorticity gradient at \npeak\ to retrieve
the values from spline interpolations (bottom panel,
\Fig{fig:hcorem25-R-vortN2Ri}).  \Fig{fig:vort-scaling} shows the
vorticity scaling for both the envelope (orange symbols in top panel)
and for the \npeak\ (bottom panel).

The simulation vorticity follows the scaling pattern of the tangential
velocity component $\propto L^{2/3}$ for heating factors $< 1000$ and
$\propto L^{1/3}$ for larger heating factors. Since the vorticity
magnitude for IGWs is essentially the horizontal vorticity component
(\Fig{fig:hcorem25-R-vortN2Ri}), this implies that the total vorticity
magnitude is dominated by the term $\partial U_\mathrm{h} /\partial r$
in \Eq{eq:igwvort}, and it confirms that the simulated IGW vorticity
magnitude represents dominantly horizontal shear motions.

The $768^3$-grid simulations generally fall below the scaling
$L^{2/3}$, and the change to $1/3$ scaling is not as clear as it is in
the $1152^3$-grid simulations. This is especially so for the lowest
heating factor, where also the $1152^3$-grid vorticity falls below the
$L^{2/3}$ scaling. This effect can be attributed to insufficient grid
resolution at the lowest \Mach\ numbers. Since the tangential velocity
component does trace the $2/3$ power law down to the lowest heating
factors, it is likely that at these low heating rates, the radial
resolution of horizontal flow features is insufficient to capture the
smaller scales, hence the calculated velocity gradients are too small.
\begin{figure}  
  \includegraphics[width=\columnwidth]{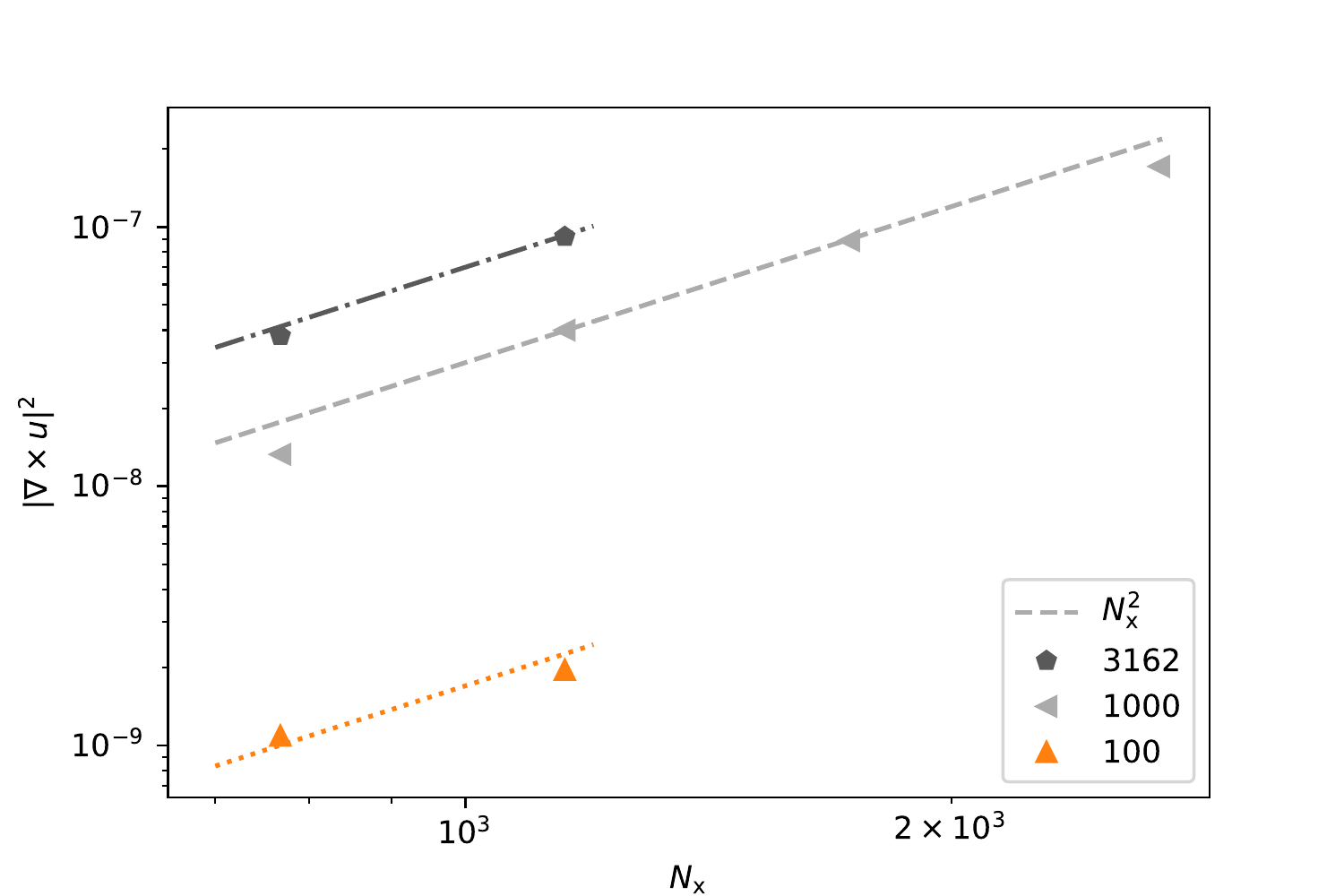}
  \caption{ \vortsq\ at $N^2$-peak radius as a function of grid
    resolution. $N_\mathrm{x}$ is the number of grid cells in one
    dimension. The legend gives the heating factor.}
  \lFig{fig:grid-vort-N2peak}
\end{figure}
The peak of the $U_\mathrm{r}$ spectra is somewhat lower for the
highest and lowest heating rates. However, for the lowest heating rate
the deviation of spherically- and time-averaged radial velocities from
the scaling laws established by the higher heating-rate runs
(\Sect{vorticity}) suggests that the lowest heating rate case shown
(M116) has insufficient grid resolution at $1152^3$ grids to resolve
the $U_\mathrm{r}$ velocity component accurately.  \cite{Saux:22}
suggested that higher heating rates with their larger convective
frequencies would excite IGWs with higher frequencies. There is no
evidence for that in these spectra at the \npeak\ location. However,
if such differences appear only near the convective frequency then
higher frequency resolution, i.e.\ longer time series than available
from these simulations may be required to detect such trend if it
exists.
\begin{figure}  
  \includegraphics[width=\columnwidth]{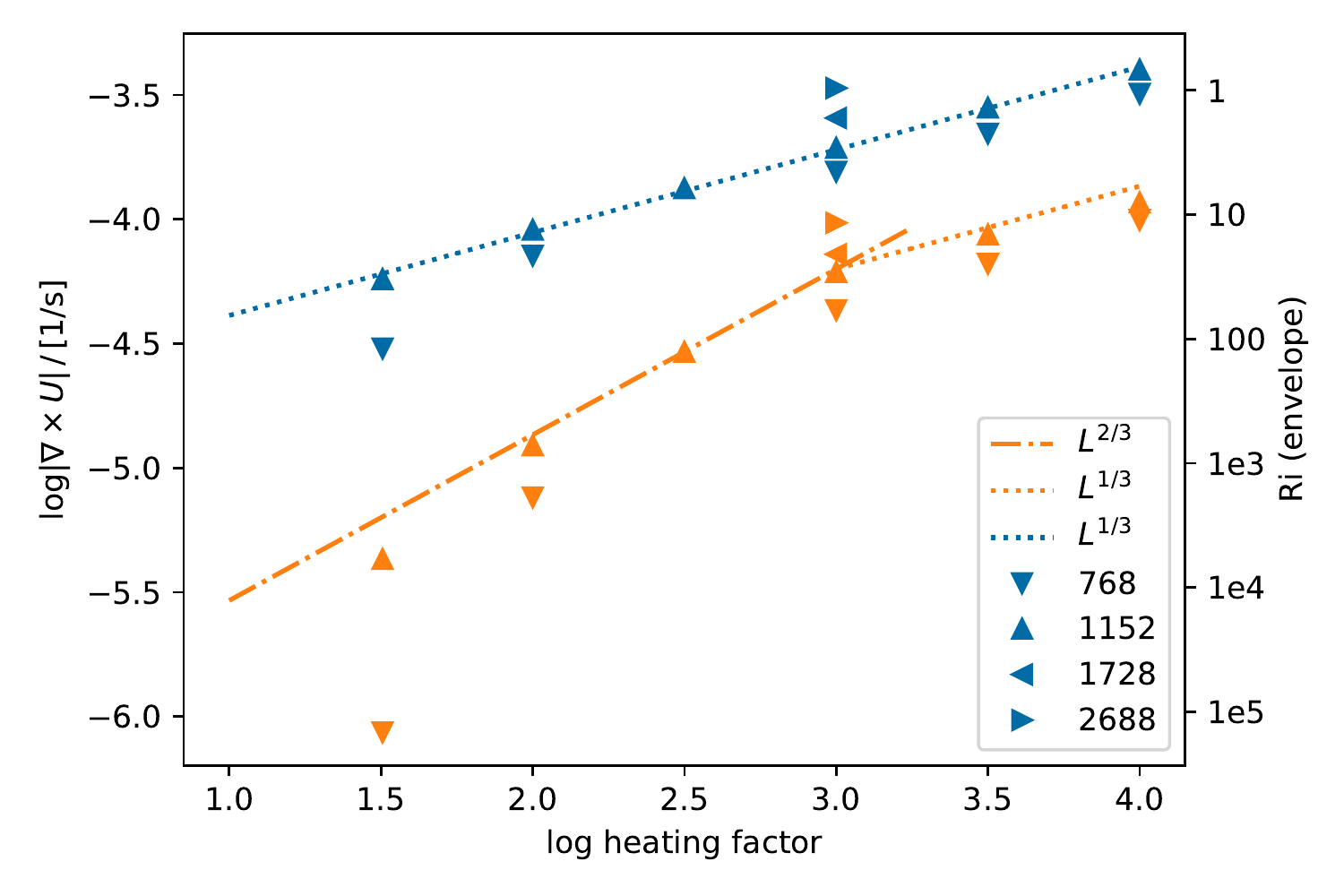}
  \includegraphics[width=\columnwidth]{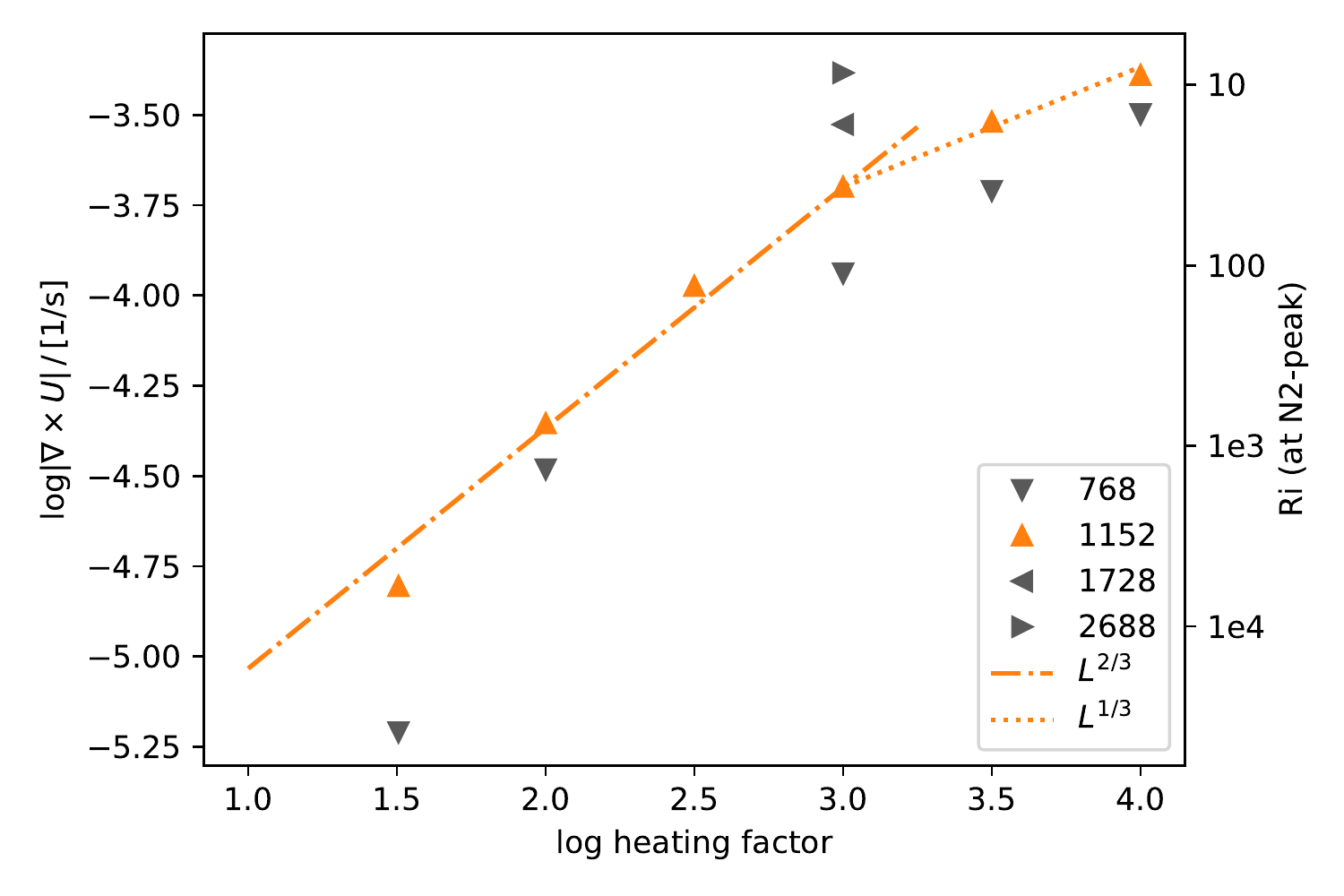}
  \caption{Vorticity as a function of heating. Top: envelope (orange)
    and convection (blue); bottom: at $N^2$-peak radius. The vorticity
    values for the envelope are obtained by time-averaging the
    spherical averages (\Fig{fig:vort_conv_calc}) over the range
    $R/\Mm \in [1800,2000]$. The convective vorticity values are the
    radial average over the range $800<R/\Mm<1100$.  On the right
    ordinate, the Richardson number according to \Eq{eq:Ri-def} is
    shown. Strictly speaking, each point would have its own value of
    $N^2$, and points can't simultaneously be on the same vorticity
    scale and on the same Ri scale. Here, the Ri axis is indicative,
    assuming for all points the average value for $N^2$ at the
    \npeak\ of all $1152^3$-grid simulations ($<N^2> =
    \unit{\natlog{1.5}{-6}}{\mathrm{1/s^2}}$ , $N^2 \in [1.1,1.9]
    10^{-6} \mathrm{1/s^2}$). }  \lFig{fig:vort-scaling}
\end{figure}

\subsection{Critical interpretation of measured IGW mixing as shear-mixing}
\lSect{digw-thermal}
In this section, we discuss the scaling relations from the simulations
for IGW mixing and vorticity under the \emph{assumption} that IGW
mixing is due to shear mixing as described in \Sect{mixing-analysis},
specifically \Eq{eq:digwtherm}. We thereby test whether, based on the
simulations presented here, we are able to confirm or rule out that
this is the case.
\subsubsection{Scaling of \digwhydro\ with $1/Ri$}
\Fig{fig:Scaling-D-core-and-env} shows that, focusing on the
$1152^3$-grid simulations, the measured \digwhydro\ mixing
coefficients follow the scaling $\propto L^{4/3}$ for $\log
\mathrm{heating\ factors} \goa 2$. This would be consistent with the
shear-mixing model \Eq{eq:digwtherm} with IGW vorticity scaling with
$L^{2/3}$. However, $\vort \propto L^{1/3}$ and not $\propto L^{2/3}$
for heating factors $\geq 1000\times$, yet \digwhydro\ does not follow
this power-law change toward a slower increase with heating. Again, as
in the previous section, we consider if this behaviour could be due to
the somewhat different \npeak\ profiles for lower heating rate runs
versus the higher heating rate runs. The possible effect of the $N^2$
dependence in \Eq{eq:digwtherm} can be eliminated by establishing the
scaling relation of \digwhydro\ with $1/Ri$:
\begin{equation}	
\digwhydro \propto \frac{1}{Ri} \approx \frac{(\nabla\times\mathbf{u})^2}{N^2} \propto L^{4/3} \, . 
\lEq{eq:D-IGW-scaling}
\end{equation}
If IGW mixing is due to shear mixing following \Eq{eq:digwtherm}, then
we should find $\digwhydro \propto \frac{1}{Ri}$ for all heating
factors.  However, as shown in \Fig{fig:hcorem25-Ri-diffusivity}, this
is not the case. Focusing again on the higher accuracy $1152^3$-grid
simulations, the scaling follows the trend expected for shear mixing
for heating factors $<1000\times$. But for larger heating factors, the
scaling power changes by a factor of two, consistent with the change
of vorticity scaling being the dominant aspect. Thus, the small
variation of $N^2$ as a function of heating is not responsible for the
change of scaling at heating factors $\geq 1000\times$. To make this
point even clearer, we show each pair of ($1/Ri$, \digwhydro) in
\Fig{fig:hcorem25-Ri-diffusivity} using both the actual $N^2$ value
measured for each simulation (filled symbols), as well as the $1/Ri$
values resulting from using an average, constant value for $N^2$ (open
symbols). This shows that while individual points may shift a bit, the
break from the scaling in \Eq{eq:D-IGW-scaling} does not depend on the
minor dependence of $N^2$ on the heating factor.

The change of scaling in \Fig{fig:hcorem25-Ri-diffusivity} seems to be
in contradiction to the hypothesis that the measured IGW mixing is due
to shear mixing according to \Eq{eq:D-IGW-scaling}. Even though the
vorticities do not follow the scaling law expected for IGWs for
heating factors $>1000\times$ because the tangential velocity does not
follow the IGW scaling relation in those cases (\Sect{vorticity}), we
may still expect \Eq{eq:D-IGW-scaling} to hold. But this is not the
case, as the break from scaling in the relation of \digwhydro\ with
$1/Ri$ (\Fig{fig:hcorem25-Ri-diffusivity}) shows. A complicating
factor could be that \Eq{eq:digwtherm} also contains the thermal
conductivity, which will be discussed in the next section.

The $768^3$-grid simulations follow the scaling $\digwhydro \propto
1/Ri$ for all but the lowest heating factor. However, as pointed out
before, $768^3$-grid vorticities less
accurately follow the $L^{2/3}$ IGW vorticity scaling and therefore
do not exhibit the change in scaling from power $2/3$ to $1/3$ as
clearly. For the lowest heating factors, mixing is larger compared to
the $L^{4/3}$ scaling (\Fig{fig:Scaling-D-core-and-env}), and more so for
smaller grids. This is consistent with numerical diffusion
contributions, to be discussed further below.

A question arises as to how sensitive the determination of $D$ and
$1/Ri$ is to the exact position of where the values are taken
(cf.\ \Fig{fig:hcorem25-R-vortN2Ri}). Mini-profiles provided for runs
M114 and M116 as short solid lines in
\Fig{fig:hcorem25-Ri-diffusivity} show the simultaneous change of both
quantities accrosss the \npeak\ feature.  For each of these lines, the
large $1/Ri$ end of the line corresponds to the inside of the
\npeak\ profile, and this is the convection side. For the broader M116
peak, the mini-profile is approximately aligned with the $\propto
1/Ri$ scaling law. The M114 mini-profile follows the heating-series
scaling relation broadly but deviates on the stable side of the
\npeak\ profile.
\begin{figure}  
  \includegraphics[width=\columnwidth]{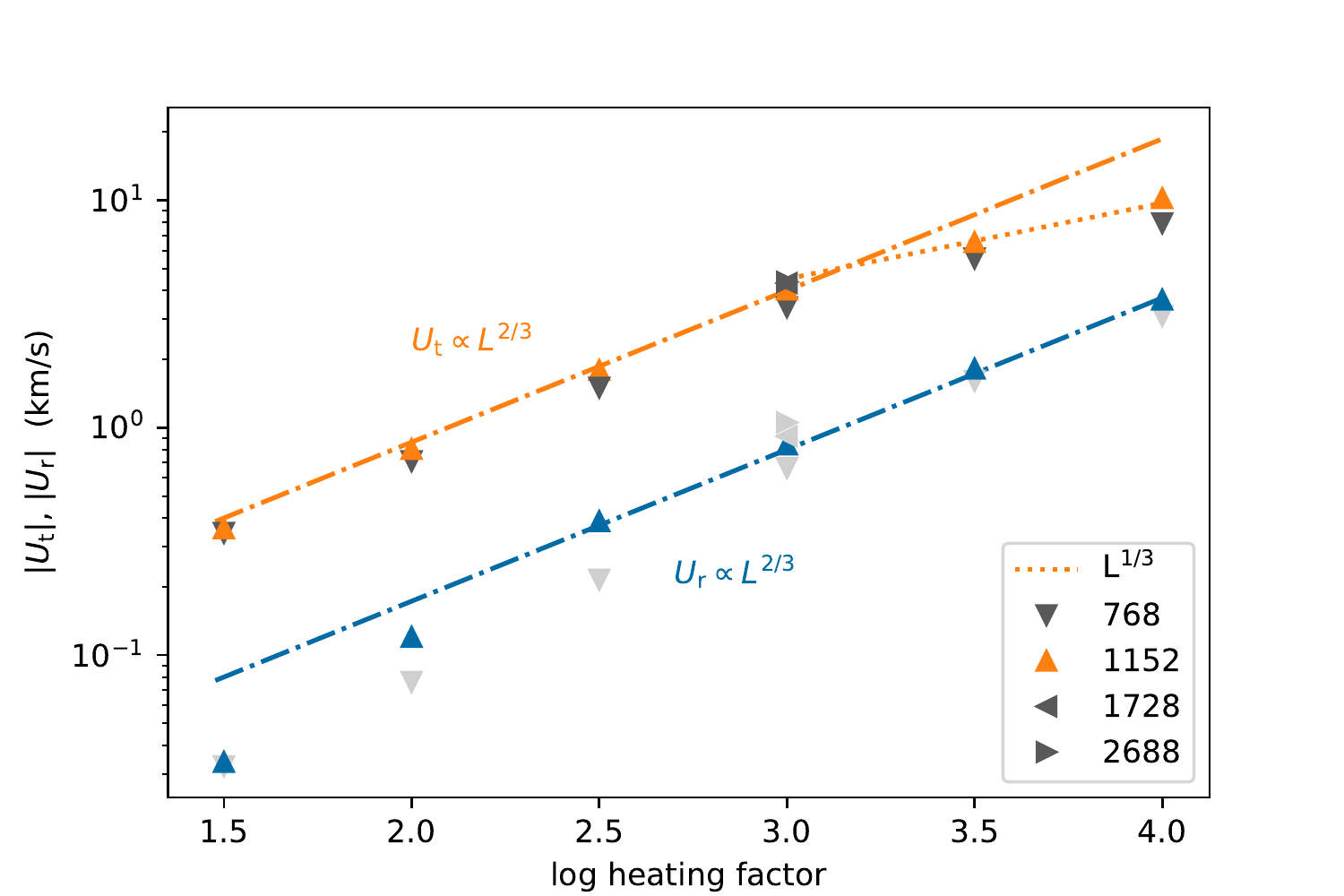}
  \caption{Tangential and radial velocity components at the
    \npeak\ radius as a function of heating factor. Upside-down
    triangles are from the $1152^3$-grid simulations. The lower
    sequence of blue and light grey symbols show the radial velocity
    component. Orange and dark grey symbols represent the tangential
    velocity component. Dash-dotted lines show scaling $\propto {2/3}$
    and the dotted line shows scaling $\propto {1/3}$.}
  \lFig{fig:hcorem25-heating-UtUr-N2peak}
\end{figure}

\subsubsection{Thermal conduction}
\lSect{s.therm-cond}
These simulations do not include thermal conduction. However, the
shear-mixing model described in \Sect{mixing-analysis} requires that
$\digwhydro \propto $ the thermal diffusivity $K$
(\Eq{eq:K_therm}). As discussed in \Sect{mixing-analysis}, thermal
loss of entropy memory enables shear flows to cause mixing for
$Ri>1/4$. Without thermal conduction, displaced fluid elements would
remember the entropy of their origin and return ultimately to that
radial position. Thus, thermal conduction would facilitate partial or
complete entropy memory loss without which mixing according to the
shear mixing mechanism outlined in \Sect{mixing-analysis} would be
impossible. Since these simulations do not include thermal conduction
explicitly, numerical diffusion of entropy would be facilitating IGW
mixing if it is caused by shear according to \Eq{eq:K_therm}. The
numerical entropy diffusion would be related to the accuracy of
\code{PPM} used to calculate the hydrodynamics, with error terms
expected to scale with the number of grid points (in one dimension)
$N_\mathrm{x}$ as $N_\mathrm{x}^{-3}$
\citep{Porter_Woodward_1994}. However, the advection of concentration
$FV$ is done in \ppmstar\ with the higher-order \code{PPB} method. Due
to this feature of \ppmstar\ treating hydrodynamic and concentration
advection at different orders, these simulations could capture mixing
due to small-scale shear at $Ri>1/4$ calculated with high-order
\code{PPB} aided by thermal conduction due to more efficient entropy
diffusion resulting from the lower order of \code{PPM}.

More specifically, in the presence of fluid shear the \code{PPM}
difference scheme has an effective numerical viscosity that has been
measured and characterized quantitatively in
\cite{Porter_Woodward_1994}.  This viscosity arises from interpolation
errors that cause a slight diffusion of mometum.  The diffusion of
entropy in such a flow should occur through the same sort of
interpolation errors and thus should be characterized in the same way
in terms not of effective viscosity but instead of effective thermal
diffusivity.  The effective viscosity of the PPM scheme in this
context consists of two terms, one that scales as $\Delta x^3/
\lambda^2$ and another scaling as $\Delta x^4/ \lambda^3$ where
$\Delta x$ is the cell width.  The second term becomes important only
for sinusoidal disturbance wavelengths, $\lambda$, that are smaller
than about $10 \Delta x$.  For reasonably resolved wavelengths
$\lambda$ the dominant term in the effective viscosity of the
\code{PPM} scheme scales as $\Delta x^3/ \lambda^2$.  This means that
for a given grid cell size $\Delta x$ the effective viscosity is
inversely proportional to the square of the disturbance wavelength.
Consequently, there is no single viscosity that one can ascribe to a
\code{PPM} simulation on a given grid, because the effective viscosity
rapidly becomes smaller as the wavelength of the disturbance of
interest increases.  In the flows discussed here, we consider shear
generated by IGWs in the region of \npeak\ at the convective boundary.
If we consider grids that all capture the spectrum of such IGWs fairly
well, then we would expect the resulting diffusion of entropy to scale
with $\Delta x^3$.  If, instead, more of the IGW spectrum at shorter
wavelengths is captured as the grid is refined, and if this newly
captured spectrum component contains sufficient power to be important
in diffusing entropy, then the effective entropy or species diffusion
could scale differently with $\Delta x$. As shown in the left column
of \Fig{fig:N2-spectrum} the overall spectrum is captured fairly well
by all grids, but it is also correct that more power appears at higher
wave numbers.

\begin{figure}
  \includegraphics[width=\columnwidth]{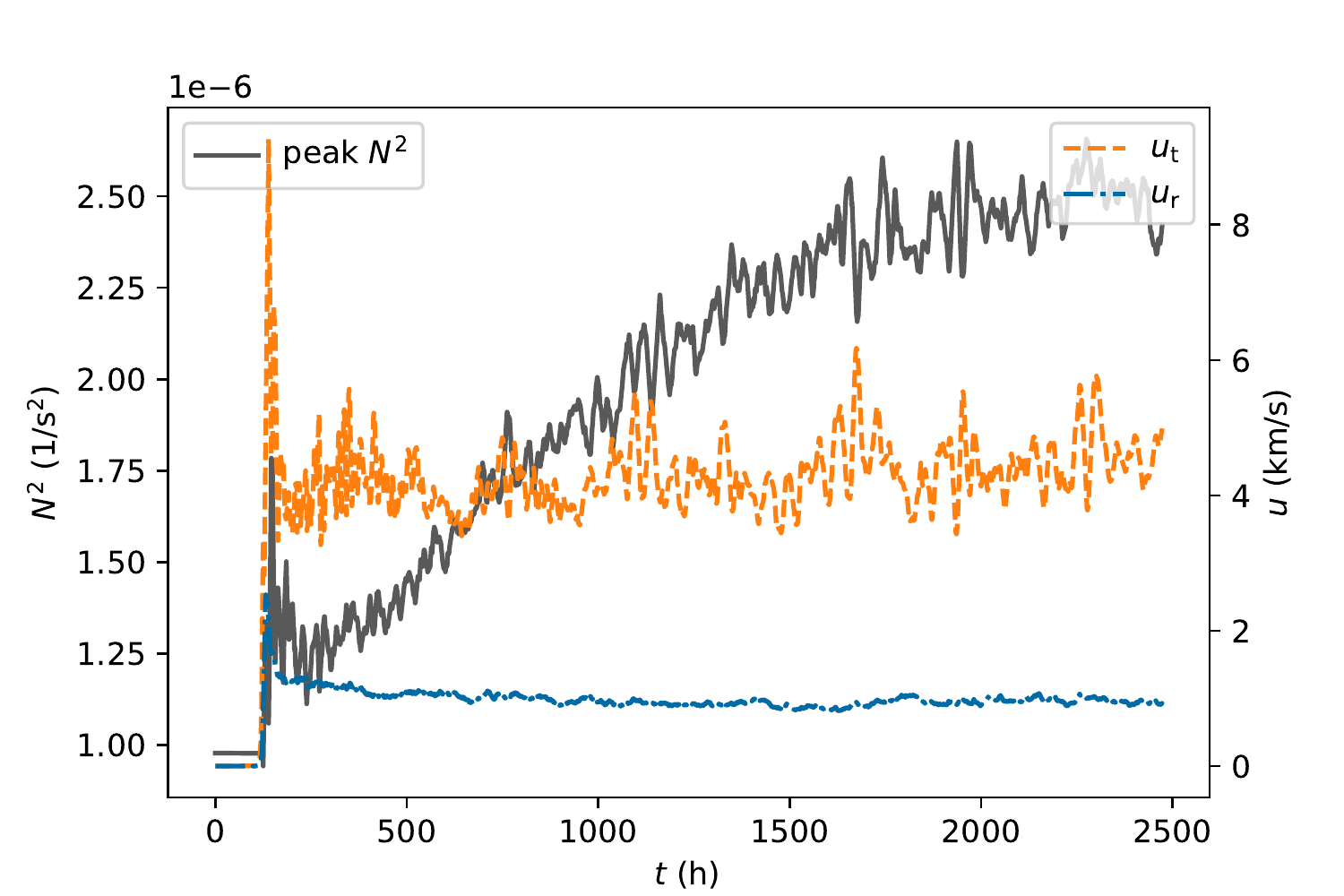}
  \caption{Tangential and radial velocity magnitude at the radius of
    the \npeak\ for run M115 ($1728$ grid, $1000\times$ heating
    factor) and the value of $N^2$ at the radius of the \npeak. $N^2$
    gradually increases over the first \unit{1500}{\hour} while the
    boundary migrates outward through the initial boundary profile
    (cf.\ \Sect{s.bnd_sph_ave} and middle panel
    \Fig{fig:RProf-N2FVUt-boundary}).}  \lFig{fig:M115-N2urut-time}
\end{figure} 
In \ppmstar\ species advection is followed with the higher-order
\code{PPB} scheme \citep{Woodward:2013uf} whereas the hydro is
computed using \code{PPM}. Formally, \code{PPB} is two orders more
accurate than \code{PPM}, so its numerical diffusion should scale with
grid resolution not as $\Delta x^3$ but as $\Delta x^5$. Ultimately,
an advection scheme can capture species flows in absolute terms only as
accurately as the underlying velocity field.  In the bottom panel of
\Fig{fig:f-bound-mix}, it can be seen that the FV gradient profile
differs slightly from the $N^2$ profile. In a hypothetical simulation
without physical heat conduction in which the numerical species and
entropy diffusion are the same, initially identical $FV$
(concentration) and entropy profiles must remain identical with time
everywhere in the simulation. In the bottom panel of
\Fig{fig:f-bound-mix} it can be seen that the $FV$ gradient profile is
slightly different from the $N^2$ profile. Despite not including
thermal diffusivity in these simulations, numerical diffusion of heat
changes the entropy profile in addition to dynamic process, which for
concentration mixing are followed with the higher accuracy of the
\code{PPB} scheme. This is an indication that in these simulations
species mixing and entropy mixing are indeed subject to different
numerical accuracy.

\begin{figure}
  \includegraphics[width=\columnwidth]{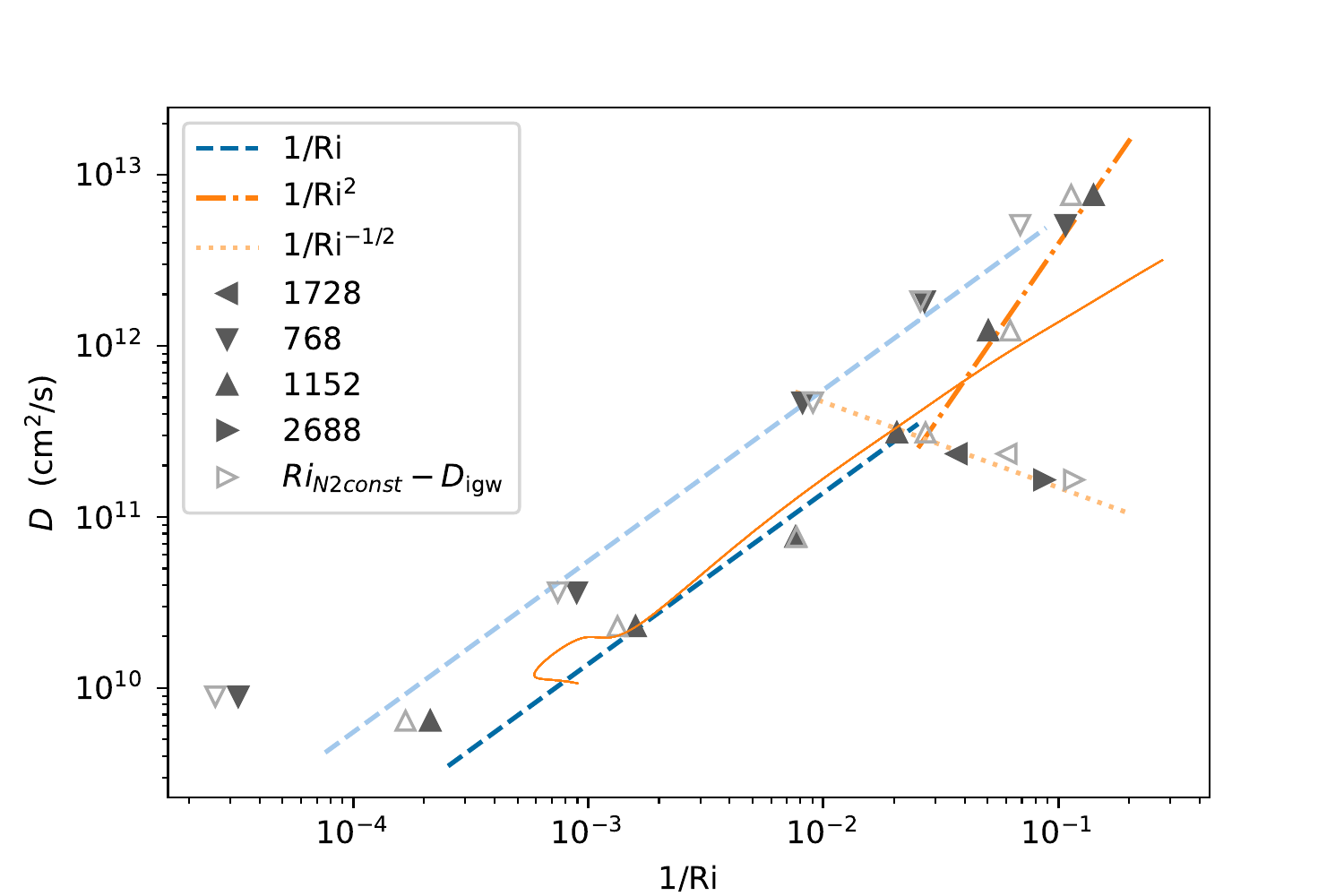}
  \caption{IGW mixing diffusion coefficient at the $N^2$-peak radius
    as in \Fig{fig:Scaling-D-core-and-env} as a function of $1/Ri =
    \frac{\vortsq}{N^2}$. In order to show the magnitude of the
    variation of $N^2$, open symbols show $1/Ri$ values calculated with
    the $N^2$ average of the $1152^3$-grid simulations (see caption of
    \Fig{fig:vort-scaling}).  A thin
      solid orange line provides the relation between $D$ and
      $1/Ri$ along the \npeak\ profile for M116 (\Fig{fig:f-bound-mix}) for the radius range $R \in \unit{[1465,1555]}{\Mm}$, where values of large $1/Ri$ correspond to small radii.}
  \lFig{fig:hcorem25-Ri-diffusivity}
\end{figure}
$\digwhydro$ is the measurement of species transport determined from
the time evolution of $FV$ profiles of the simulations
(\Sect{digw-dynamic}). \Fig{fig:hcorem25-Ri-diffusivity} shows that
$\digwhydro \propto (1/Ri)^{-1/2}$ as a function of grid resolution
for heating factor $1000$. This suggests to analyze the dependence of
the species and entropy diffusivity on grid resolution a bit
further. \Fig{fig:grid-D} shows that $\digwhydro \propto
N_\mathrm{x}^{-1}$, where $N_\mathrm{x}$ is the number of grid points
in one direction.  This would be consistent with the case of numerical
diffusivity proportional to $\Delta x$ mentioned above.  $N^2$ does
not depend significantly on $N_\mathrm{x}$. However, $\vortsq \propto
N_\mathrm{x}^2$ (\Fig{fig:grid-vort-N2peak}), and if IGW mixing is due
to shear according to \Eq{eq:digwtherm} the scaling of
\digwhydro\ with grid resolution can also be understood as a
consequence of numerical entropy diffusion scaling with
$N_\mathrm{x}^{-3}$.  Using the measured \digwhydro\ and \vort\, a
hydrodynamic IGW diffusivity can be determined according to
\begin{equation}
  \khydro = \frac{\dhydro {N^2}}{\eta (\nabla\times\mathbf{u})^2} \, .
  \lEq{eq:khydro}
\end{equation}
This would be a reflection of the numerical \code{PPM} entropy
diffusivity \emph{if} the measured \digwhydro\ is due to shear mixing
according to \Eq{eq:digwtherm}, and using $\eta = 1$. Then the scaling
for \khydro\ with $N_\mathrm{x}$ follows from that of vorticity and
\digwhydro\ and is $\propto N_\mathrm{x}^{-1} / N_\mathrm{x}^{2} =
N_\mathrm{x}^{-3}$ as shown in
\Fig{fig:hcorem25-grid-ppm-entropy-diffusivity}. This corresponds to
the expected scaling of the \code{PPM} error terms if all grids
resolve the relevant scales reasonably well
\citep{Porter_Woodward_1994}.

 \khydro\ is a possible expression of the entropy diffusion in the
 sense that \emph{if} IGW mixing is due to shear mixing according to
 \Eq{eq:digwtherm}, then the \code{PPM} entropy diffusion has the same
 cumulative effect on entropy memory loss of perturbed fluid elements
 that a thermal diffusivity of this magnitude would have, given the
 measured mixing, the vorticity and $N^2$ of the simulation. However,
 as explained earlier in this section numerical entropy diffusivity
 has a very different dependence on the spatial spectrum than
 radiative diffusion would have. Therefore, the values and scaling
 relations determined for \khydro\ are specific to the particular
 setup and are \emph{not} numerically equal to a numerical
 viscosity. With these caveats in mind, we can then ask what the grid
 size would be so that for a given simulation of this setup, the
 effect of thermal conductivity $K$ is approximately as large as the
 measured \khydro. In order to maintain the same thermal equilibrium
 stratification, $K$ would be added $\propto L$ in simulations with
 radiative diffusion. The $K$ values according to this scaling have
 been added as horizontal dotted lines for three heating factors to
 \Fig{fig:hcorem25-grid-ppm-entropy-diffusivity}. Assuming $\eta = 1$
 simulations with $N_\mathrm{x} \goa 1400$ would satisfy $\khydro \loa
 K$ for $1000\times$ heating factors. At this or finer grids, a
 physical diffusivity scaled proportionally to the heating factor
 would dominate over numerical entropy mixing \emph{in its impact on
 IGW mixing if that mixing is due to shear mixing.} Thermal
 diffusivity would have a number of other effects on the dynamic
 evolution of the overall simulation, for example in establishing the
 quasi-equilibrium state. The grid dependence of those effects would
 be different for numerical entropy diffusion compared to thermal
 diffusion.

Using a larger heating rate would in principle require a smaller
$N_\mathrm{x}$, except that the measured mixing does not follow the
shear mixing scaling in that case, thus making application of
\Eq{eq:khydro} questionable.

Considering the complex nature of the flow under consideration this
analysis of cumulative effects of different types of error terms in
the framework of a specific physics model may necessarily remain
somewhat inconclusive. However, having established the dependencies of
various quantities related to cause and effect of IGW mixing on grid
resolution in these simulations without thermal conduction provides a
valuable reference point for simulations that do include thermal
diffusion.
\begin{figure}  
  \includegraphics[width=1.02\columnwidth]{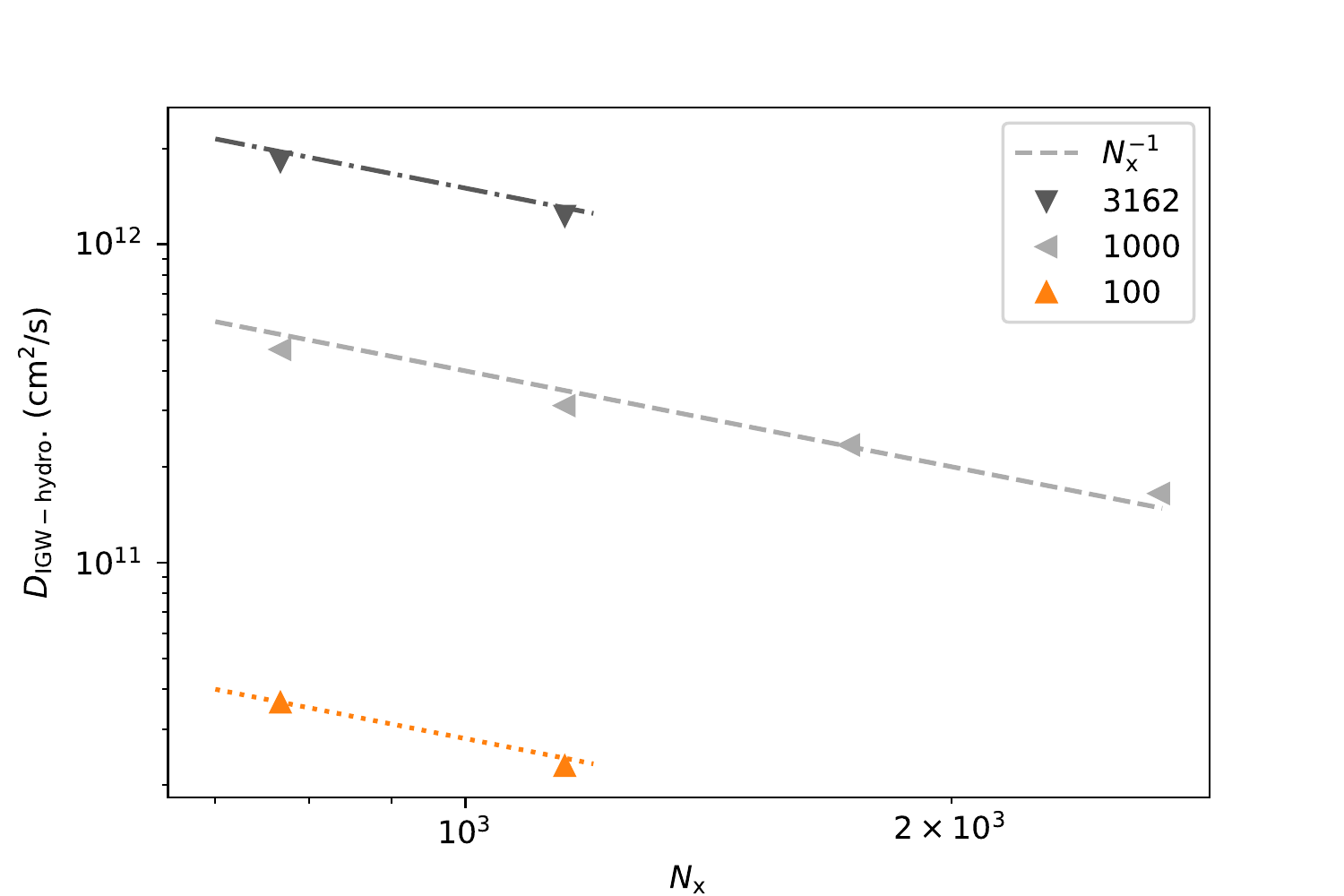}
  \includegraphics[width=0.95\columnwidth]{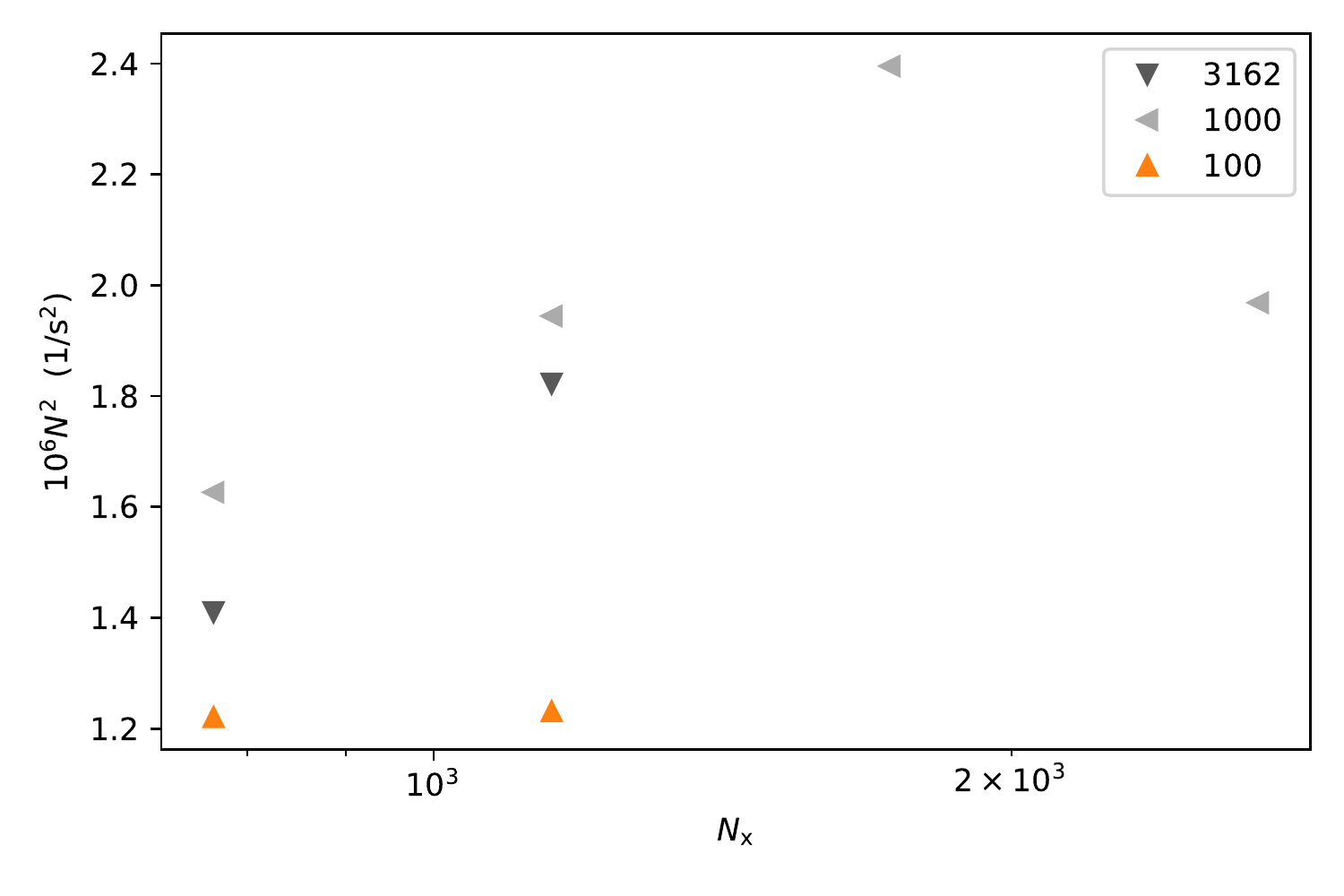}
  \caption{Top: \digwhydro\ at $N^2$-peak radius as a function of grid
    resolution. $N_\mathrm{x}$ is the number of grid cells in one
    dimension. A power law with exponent $-1$ is also shown. The power
    law fit gives an exponent $-0.85$. Bottom: $N^2$ at peak.  The
    lower $N^2$ for the $2688^3$-grid $1000\times$ run (M111) is due
    to the shorter duration of that run, in which the initial
    \npeak\ profile still dominates over the dumps averaged
    (cf.\ \Fig{fig:M115-N2urut-time}).}  \lFig{fig:grid-D}
\end{figure}
\begin{figure}  
  \includegraphics[width=\columnwidth]{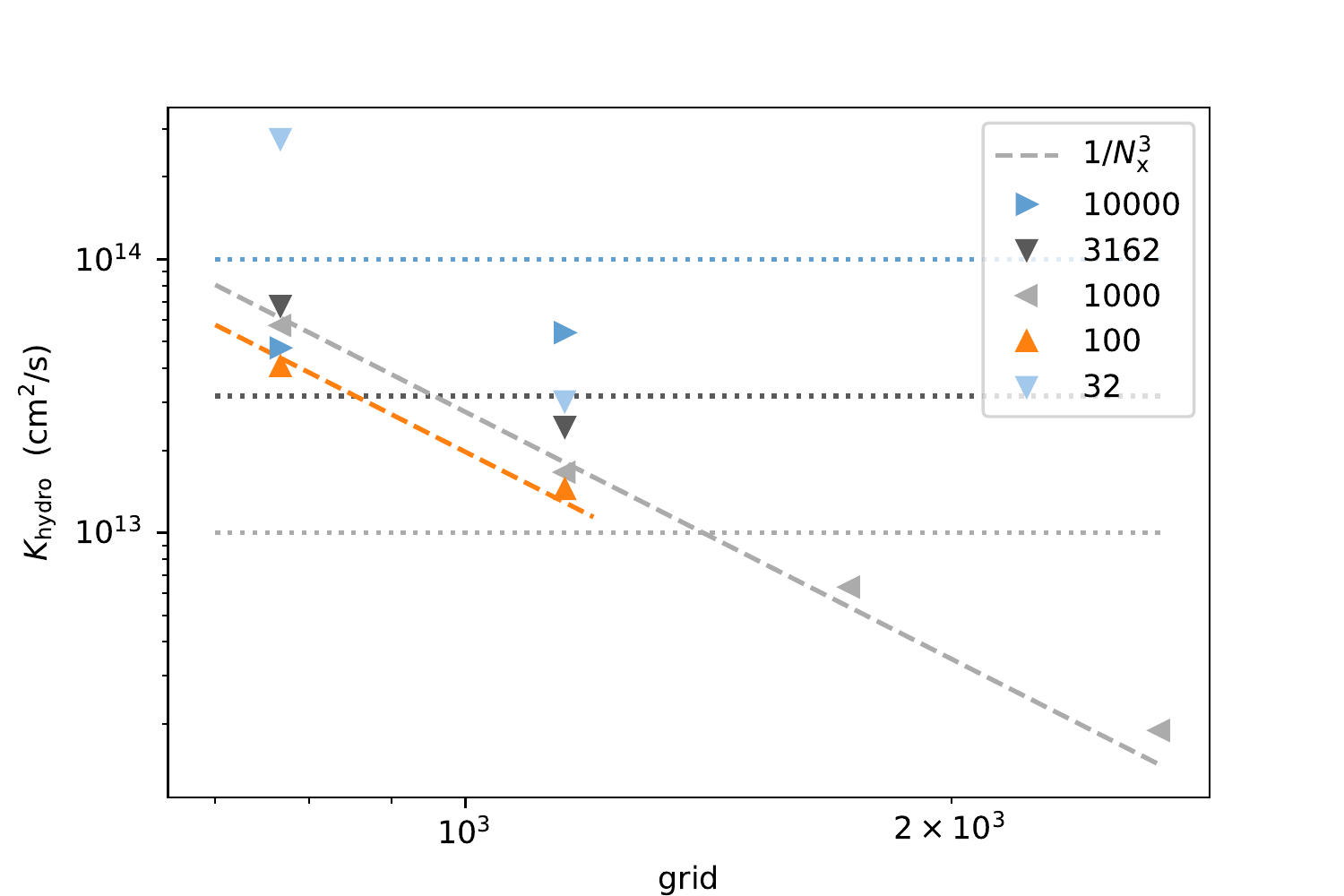}
  \caption{IGW effective entropy diffusivity in terms of
    \khydro\ calculated from \Eq{eq:khydro} under the assumption that
    measured IGW mixing is due to shear mixing, and using the
    simulation values \digwhydro\, \vort\ shown in \Fig{fig:grid-D},
    and $\eta = 1$. Dotted horizontal lines indicate the radiative
    diffusivity when scaled as $K \propto L$, from bottom to top for
    heating factors $1000\times$, $3162\times$, $10000\times$. }
  \lFig{fig:hcorem25-grid-ppm-entropy-diffusivity}
\end{figure}

\subsection{Discussion}
The main results and findings of this section are as follows:
\begin{itemize}
  \item We measured convective mixing and mixing due to IGWs for
    heating factors from $31.6\times$ to $10,000\times$, for at least two grid
    resolutions in each case, and for four grid resolutions for
    heating factor $1000\times$. These two sets of mixing coefficients
    follow distinctly different scaling relations $\dconvhydro \propto
    L^{1/3}$ and $\digwhydro \propto L^{4/3}$. This confirms their
    distinctly different underlying mixing physics.
  \item The convective mixing results are consistent with the commonly
    adopted relation $\dconvhydro \propto l_\mathrm{mix}
    v_\mathrm{conv}$. Our simulations support a new exponential model
    (\Eq{e.convbound}) of how the mixing-length parameter decreases
    toward the convective boundary of main-sequence core convection in
    a massive star. This mixing-length parameter can be used to relate
    the MLT convective velocity to the
    species-mixing diffusion coefficient, instead of using a constant
    MLT mixing-length parameter throughout the entire convection zone.
  \item We probe to what extent the measured IGW mixing is consistent
    with the predictions of shear-induced mixing
    (\Sect{mixing-analysis}, \Eq{eq:digwtherm}). The total vorticity
    magnitude of IGWs is dominated by the horizontal component, which
    in turn is dominated by the derivative of the horizontal velocity
    component in the radial direction. It follows a scaling law
    $\vortsq \propto L^{4/3}$ for heating rates up to $1000\times$,
    but $\propto L^{2/3}$ above, likely because the higher power of
    IGW velocity scaling relative to convective velocity scaling leads
    to IGW velocities approaching those of the convective core. No
    matter what the cause of this change in scaling the measured
    \digwhydro\ values should remain $\propto 1/Ri \propto \vortsq$
    according to the shear-mixing prediction of IGWs, which they do
    not (\Fig{fig:hcorem25-Ri-diffusivity}). This finding questions
    the interpretation of IGW mixing measured in these simulations as
    due to shear mixing according to \Eq{eq:digwtherm}.
  \item In these simulations without radiative conduction, the role of
    $K$ in \Eq{eq:digwtherm} would be that of numerical entropy
    diffusion, which can be expressed as an equivalent
    \khydro\ (\Eq{eq:khydro}). This inferred diffusivity follows the
    scaling as a function of grid resolution that is consistent with
    the expected \code{PPM} error term if the relevant spectrum of
    IGWs is reasonably resolved on all grids.
 \end{itemize}

This leads to the question of what can be concluded about the IGW
mixing at nominal heating in a real star. By conducting a grid of
simulations as a function of heating factor and grid resolution we
have measured mixing due to IGWs in the \npeak\ region. The
measured scaling for $\digwhydro \propto L^{4/3}$ does not explicitly
take into account the expected dependence on $K$. If we use the
scaling relation as it derives from the simulations
(\Fig{fig:Scaling-D-core-and-env}), then extrapolating the
$1000\times$ measured \digwhydro\ values to nominal heating yields
$D_\mathrm{IGW} \approx \unit{\natlog{2.5}{7}}{cm^2/s}$. Such a large
IGW mixing value would lead to significant changes of the evolution of
a \unit{25}{\Msun} star, in the Hertzsprung-Russell diagram and in
terms of internal mixing that would lead to surface enrichment of He
and CNO elements.

However, if we assume that IGW mixing is due to thermally-enhanced
shear-mixing according to \Eq{eq:digwtherm} (adopting for this
estimate $\eta = 1$), we can arrive at an estimate in the following
way. The dependence of \khydro\ on the heating factor is rather small
(\Fig{fig:hcorem25-grid-ppm-entropy-diffusivity}). We tentatively
interpret \Fig{fig:hcorem25-grid-ppm-entropy-diffusivity} to suggest
that a simulation with a heating factor of $1000\times$ and a grid
size $N_\mathrm{x}$ somewhere between the $1152^3$ and $1728^3$ grid
would be equivalent to a simulation with a radiative diffusivity $K$
that has been scaled $\propto L$, because at that $N_\mathrm{x}$ the
inferred \khydro\ is approximately the same as would be the scaled
radiative diffusivity. A simulation with such a $N_\mathrm{x}$
therefore mimics a simulation with radiative diffusion included at the
scaled rate. Then, interpolating between the measured diffusivities of
the $1152^3$- (M114) and $1728^3$-grid (M115) runs, the resulting IGW
diffusivity at $1000\times$ heating factor would be $\approx
\unit{\natlog{3}{11}}{cm^2/s}$. Adopting from \Sect{vorticity} that
$\vortsq \propto L^{4/3}$ and that $N^2$ does not contribute
significantly to the scaling of \digwtherm, and considering that we
assumed $K \propto L$, the scaling for IGW shear mixing would be
$\digwtherm \propto L^{7/3}$ and $ \unit{\natlog{3}{11}}{cm^2/s}$ at
$1000\times$ heating factor would correspond to $\approx
\unit{\natlog{3}{4}}{cm^2/s}$ at nominal heating. This would be an
upper limit, since it is possible that $\eta < 1$.

Alternatively, scaling $Ri \approx N^2/\vortsq \approx 50$ at
\npeak\ (\Fig{fig:hcorem25-R-vortN2Ri}) using $\vortsq \propto
L^{4/3}$ and adopting for the nominal thermal diffusivity $K \approx
\unit{10^{10}}{cm^2/s}$, the estimate for IGW diffusivity is $\digw
\approx K/Ri = \unit{\natlog{2}{4}}{cm^2/s}$, again adopting $\eta =
1$. Such a low value would have a limited and local effect on mixing in the
boundary region immediately above the convection-dominated core but
would not contribute to mixing from the core to the surface over the
main-sequence lifetime of a \unit{25}{\Msun} star.

\section{Conclusions}
\lSect{conclusions} Our $4\pi$ 3D simulations include the entire
convective core and an additional $\approx \unit{1000}{Mm}$ in radius
of stably stratified star above the convective core, amounting in
combination to $\approx 50\%$ of the radial extent of a H-burning
$25\Msun$ star. Convective flows are large-scale, and the largest
coherent structure is a drifting dipole in which flows are passing
through the centre to the convective boundary and returning along the
boundary to the antipode. There, mutually opposing horizontal pressure
gradients of the converging flow force the flow inward and create
highly unstable boundary-separation wedges.

It is interesting to compare these results with the picture of
convection on which MLT is based. The convection zone
contains radially just over two pressure-scale heights. The notion of
a dominant fluid element descending for a distance of about a mixing
length toward the centre, and fluid elements rising from the centre to
the boundary, can indeed be observed in the 3D simulations. But in the
simulations, these same fluid flow elements simply stream right
through the centre and do not turn around there. Our simulations show
instead large-scale flows that flow directly through the central
region but that do not originate nor terminate there, as in a flower
petal pattern.

The power spectrum shows the familiar turbulent cascade in the inner
regions but the spectrum of the radial velocity component flattens
toward the boundary, while the spatial spectrum of the tangential
velocity component remains close to $\propto l^{-5/3}$. Interactions
of a broad spectrum of equally powerful convective motions excite IGWs
in the stable layer. These IGWs display eigenmodes in good agreement
with the predictions of \code{GYRE} models based on the spherically averaged
radial structure of the 3D simulations. The transition from
convective-dominated to wave-dominated flows can be characterized from
the 3D data through the higher-order moments of the PDF of the radial
velocity on the sphere of a given radius. The product of the skew and
excess kurtosis has a maximum at the dynamic boundary, and for IGWs
the PDF is Gaussian.

Our simulations show an unrealistically high mass-entrainment rate,
similar to previous simulations
\citep{Meakin:2007dj,Gilet:2013bj}. The reason for these high
entrainment rates is that the underlying MESA model, from which the
initial setup for the 3D simulation is derived, is not in
dynamic-thermal equilibrium \citep{anders:21}. Simulations with
radiative diffusion included, and with a sufficiently large
nearly-adiabatic penetration zone, show a realistic mass-entrainment
rate and will be presented in a forthcoming paper.

In \Sect{mixing-1D}, we showed that we measure mixing due to IGWs in
the \npeak\ region. In order to determine IGW mixing at nominal
heating, it is essential to identify the actual physical mechanism of
IGW mixing in order to apply the correct scaling relation. In this
paper, we have explored the shear-mixing model to this end
(\Sect{mixing-analysis}). The simulations presented here do not
provide conclusive evidence for this physical mixing process. If IGW
mixing is due to shear, as described by \Eq{eq:digwtherm}, then the
simulations presented in this paper suggest that $\digw \approx 2$ to
$\unit{\natlog{3}{4}}{cm^2/s}$ in the \npeak\ region above the
convective core. Such mixing rates would at most have local impact on
mixing but would not lead to mixing that alters the surface
composition in a \unit{25}{\Msun} star over the main-sequence lifetime.
However, other physical mixing mechanisms are possibly
responsible for IGW mixing. For example, in a forthcoming paper, we
will explore a process that we call advective stochastic mixing (ASM), which
would scale as $\digw \propto U_\mathrm{r}^2/N$ and may depend less 
or not at all on thermal diffusivity. This model would also be compatible with
the scaling relations established here but would lead to larger IGW mixing
efficiencies when extrapolated to nominal heating.

\section*{Acknowledgements}
FH acknowledges funding through an NSERC Discovery Grant, and PRW
acknowledges funding through NSF grants 1814181 and 2032010.  Both
have been supported through NSF award PHY-1430152 (JINA Center for the
Evolution of the Elements).  RA acknowledges funding as a CITA
National Fellow at the University of Victoria. The simulations for
this work were carried out on the NSF Frontera supercomputer operated
by TACC at the University of Austin, Texas and on the Compute Canada
Niagara supercomputer operated by SciNet at the University of
Toronto. The data analysis was carried out on the Astrohub virtual
research environment (https://www.ppmstar.org) hosted on the Computed
Canada Arbutus Cloud at the University of Victoria.  We also wish to
thank Conny Aerts and Dominic Bowman for very inspiring
discussions. This work has benefited from scientific interactions at
the KITP program "Probes of Transport in Stars" in November 2021, and
therefore supported in part by the National Science Foundation (NSF)
under Grant Number NSF PHY-1748958. We specifically like to
acknowledge Daniel Lecoanet, Matteo Cantiello, Ben Brown and Evan Anders for
interesting discussions at KITP.

\section*{Data availability}
3D and spherically averaged 1D simulation outputs are available at
\url{https://www.ppmstar.org} along with python notebooks that have
been used to create plots in this paper.

\bibliographystyle{mnras}

\bsp	%

\appendix
\section{Constructing the 1D base state}
\lSect{s.append-constructbasestate}

In order to construct the base state from a \code{MESA} profile, the entropy and pressure is calculated from the \code{MESA} temperature, density, and mean molecular weight profiles.
\begin{eqnarray}
P_\mathrm{gas}  &=& \frac{\mathcal{R}}{\mu}\rho T  \\
S_\mathrm{gas}  &=& c_V \log T  - \frac{\mathcal{R}}{\mu} \log \rho  + \mathrm{const.} 
\lEq{eos}
\end{eqnarray}

With a maximum radius of $R_\mathrm{max}=2700\Mm$, for these
simulations a grid cell is $\Delta x = 7.03\Mm$ on a
$768^3$-grid. This is usually the smallest grid used in our
simulations. The following filtering procedure is performed on a ten
times finer radial grid. The entropy on this filtering grid is
obtained by interpolating the entropy calculated with \Eq{eos}
from the \code{MESA} $T$, $\rho$, and $\mu$ using
\code{scipy.interpolate.PchipInterpolator}.
\begin{figure}
  \includegraphics[width=\columnwidth]{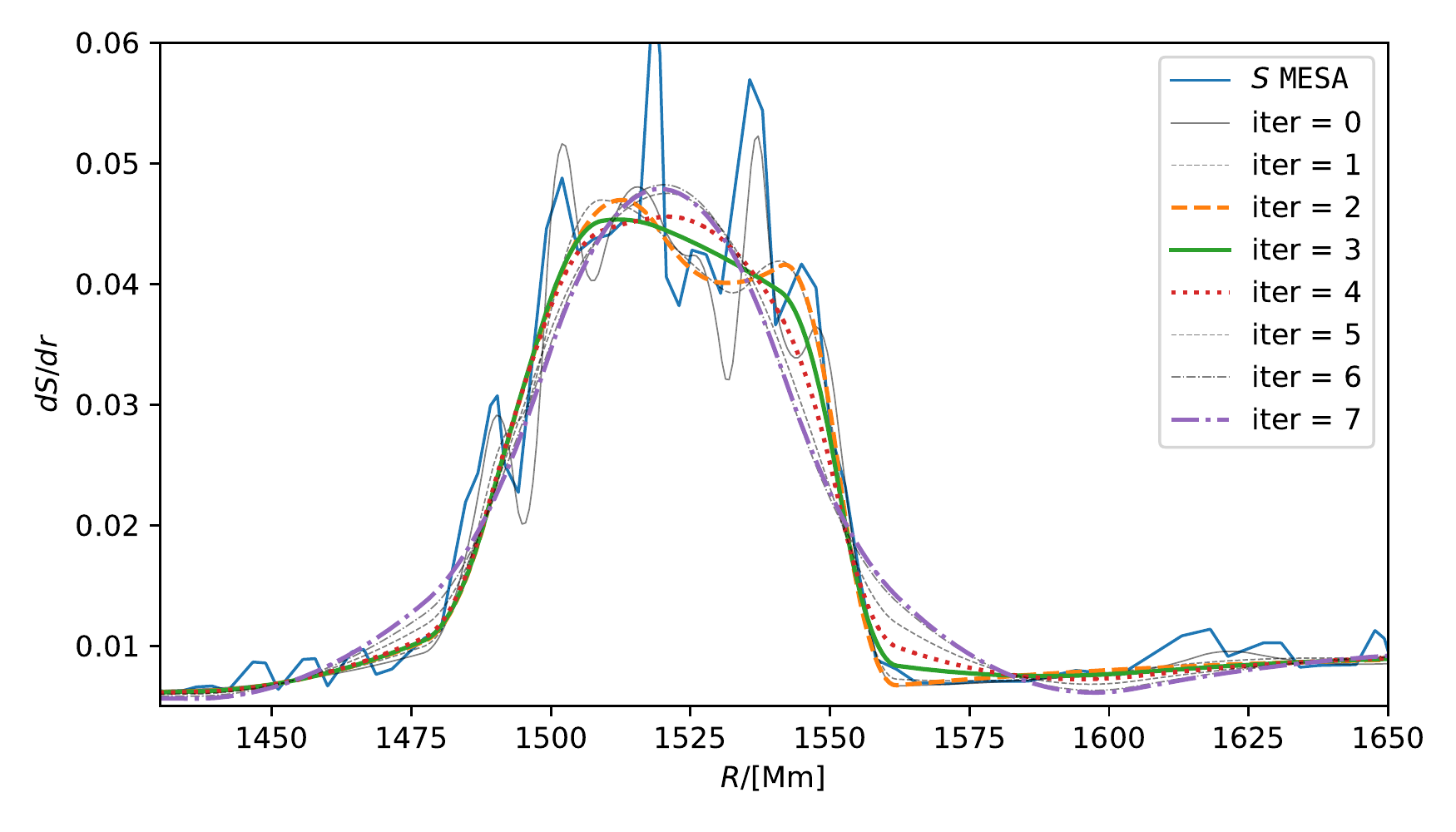}
  \includegraphics[width=\columnwidth]{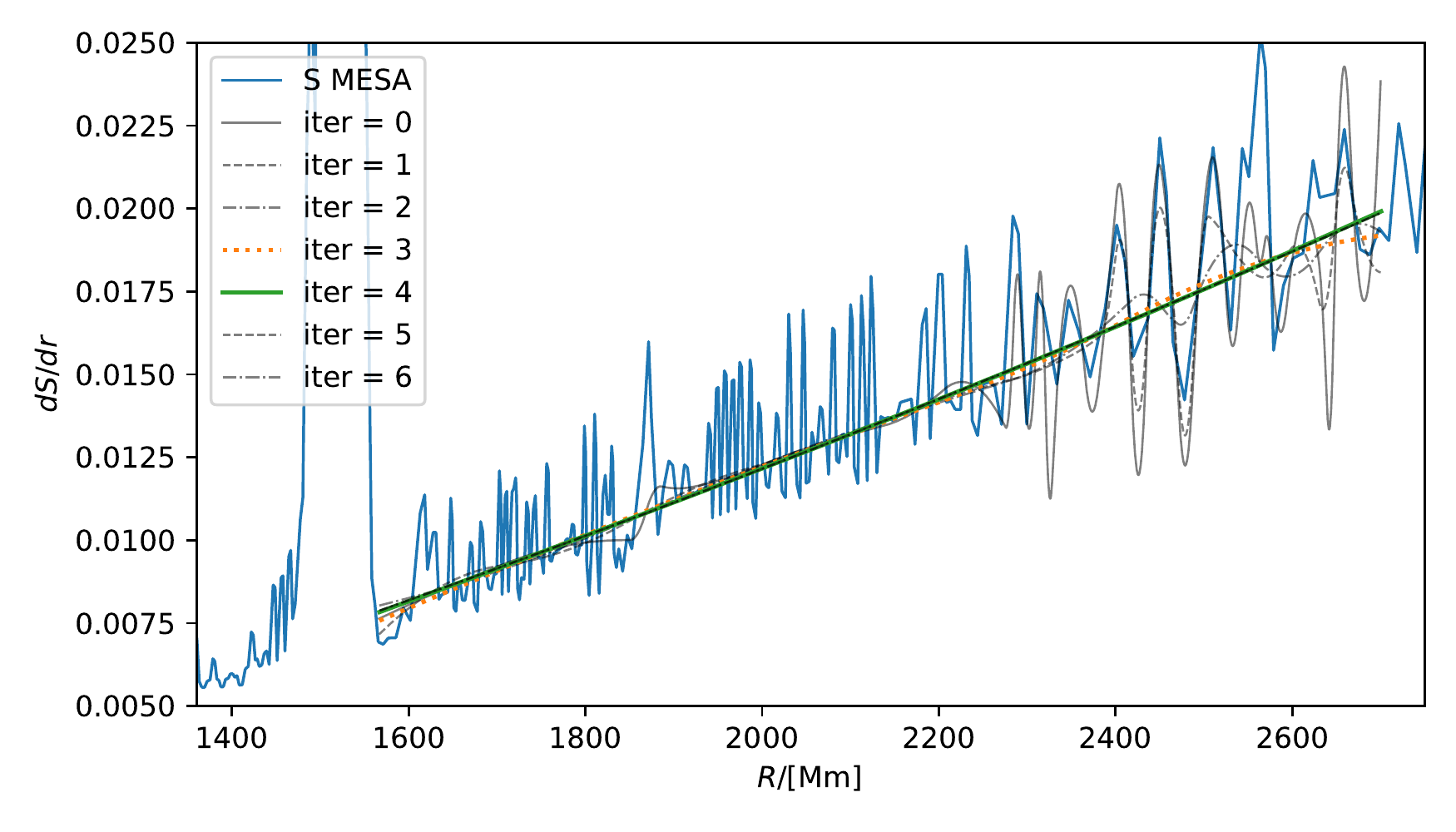}
  \caption{Filtering of \code{MESA} radial entropy derivative through recursive spline fitting, done separately for the core-envelope transition (top) and the envelope (bottom).}
  \lFig{dSdr-basestate}
\end{figure}
\begin{figure}
  \includegraphics[width=\columnwidth]{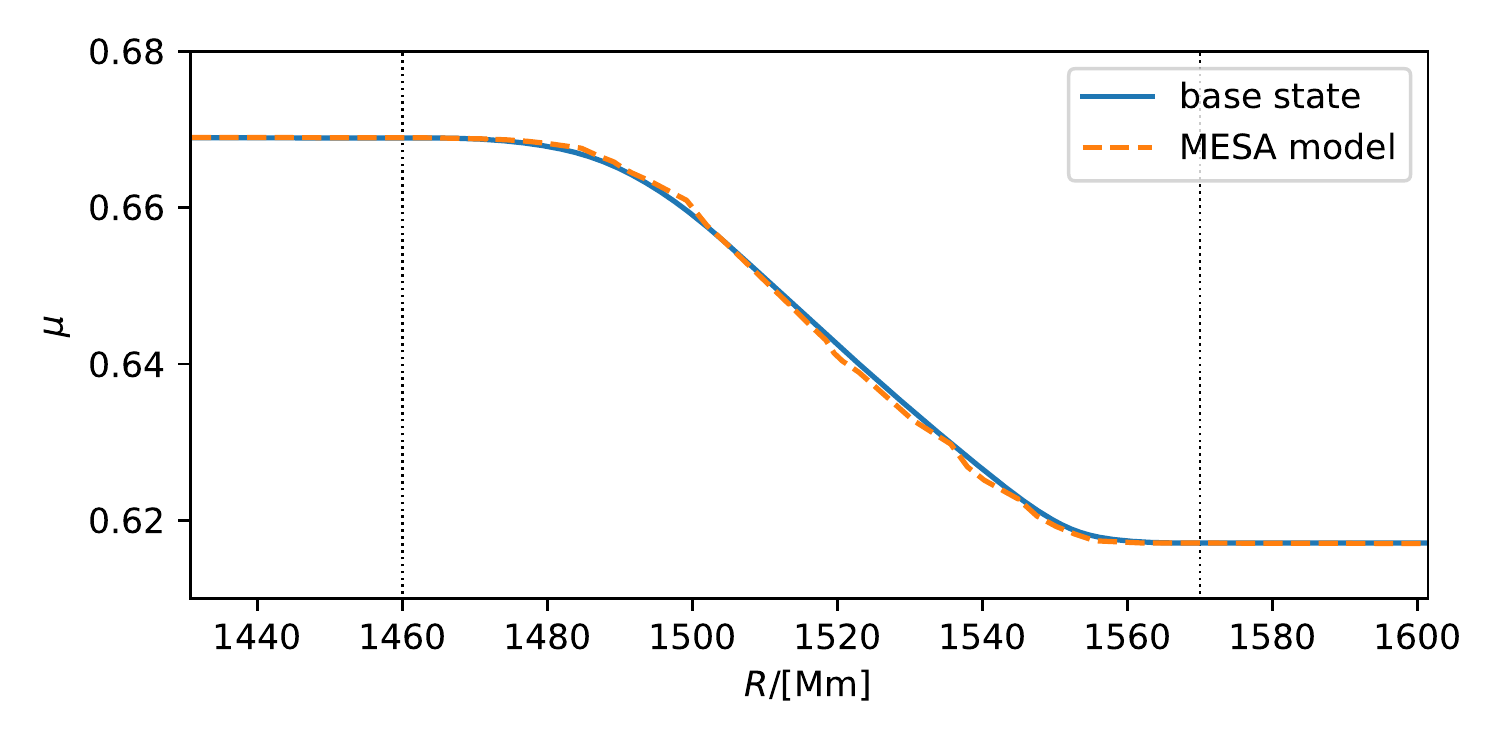}
  \includegraphics[width=\columnwidth]{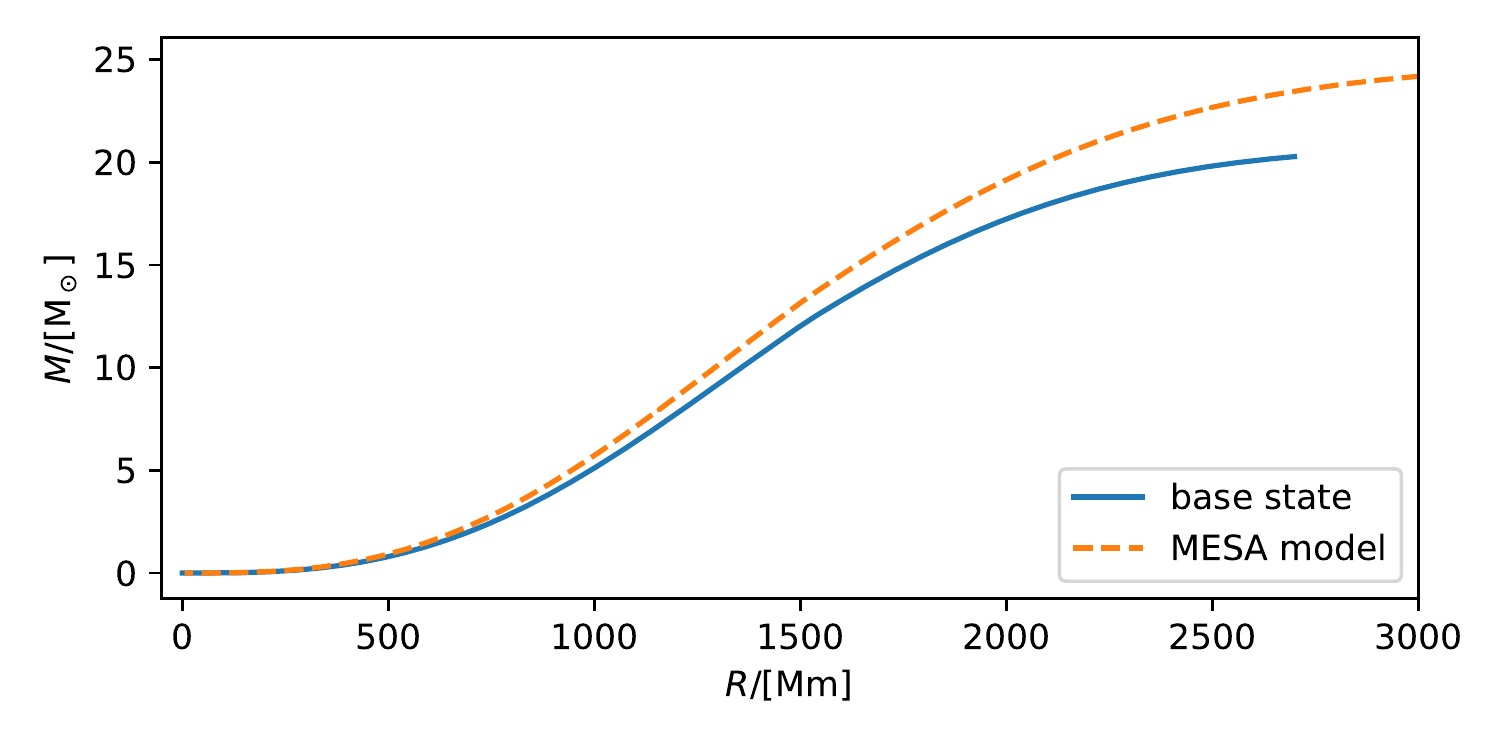}
  \includegraphics[width=\columnwidth]{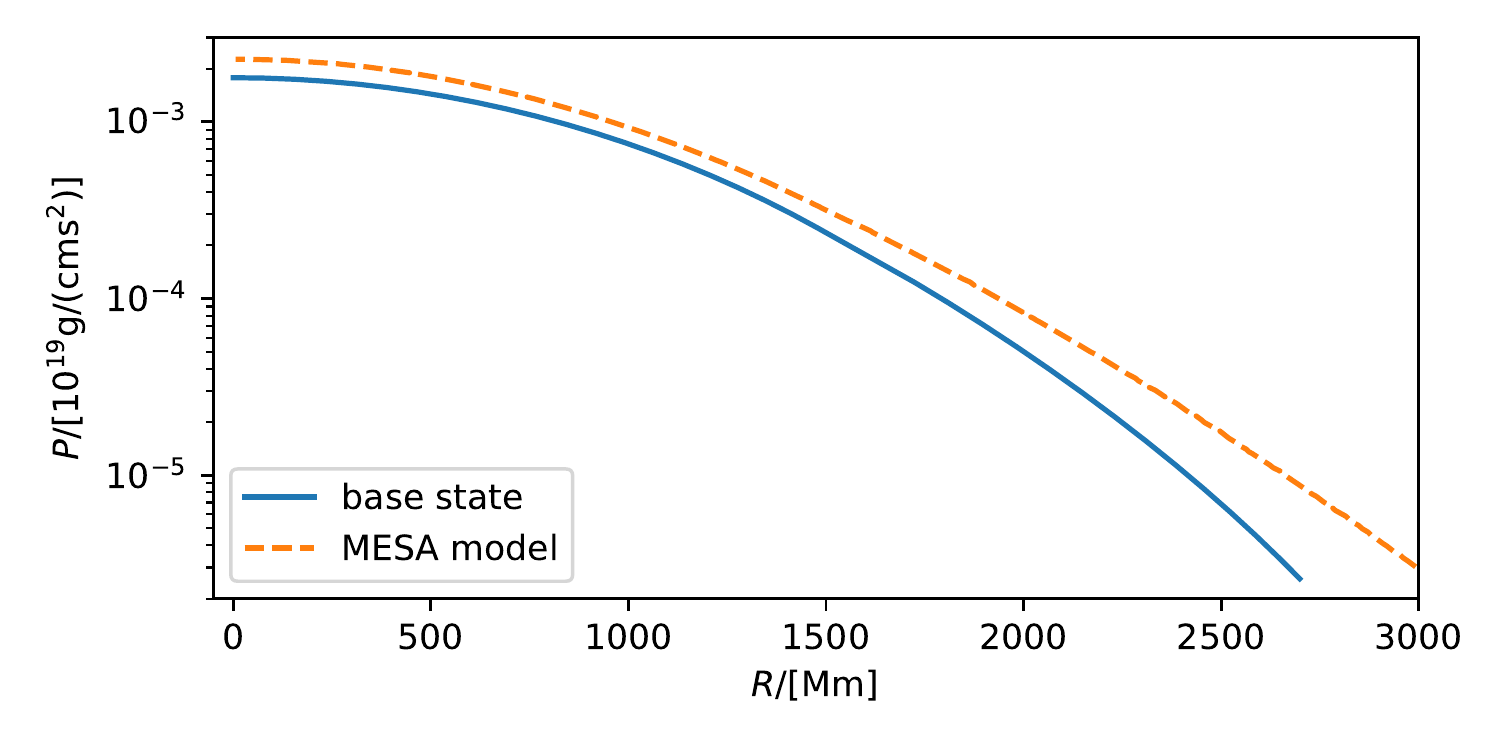}
  \includegraphics[width=\columnwidth]{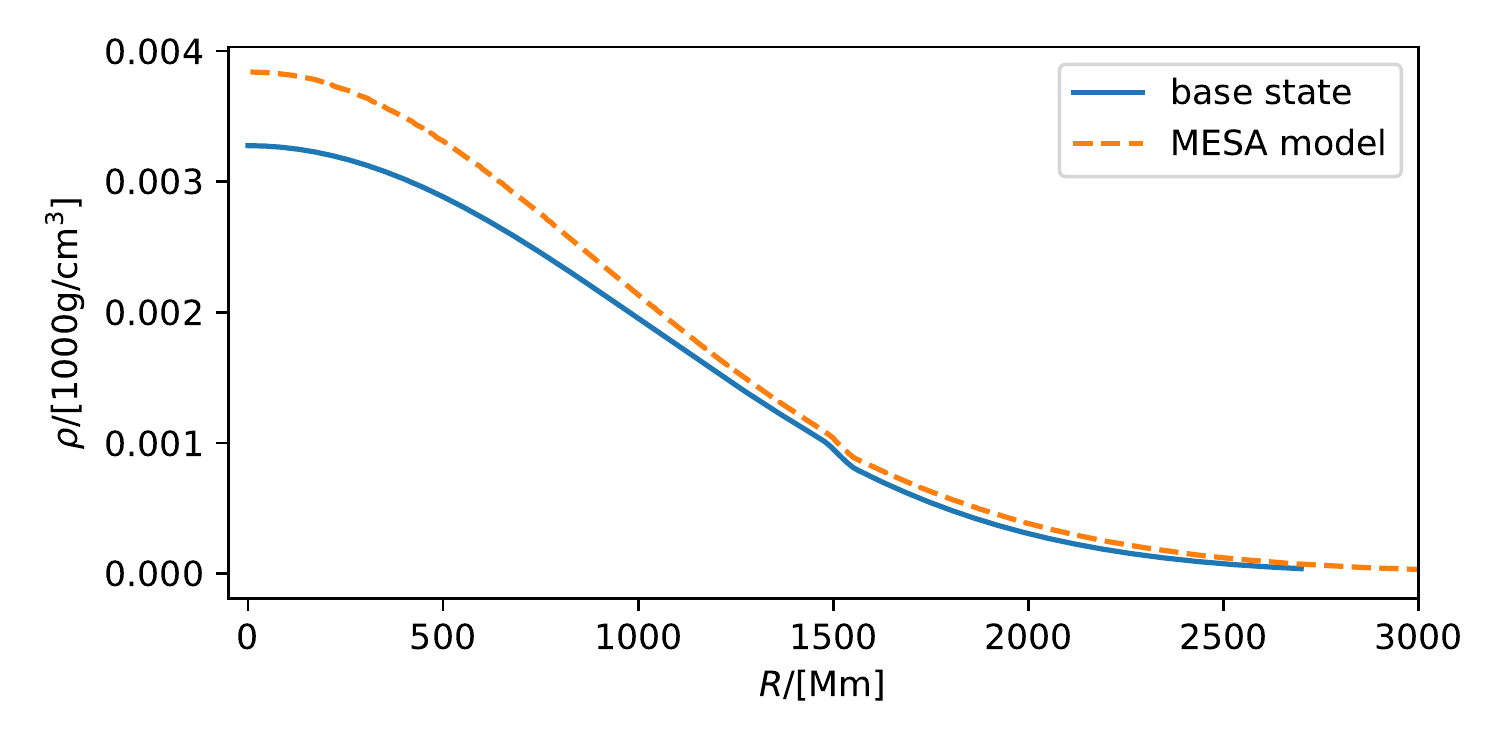}
  \includegraphics[width=\columnwidth]{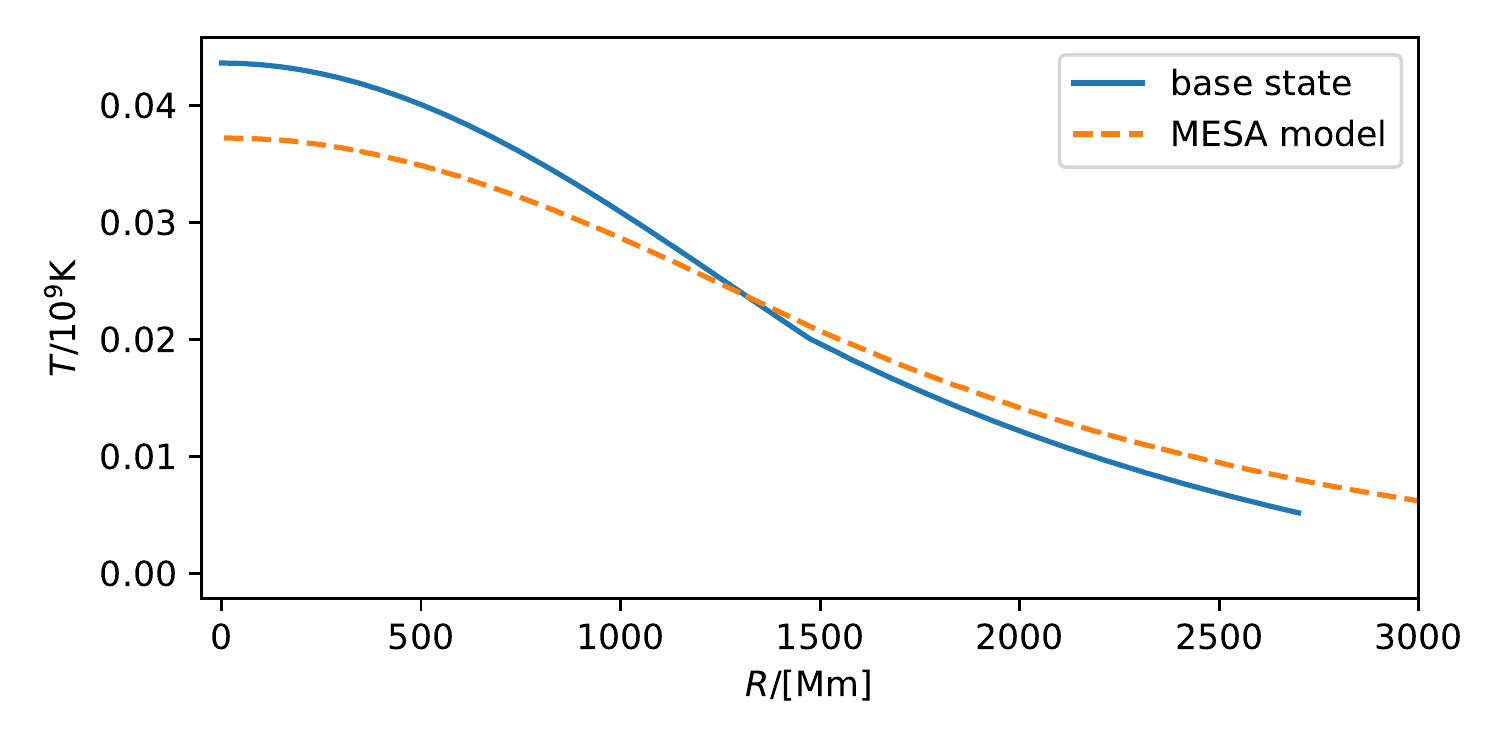}
  \caption{Comparison of adopted base state for the 3D simulations and
    the \code{MESA} radial profile. From top to bottom: mean molecular
    weight, mass, pressure, density, temperature. Except for mass,
    quantities are given in their code units. }
  \lFig{base-MESA-compare}
\end{figure}         

In order to remove short wavelength noise in the radial \code{MESA}
profile and obtain profiles with smooth first derivatives and
continuous second derivatives, we have tried a number of strategies to
filter the \code{MESA} entropy profile, including moving averages with
different filter width, fitting the boundary with up to three
Gaussians, or fitting with up to 20th-degree polynomials. None of
these methods was particularly satisfactory. We found that the best
way to preserve the overall shape of the radial entropy gradient
without introducing short wavelength oscillations or ringing is with
recursive spline fitting.

For up to $\approx 10$ iterations, the grid is moved by a small amount
and the entropy is interpolated onto this moved grid using the
\code{SciPy} method \code{interpolate.splrep}, with small values for
the smoothing condition $s$ and the spline function defined for the
previous grid. The best result is achieved when moving the grid by a
single-digit percentage $p_\mathrm{mv}$ of the interpolation grid. The
number of iterations $n_\mathrm{iter\_filt}$ is selected by visual
inspection with the goal to filter small wavelength noise from the
\mesa\ profile, retain the overall stratification, and not introduce
new spatial oscillations not present in the \mesa\ profile.

To filter the core-envelope transition $s=0.04$, $p_\mathrm{mv}=5\%$
and $n_\mathrm{iter\_filt}=4$ (\texttt{iter=3} in
\Fig{dSdr-basestate}), and for the envelope $s=0.1$,
$p_\mathrm{mv}=20\%$ and $n_\mathrm{iter\_filt}=5$
(\texttt{iter=4}). The exact values are sensitive to the specific
underlying \code{MESA} model and details of the interpolation
grid. The entropy gradient in the core is set to zero and a radius
range is determined by visual inspection in which the core solution is
stitched to the transition region ($R_\mathrm{stich\_env} \in
[1460,1490]$) and in turn to the envelope profile
($R_\mathrm{stich\_env} \in [1550,1570]$). The transition from one
solution to the other in the stitched region is modelled with
$\sin(x)$, $x\in[-pi/2, \pi/2]$, appropriately scaled.

For the $\mu$ profile, we model the transition to have in the
core-envelope transition the same radial gradient profile as the
entropy normalized by $\Delta \mu/\Delta S$, where each of these
quantities is the integral across the transition. The $\mu$ gradient
has to approach zero at the top of the core-envelope transition, while
the entropy gradient stays positive. This difference is accommodated
by adding another stitching radius range ($[1551,1560]$) at the top of
the core-envelope transition.  The resulting $\mu$ profile is shown in
comparison with the MESA profile in \Fig{base-MESA-compare} along with
the other state variables and the enclosed mass.

\section{Derivation of Ri}
\lSect{Richardson}
Here we show that \Eq{eq:Ri-def} is equivalent to the definition of Ri. In
\Eq{eq:Ri-def}, $N^2$ can be expressed as
\begin{equation}
    N^2 = \frac{g \delta}{H_P} \left( \nabla_{\rm ad} - \nabla + 
           \frac{\phi}{\delta} \nabla_{\mu} \right),
\end{equation}
where $\delta \equiv -\left. \frac{\partial \ln \rho}{\partial \ln T} \right|_{P,\mu}$,
$\phi \equiv \left.\frac{\partial \ln \rho}{\partial \ln \mu}\right|_{P,T}$, and the 
different gradients have their standard meanings. Using those definitions, we have
\begin{equation}
N^2 = \frac{g}{\rho}  \left( \left.\frac{\partial \rho}{\partial T} \right|_{P,\mu}   \left.\frac{\partial T}{\partial P}\right|_{S,\mu} \frac{dP}{dz} - \left.\frac{\partial \rho}{\partial T} \right|_{P,\mu} \frac{dT}{dz} -   \left.\frac{\partial \rho}{\partial \mu}\right|_{P,T} \frac{d \mu}{dz}  \right).
\lEq{eq:N2expand}
\end{equation}
The first term can be written as
\begin{equation}
\left.\frac{\partial \rho}{\partial T} \right|_{P,\mu}  \left.\frac{\partial T}{\partial P}\right|_{S,\mu} \frac{dP}{dz} = - \left.\frac{\partial \rho}{\partial S}\right|_{P,\mu} \left.\frac{\partial S}{\partial P}\right|_{T,\mu} \frac{dP}{dz},
\lEq{eq:N2partial1}
\end{equation}
where we have made use of the Maxwell relations 
$\left.\frac{\partial \rho}{\partial T}\right|_{P,\mu} = \rho^2 \left.\frac{\partial S}{\partial P}\right|_{T,\mu}$
and
$\left.\frac{\partial T}{\partial P}\right|_{S,\mu}=-\frac{1}{\rho^2} \left.\frac{\partial \rho}{\partial S}\right|_{P,\mu}$. The second term of \Eq{eq:N2expand} can be expresed as
\begin{equation}
     - \left.\frac{\partial \rho}{\partial T} \right|_{P,\mu} \frac{dT}{dz} = - \left.\frac{\partial \rho}{\partial S}\right|_{P,\mu} \left.\frac{\partial S}{\partial T}\right|_{P,\mu} \frac{dT}{dz},
\end{equation}
and the third term as 
\begin{equation}
    - \left.\frac{\partial \rho}{\partial \mu}\right|_{P,T}\frac{d\mu}{dz} = - \left.\frac{\partial \rho}{\partial S}\right|_{P,T} \left.\frac{\partial S}{\partial \mu}\right|_{P,T} \frac{d\mu}{dz}.
\lEq{eq:N2partial2}
\end{equation}
Now, since 
\begin{equation}
    \frac{dS}{dz} = \left.\frac{\partial S}{\partial P}\right|_{T,\mu} \frac{dP}{dz} + \left.\frac{\partial S}{\partial T}\right|_{P,\mu} \frac{dT}{dz} + \left.\frac{\partial S}{\partial \mu}\right|_{P,T} \frac{d\mu}{dz},
\end{equation}
substituting Eqs.~(\Eqref{eq:N2partial1})--(\Eqref{eq:N2partial2}) into \Eq{eq:N2expand} yields
\begin{equation}
    N^2 = \frac{-g (\partial \rho / \partial S)_P (dS/dz)}{\rho}.
\end{equation}
We can then see that \Eq{eq:Ri-def} is equivalent to the definition of Ri \citep[Eqn.\, 8.13,][]{shu:92}
\begin{equation}
{\rm Ri} \equiv \frac{-g (\partial \rho / \partial S)_P (dS/dz)}{\rho (dU/dz)^2}
 = \frac{N^2}{(dU/dz)^2}.
\end{equation}

\end{document}